\documentclass[journal]{IEEEtai}

\usepackage[colorlinks,urlcolor=blue,linkcolor=blue,citecolor=blue]{hyperref}

\usepackage{color,array}

\usepackage{graphicx}

\usepackage[switch]{lineno}
\usepackage{cite, caption}
\usepackage{amsmath,amssymb,amsfonts}
\usepackage{dsfont}
\usepackage{algorithmic}
\usepackage{graphicx}
\usepackage{textcomp}
\usepackage{multirow}
\usepackage{hhline}

\begin{document}

\title{Pathologist-Like Explanations Unveiled: an Explainable Deep Learning System for White Blood Cell Classification} 

\author{Aditya Shankar Pal, Debojyoti Biswas, Joy Mahapatra, Debasis Banerjee, Prantar Chakrabarti and Utpal Garain, \IEEEmembership{Member, IEEE}
\thanks{Aditya Shankar Pal, Debojyoti Biswas, Joy Mahapatra, and Utpal Garain are with the Indian Statistical Institute, Kolkata 700108, India (e-mail: adityashankarpal\_r@isical.ac.in, debojyotibiswas11@gmail.com, joymahapatra90@gmail.com, utpal@isical.ac.in). }
\thanks{Debasis Bandyopadhyay is with the Drs Tribedi \& Roy Diagnostic Laboratory, Kolkata, India (e-mail: debasisbanerjee.haematology@gmail.com)}
\thanks{Prantar Chakrabarti is with the Zoho Corporation, India and Centre for Digital Healthcare Technologies, MAKAUT, India (e-mail: prantar@gmail.com)}
\thanks{ \emph{Corresponding author: Aditya Shankar Pal}}}

\maketitle

\begin{abstract}
White blood cells (WBCs) play a crucial role in safeguarding the human body against pathogens and foreign substances. Leveraging the abundance of WBC imaging data and the power of deep learning algorithms, automated WBC analysis has the potential for remarkable accuracy. However, the capability of deep learning models to explain their WBC classification remains largely unexplored. In this study, we introduce \emph{HemaX}, an explainable deep neural network-based model that produces pathologist-like explanations using five attributes: granularity, cytoplasm color, nucleus shape, size relative to red blood cells, and nucleus to cytoplasm ratio (N:C), along with cell classification, localization, and segmentation. HemaX is trained and evaluated on a novel dataset, LeukoX, comprising 467 blood smear images encompassing ten (10) WBC types. The proposed model achieves impressive results, with an average classification accuracy of 81.08\% and a Jaccard index of 89.16\% for cell localization. Additionally, HemaX performs well in generating the five explanations with a normalized mean square error of 0.0317 for N:C ratio and over 80\% accuracy for the other four attributes. Comprehensive experiments comparing against multiple state-of-the-art models demonstrate that HemaX's classification accuracy remains unaffected by its ability to provide explanations. Moreover, empirical analyses and validation by expert hematologists confirm the faithfulness of explanations predicted by our proposed model.
\end{abstract}

\begin{IEEEImpStatement}
This study addresses the issue of explainability in white blood cell classification. The robust performance of HemaX in WBC classification, along with its ability to provide explanations, is expected to enhance pathologists' confidence in the reliability of this system. This study differs from previous research, which primarily evaluated Artificial Intelligence (AI) systems for WBC classification while often keeping their inner workings opaque. HemaX's key advantage lies in its unified approach, enabling it to seamlessly perform multiple tasks without requiring fundamental architectural changes. This sets HemaX apart from previous systems and underscores its versatility. The creation of the LeukoX dataset introduces significant enhancements to existing WBC datasets, which previously focused solely on classification and segmentation challenges. This new dataset for the explainable WBC classification task promotes innovation and advancement in the use of deep learning models for both medical and technological purposes.
\end{IEEEImpStatement}


\section{Introduction}
\label{sec:introduction}
\IEEEPARstart{W}{hite} blood cells (WBCs), also known as leukocytes, are a diverse group of nucleated blood cells responsible for defending the body against pathogens and foreign substances. The normal range of leukocytes in the blood typically falls between 4,000 to 10,000 per microliter~\cite{walker1990clinical}. Deviations from these levels can indicate various illnesses such as lymphoma, leukemia, and viral infections, guiding treatment decisions. Therefore, the analysis of leukocytes is crucial for maintaining and diagnosing health conditions.\par
Conventional blood smear analysis requires pathologists to meticulously examine each region under a microscope, making it a laborious and time-consuming task prone to human errors~\cite{rezatofighi2011automatic}. Machine learning-based approaches for leukocyte classification have emerged to address these challenges~\cite{gomez2008feature,ushizima2005support,kulkarni2013classification}. These traditional machine learning-based techniques typically involve several stages: pre-processing, segmentation, feature extraction, and classification~\cite{gupta2019optimized,tavakoli2021new,khashman2008ibcis,wang2016spectral}. This framework suffers from two major drawbacks: (i) human-engineered feature extraction method and (ii) any errors in the early modules badly affect the final classification accuracy.\par
In recent years, deep learning models have made manual feature extraction obsolete and demonstrated promising results in medical image analysis, showing improved accuracy and robustness~\cite{li2023rethinking,abubaker2022detection,mitani2020detection}. Numerous deep learning-based models have also been developed for white blood cell analysis~\cite{shahin2019white,yao2021classification,al2021improving,zhou2023cuss,li2023deep,tahiri2023white,leng2023deep,tougaccar2020classification,han2023one}. However, a notable limitation of deep learning models is their inability to offer human-interpretable explanations for their predictions~\cite{rawal2021recent,tizhoosh2018artificial,wu2021interpretable,Niazi2019-td}. Consequently, clinicians may lack confidence in a diagnostic system that cannot provide explanations when required~\cite{esteva2019guide}.\par
\begin{figure}[ht!]
\centering
\includegraphics[width=\linewidth]{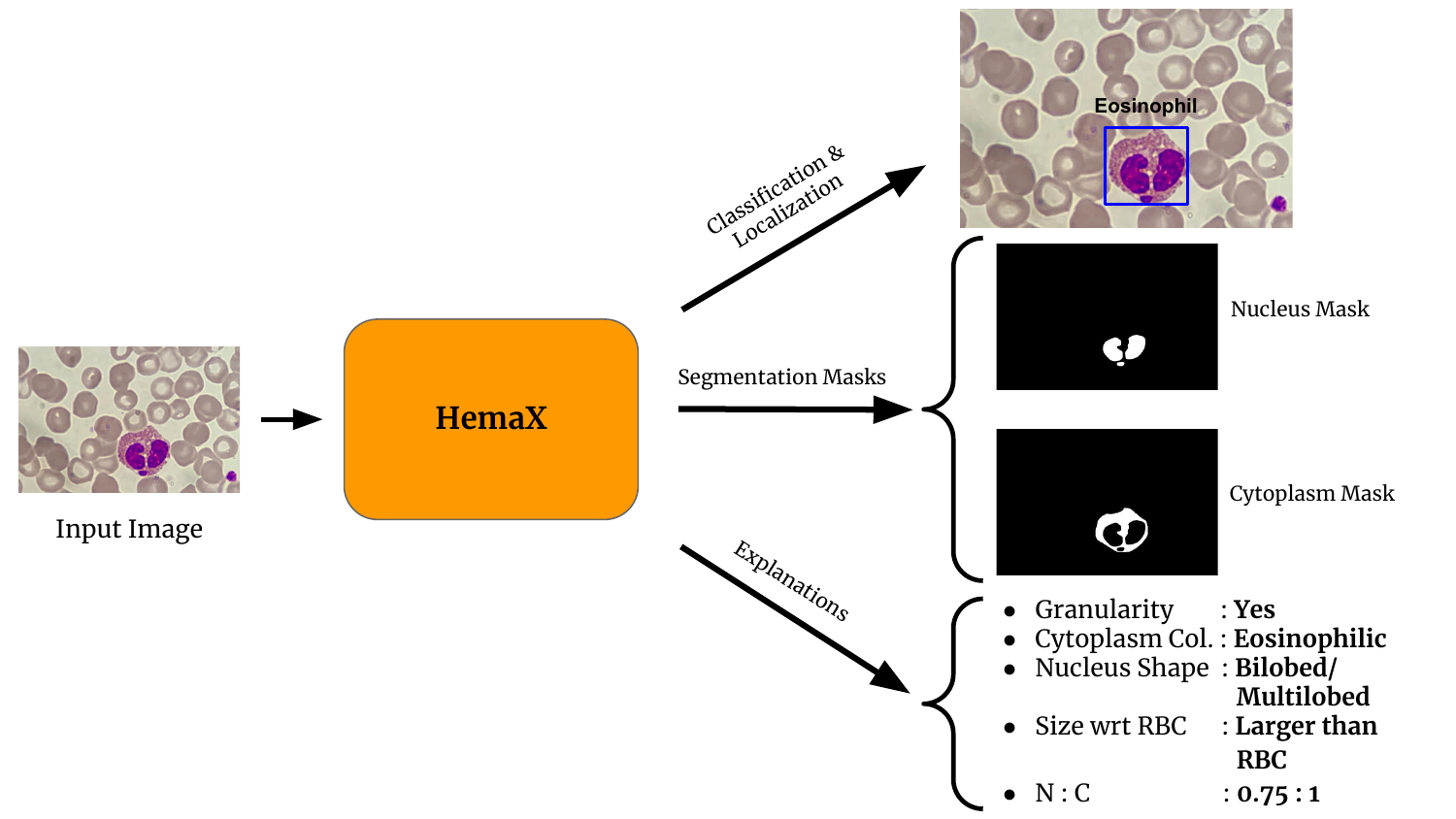}
\caption{Objectives achieved by HemaX}
\label{fig:HemaX_task}
\end{figure}

In this study, we propose an explainable end-to-end model, called \textit{HemaX}, to bridge the gap between the deep learning paradigm and medical professionals by providing them with insights into the model's decision-making process. Unlike existing models, \textit{HemaX} can simultaneously produce pathologist-like explanations for five attributes, along with cell classification and localization. This unique feature sets \textit{HemaX} apart and shows its potential to improve the accuracy and efficiency of blood cell categorization, thus assisting medical practitioners in making more precise diagnoses. Figure~\ref{fig:HemaX_task} illustrates the capabilities of \textit{HemaX} upon receiving an input image. It efficiently locates the cell, provides segmentation with cytoplasm and nucleus masks, and generates a comprehensive report, offering cell type information and values for the five attributes commonly used by pathologists to classify WBCs. 

The proposed model is trained and evaluated on a newly developed explainable dataset focused on leukocytes. While many publicly available datasets on WBC imaging data address classification and segmentation, they often lack in appropriate annotations for tasks like generating/ predicting explanations. To address this gap, we introduce \textit{LeukoX}, a novel dataset created in collaboration with a renowned diagnostic laboratory (Section~\ref{sec:leukox}). Experiments on the \textit{LeukoX} dataset demonstrate that \textit{HemaX} achieves a classification accuracy, precision, and $F_1$ score of $0.8108, 0.8351,$ and $0.8089$, respectively, outperforming other competing classifiers such as Faster RCNN~\cite{ren2015faster}, CUSS-Net~\cite{zhou2023cuss}, and Tavakoli's algorithm~\cite{tavakoli2021new} significantly. We have also shown the generalizability of LeukoX on two publicly available datasets namely LISC~\cite{rezatofighi2011automatic} and Raabin~\cite{kouzehkanan2022large}. In brief, the significant contributions presented in this paper are as follows:
\begin{enumerate}
    \item We propose a transformer encoder-decoder based model, \textit{HemaX}, for simultaneously classifying, segmenting and explaining the cell types of white blood cells. 
    \item We compiled the colored images of peripheral blood smear from 10 cell types into a new dataset, LeukoX. The LeukoX dataset contains information regarding the cell type, bounding box location, segmentation mask and pathologist-like explanations for five attributes associated with each cell present in it.
    \item Experiments conducted on LeukoX, LISC and Raabin dataset show that \textit{HemaX} performs above par compared to the other state-of-the-art classifiers.
    \item Empirical analysis and expert evaluation further validate the results and establish the faithfulness of \textit{HemaX}.
\end{enumerate}

The remaining portion of the paper is structured as follows. The development of the LeukoX dataset is outlined in Section~\ref{sec:leukox}. A comprehensive description of the proposed HemaX model, along with a technique for establishing faithfulness, is provided in Section~\ref{sec:methodology}. Subsequently, Section~\ref{sec:results} offers insights into the baseline implementation, the process of selecting the best HemaX model, and its comparison across the LeukoX, LISC, and Raabin datasets with various state-of-the-art models. Additionally, the section also engages in a discussion of the results, and establishes the faithfulness of the explanations predicted by the proposed model as well as their verification by an expert hematologist. Finally, the paper is concluded in Section~\ref{sec:conclusion}.

\section{LeukoX Dataset}\label{sec:leukox}
We have developed \textit{LeukoX}, a dataset containing peripheral blood smear (PBS) images in collaboration with Drs. Roy and Tribedi Diagnostic Laboratory Kolkata\footnote{https://www.tribediandroy.com/}. The dataset is prepared under ethical clearance from the Ethics Committee, Fortis Hospital\footnote{https://www.fortishealthcare.com/india/fortis-hospital-in-anandapur-kolkata-west-bengal}. The dataset comprises peripheral blood smear (PBS) images stained with the Leishman Giemsa procedure, acquired at $100x$ magnification using the wide-angle camera with a 12-megapixel f/1.8 aperture on the iPhone XR. \textit{LeukoX} contains cells from ten distinct WBC types: \textit{neutrophils}, \textit{lymphocytes}, \textit{eosinophils}, \textit{monocytes}, \textit{basophils}, \textit{band cells}, \textit{metamyelocytes}, \textit{myelocytes}, \textit{promyelocytes}, and \textit{blast cells}. The images come in two resolutions: $4032 \times 3024$ and $1024 \times 768$. The dataset includes 50 PBS images for each of the nine WBC types and 17 PBS images for blast cells. The cells are manually annotated by an expert hematologist, and segmentation masks for the nucleus and cytoplasm regions are generated using the Computer Vision Annotation Tool (CVAT)\footnote{https://github.com/opencv/cvat}. Additionally, each sample is annotated with five attributes characterizing the cell type: granularity (`yes' or `no'), cytoplasm color (`eosinophilic' or `basophilic'), nucleus shape (`horseshoe-shaped/kidney bean', `bilobed/multilobed', or `round/oval'), size with respect to red blood cells (RBCs) (`larger', `nearly similar', or `smaller'), and the nucleus to cytoplasm ratio (N:C ratio). The N:C ratio is determined by calculating the number of pixels in the nucleus region to the total number of pixels in the cytoplasm region. \par
The construction of LeukoX dataset introduces several noteworthy strengths. First, it capitalizes on the capabilities of smartphone cameras to capture WBC images, aligning with the pursuit of cost-effective diagnostics as a viable avenue. The widespread adoption of smartphones in society has precipitated their integration into the realm of medical diagnostics, marking an expanding trend~\cite{ozcan2014mobile,kwon2016medical}. The ubiquity and affordability of high-pixel-density smartphone cameras have garnered recognition for their imaging capabilities across diverse scientific domains~\cite{kim2017smartphone,jung2015smartphone,de2020smartphone,dutta2015dye}. Second, the dataset developed as part of this research, offered under open data licensing, contributes further to the realm of explainable artificial intelligence (XAI) in medical imaging. The availability of such a dataset could potentially foster innovation, contributing to the advancement of both medical and technological applications. 

\section{Methodology} \label{sec:methodology}
In this section, we present our novel system, HemaX and set forth a technique for establishing the faithfulness of explanations.

\subsection{\textit{HemaX}} \label{section:proposed-model}
\begin{figure*}[ht!]
\centering
\includegraphics[width=0.8\linewidth]{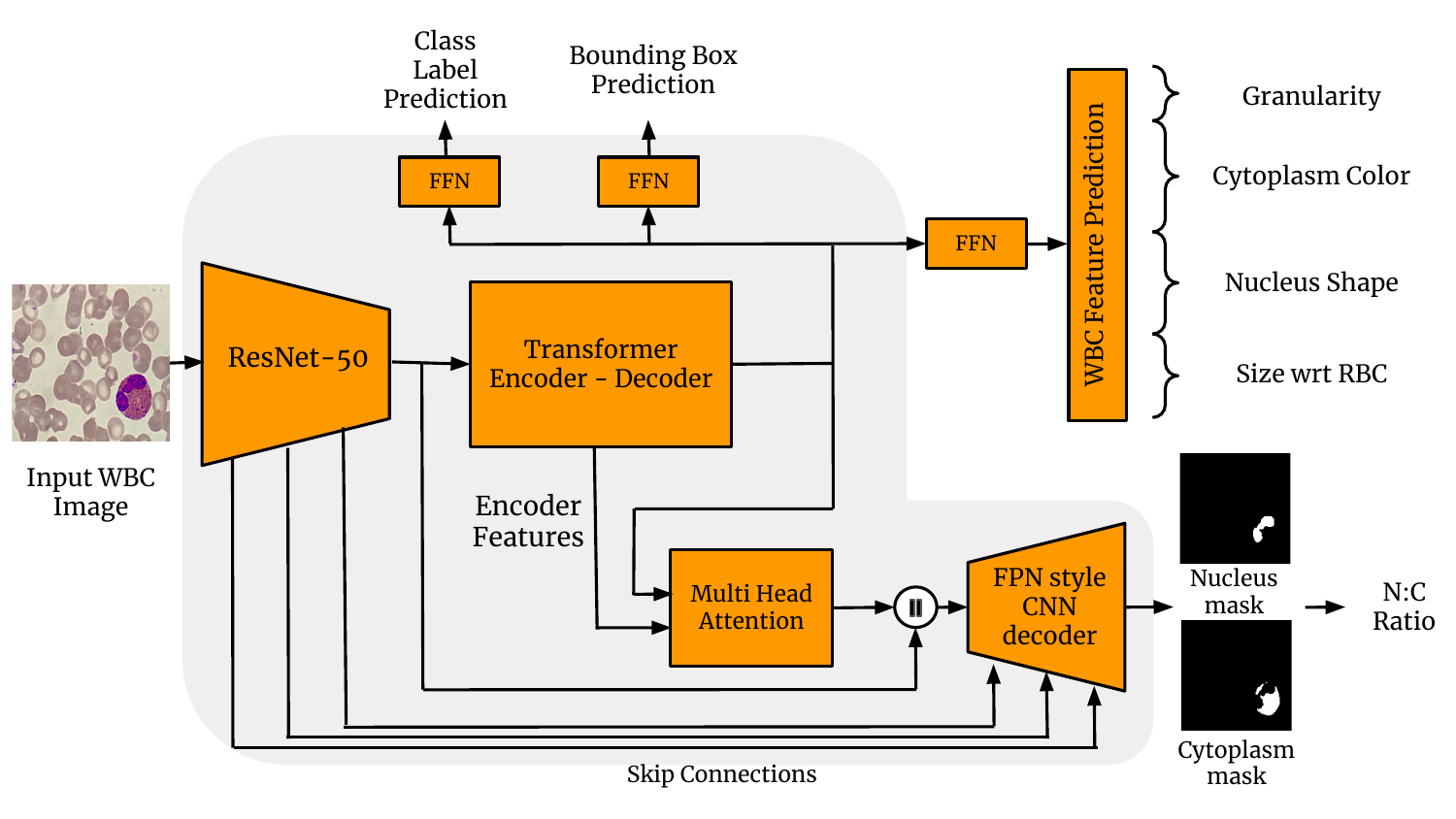}
\caption{Architecture of HemaX, an end-to-end model for explainabilty of WBC. The transformer encoder-decoder attends to the image features extracted by the ResNet-50 backbone and forwards it to the FFNNs. The FFNNs perform classification, localization, and feature prediction simultaneously. The segmentation maps are computed separately for the nucleus and cytoplasm region. The highlighted region in the figure represents the DETR panoptic segmentation architecture.}
\label{HemaX}
\end{figure*}
Transformer-based models have shown impressive and elegant results in several medical image analysis tasks~\cite{chen2023joint,kiyasseh2023vision,phan2022sleeptransformer,shokouhmand2023diagnosis,wang2022medical,you2021aligntransformer}. 
The proposed HemaX model extends the DETR panoptic segmentation~\cite{carion2020end}.
HemaX (Figure~\ref{HemaX}) yields three outputs similar to DETR (class labels, bounding box coordinates and a segmentation mask) as well as an extra branch that outputs explanations corresponding to the predicted class.
HemaX adopts the components of DETR: a ResNet-50 model pre-trained on Imagenet weight as the backbone convolutional neural network (CNN) encoder, six transformer encoder-decoder blocks, and the two prediction heads for class prediction and bounding box detection. To obtain the segmentation mask we have modified the feature pyramid network (FPN) based CNN decoder to predict a two-channel segmentation mask for each of the $N$ object queries. The two channels contain the predicted cytoplasm and nucleus mask, respectively. The N:C ratio is computed based on the number of pixels occupied by the nucleus and cytoplasm in their corresponding segmentation masks. We have also redirected the transformer decoder's output to a feed-forward network (FFN) that predicts other four explanations---granularity, cytoplasm color, nucleus shape, and size with respect to red blood cells.\\
\hspace*{0.5cm}\textit{HemaX} employs a composite loss function$(\mathcal{L})$, consisting of four losses (prediction loss, bounding box loss, segmentation loss and \emph{explanation loss}). For a given image, \textit{HemaX} produces N predictions where N is larger than the number of groundtruthed objects in the image. 
Initially, the groundtruthed set of objects is optimally matched to a predicted set with a permutation index $\sigma_i$ ($\sigma \in \mathfrak{S}_N$) using Hungarian algorithm\cite{carion2020end,kuhn1955hungarian}. The class prediction and the similarity between predicted and groundtruthed boxes are both taken into consideration by the matching cost. Following this, we calculate the composite loss of the model.  
The composite loss function $\mathcal{L}$ is given in Equation~\ref{eqn:overall_loss} where $\varnothing$ denotes the empty class.
\begin{equation}\label{eqn:overall_loss}
    \mathcal{L} = \sum_{i=1}^{N}\{\mathcal{L}_p + \mathds{1}_{\{p_i \neq \varnothing\}} \mathcal{L}_b + \mathds{1}_{\{p_i \neq \varnothing\}} \mathcal{L}_s + \mathds{1}_{\{p_i \neq \varnothing\}} \mathcal{L}_e \}
\end{equation}
The first term of the loss is the prediction loss $\mathcal{L}_p$. Let $c_i$ be the true object class and $\hat{p}_{\sigma_i}(c_i)$ be the probability of class $c_i$ for the prediction with index $\sigma_i$. The prediction loss is given by $\mathcal{L}_p = -\log\hat{p}_{\sigma_i}(c_i)$. The second term, i.e., the bounding box loss $\mathcal{L}_b$ is the linear combination of two losses: L1 and generalized intersection-over-union(GIoU) loss~\cite{rezatofighi2019generalized}. The bounding box loss between the true bounding box $b_i$ and the predicted bounding box $\hat{b}_{\sigma_i}$ is defined as $\mathcal{L}_b(b_i,\hat{b}_{\sigma_i}) = w_{GIoU}\mathcal{L}_{GIoU}(b_i,\hat{b}_{\sigma_i}) + w_{L1}|| b_i - \hat{b}_{\sigma_i} ||_1$, where $w_{GIoU}$ and $w_{L1}$ denote the weights of GIoU loss and L1 loss respectively. The third term, segmentation loss $\mathcal{L}_s$, is the weighted sum of two losses namely focal loss(FL) and S{\o}rensen-Dice loss(dice)~\cite{jadon2020survey}. The segmentation loss between actual mask $s_i$ and predicted mask $\hat{s}_{\sigma_i}$ is measured by $\mathcal{L}_s(s_i, \hat{s}_{\sigma_i}) = w_{dice}\mathcal{L}_{dice}(s_i, \hat{s}_{\sigma_i}) + w_{FL}\mathcal{L}_{FL}(s_i, \hat{s}_{\sigma_i})$, where $w_{dice}$ and $w_{FL}$ denote the weights of S{\o}rensen Dice loss and focal loss respectively. 
To compute the fourth term, i.e, explanation loss, let $E = (a_1, a_2, a_3, a_4)_{i=1}^{N}$ be the groundtruthed explanation, where $a_k$ denotes an attribute value and $A_k$ denotes the range of values for the $k$-th attribute. Note that four attributes namely, granularity, cytoplasm color, nucleus shape and size w.r.t red blood cells are considered here and the fifth attribute (i.e., N:C ratio) of the explanation is computed from the segmentation mask and it is not part of $\mathcal{L}_e$. Let $\hat{E}_{\sigma_i}= (\hat{a}_1, \hat{a}_2, \hat{a}_3, \hat{a}_4)_{i=1}^{N}$ denote the set of predicted explanations. The explanation loss $\mathcal{L}_e(E_i, \hat{E}_{\sigma_i})$ is given as the weighted sum of attribute-wise cross-entropy loss (Equation~\ref{eqn:feature_loss}) where $w_i$ represents the weight corresponding to the $i$-th attribute. 
\begin{equation}\label{eqn:feature_loss}
\mathcal{L}_e(E_i, \hat{E}_{\sigma_i})=\sum_{i=1}^{4} w_i \Bigg(-\sum_{a_j \in A_i} a_j \log\hat{a}_j\Bigg)
\end{equation}

\subsection{Faithfulness of HemaX's predictions} \label{faithfulness}
Aside from the evaluation of performance using various metrics, we also emphasise on generating `faithful' explanations aligned with the classifications. Our evaluation of explanation faithfulness involves comparing the model-induced association $\hat{A}$ (between the classification output and the explanation produced/induced by our model) with the ground truth association $A$ (between the classification target and the true explanation from the dataset).
This assessment of faithfulness draws inspiration from the concept of strong rules proposed by Agrawal et al.~\cite{Agrawal1993Association}. Associations are characterized through conditional probability distributions ($\operatorname{Pr}(\cdot \vert \cdot)$) of explanations (as described in Equation\ref{eqn:ass_1} and~\ref{eqn:ass_2}). 

\begin{align}
    \hat{A} &\triangleq p_{\hat{A}} = \operatorname{Pr}(E|C=\hat{c}) \label{eqn:ass_1}\\
    A &\triangleq p_A = \operatorname{Pr}(E|C=c) \label{eqn:ass_2}
\end{align}

\noindent where, $E$ is the explanation, $\hat{c}$ is the predicted output from \textit{HemaX} and $c$ is the ground truth. 
When dissimilar associations ($p_{\hat{A}}\not\approx p_A$) arise, the faithfulness of explanations produced by our model is significantly compromised.

\section{Results and Discussion}\label{sec:results}
In this section, we provide the details regarding baselines, datasets, training particulars and the evaluation metrics used. Additionally, we outline the process of selecting the best HemaX model and its subsequent comparison with the other competing systems. We also present the plots to establish the faithfulness of explanations as well as show the independence between predicted cell type and explanation.
\begin{figure*}[ht!]
\begin{center}
\begin{minipage}{0.7\textwidth}
\centering
\includegraphics[width=1\linewidth]{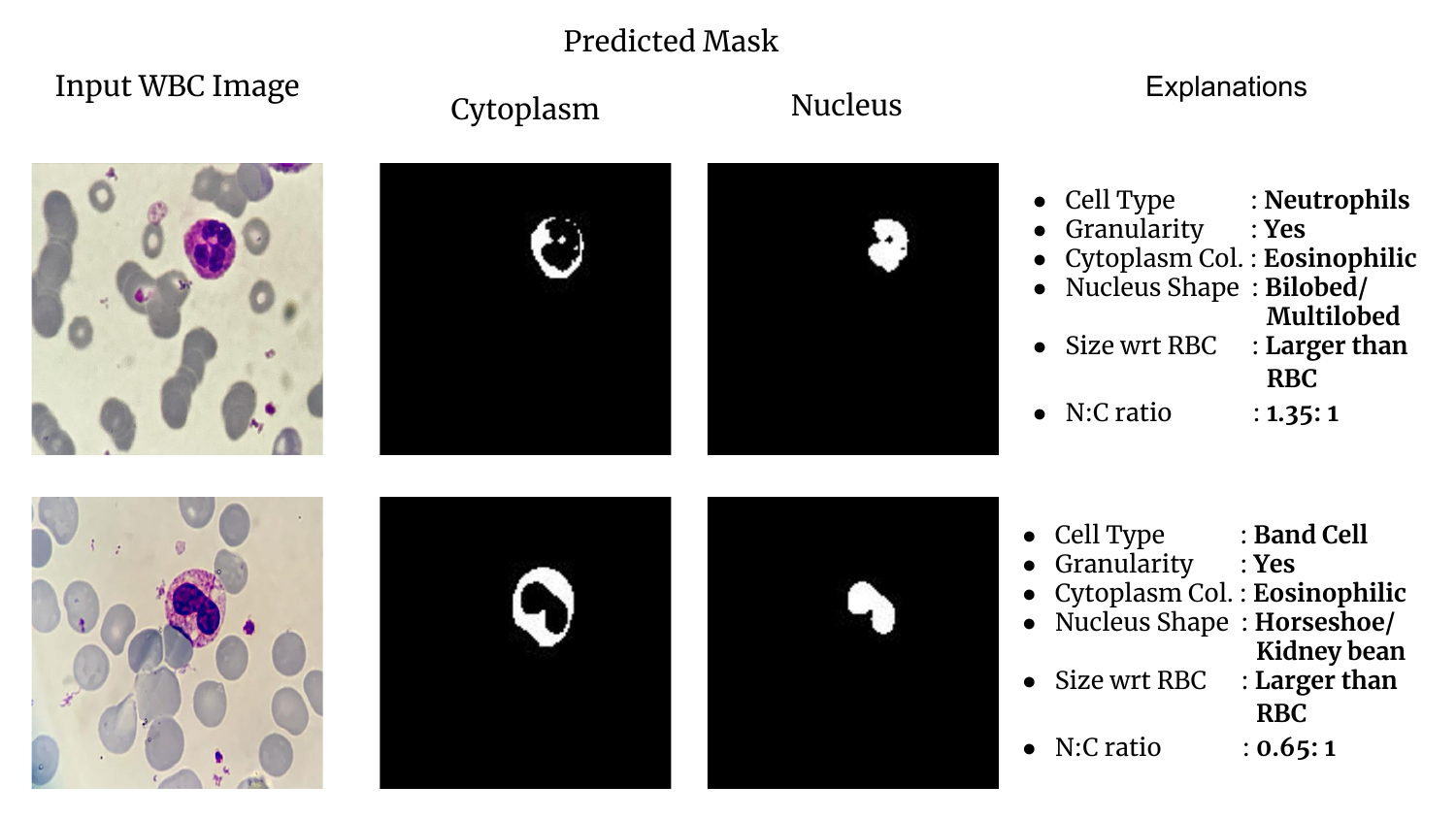}
\caption{Output on two WBC images using HemaX with no hidden layer.}
\label{fig:HemaX_ouput}
\end{minipage}
\end{center}
\end{figure*}
\subsection{Baseline Implementation Details}
The proposed network, HemaX, and all the other networks (Faster RCNN and CUSS-Net) are implemented in PyTorch and are trained using a single Nvidia RTX A6000 GPU with 48GB memory. HemaX is trained using AdamW optimizer~\cite{loshchilov2017decoupled} with a weight decay of $1e-4$ for $N=10$ object queries. The learning rate of the backbone CNN encoder, initialized with ResNet 50 model pre-trained on the ImageNet dataset, is set at $1e-5$, whereas the remainder of the model is trained using a learning rate of $1e-4$. Faster RCNN is trained by initializing its backbone with a ResNet 50 model pre-trained on the ImageNet dataset. The network is trained using the Stochastic Gradient Descent (SGD) optimizer with a learning rate of $5e-3$. CUSS-Net is trained using AdamW optimizer with a learning rate of $1e-4$. In our case, we have used CUSS-Net without the unsupervised-based strategy module. All three networks: HemaX, Faster RCNN, and CUSS-Net, are trained for $200$ epochs with a batch size of $32$ and the model corresponding to the lowest validation loss is chosen for evaluation. Tavakoli's algorithm is implemented by selecting the polynomial kernel for the support vector machine (SVM) classifier. 

\subsection{Datasets and Training Details} \label{sec:dataset_training}
In our study, we primarily utilized three datasets. Let's begin with the LeukoX dataset, the details of which have already been discussed in the Section~\ref{sec:leukox}. The $467$ images in the LeukoX dataset were divided randomly into training and test sets, with an $80\%$-$20\%$ ratio.\par
Moving on to the \emph{LISC} dataset~\cite{rezatofighi2011automatic}, it comprises $250$ blood smear images annotated by a single expert. These slides were stained using the Gismo-Right technique and captured using a light microscope with a $100$ magnification. For our experiments, we partitioned the LISC images into training and test sets, using a $90\%$-$10\%$ ratio.\par

The \emph{Raabin} dataset~\cite{kouzehkanan2022large} encompasses $1145$ images annotated with nuclei and cytoplasm details. Blood films were stained using the Giemsa technique and captured using an Olympus CX18 microscope with a $100$ magnification. Cameras from Samsung Galaxy S5 and LG G3 were used for image capture. Out of the total images, $233$ were designated as test images, while the rest were allocated for training purposes.\par

For assessing our models' performance, we employed five-fold cross-validation. Additionally, to expand the training dataset and introduce variety, we utilized standard data augmentation techniques, including scaling, rotation, translation, horizontal flipping, and vertical flipping.\par

\subsection{Evaluation Metrics}\label{sec:eval_mets}
We assess the performance of \textit{HemaX} across various aspects, including classification, bounding box localization, segmentation mask, and explanation prediction. For the classification evaluation, we utilize precision, $F_1$-score, and accuracy metrics. To evaluate bounding boxes, we calculate the Jaccard index~\cite{jaccard1901distribution}, averaging it over each bounding box instance. For the segmentation masks, we use the S{\o}rensen dice coefficient~\cite{Srensen1948AMO} to evaluate each object instance's quality and then average it over all object instances. Regarding the quality of explanations, we employ the accuracy measure for categorical attributes (e.g., granularity, cytoplasm color, nucleus shape, and size in relation to red blood cells). Additionally, we use the mean square error (MSE) to assess the accuracy of the N:C ratio explanation.

\subsection{Variants of HemaX} \label{sec:var_hmx}
\begin{table}[ht!]
\begin{minipage}[c]{\linewidth}
\resizebox{\textwidth}{!}{
\begin{tabular}{|c|c|c|c|}
\hline
\multirow{2}{*}{\textbf{Metrics}} &  \multicolumn{3}{c|}{\textbf{Number of Hidden Layers}} \\
\hhline{~---}
 &$0$ &$2$ &$4$ \\ 
\hline
\hline
\textbf{Dice Score}  & \textbf{0.9270 $\pm$  0.0020} & 0.9213 $\pm$ 0.0056 & 0.9266 $\pm$ 0.0022 \\
\hline
\textbf{Jaccard Ind.}& \textbf{0.8916 $\pm$ 0.0160}  & 0.8878 $\pm$ 0.0084 & 0.8871 $\pm$ 0.0130 \\
\hline
\textbf{Class. acc.} & \textbf{0.8108 $\pm$ 0.0161}  & 0.7978 $\pm$ 0.0258 & 0.8065 $\pm$ 0.0180 \\
\hline
\textbf{Precision}   & 0.8351 $\pm$ 0.0088           & 0.8225 $\pm$ 0.0246 & \textbf{0.8391 $\pm$ 0.0178}  \\ 
\hline 
\textbf{F1 score}    & \textbf{0.8089 $\pm$ 0.0162}  & 0.7939 $\pm$ 0.0253 & 0.8060 $\pm$ 0.0185   \\
\hline
\end{tabular}}
\caption{The scores for segmentation and classification metrics on variants of HemaX.} \label{tab:cls_var_hmx}
\end{minipage}
\begin{minipage}[c]{\linewidth}
\resizebox{\textwidth}{!}{
\begin{tabular}{|c|c|c|c|}
\hline
\multirow{2}{*}{\textbf{Attributes}} &  \multicolumn{3}{c|}{\textbf{Number of Hidden Layers}} \\
\hhline{~---}
 &$0$ &$2$ &$4$ \\ 
\hline
\hline
\textbf{Gran. acc.}         & \textbf{0.9355 $\pm$ 0.0096} & 0.9204 $\pm$ 0.0086 & \textbf{0.9355 $\pm$  0.0167} \\
\hline
\textbf{Cytoplasm col. acc.}& 0.8258 $\pm$ 0.0219 & \textbf{0.8366 $\pm$  0.0322} & 0.8172 $\pm$  0.0353 \\
\hline
\textbf{Nucleus shape acc.} & \textbf{0.8495 $\pm$ 0.0326} &  0.8430 $\pm$ 0.0277  & 0.8194 $\pm$ 0.0315 \\
\hline
\textbf{Size wrt RBC acc.}  & 0.8710 $\pm$ 0.0180 &  0.8753 $\pm$ 0.0146 & \textbf{0.8839 $\pm$ 0.0230}  \\ 
\hline
\textbf{N:C MSE}            & 0.0317 $\pm$ 0.0028 & 0.0316 $\pm$ 0.0023 & \textbf{0.0312 $\pm$ 0.0028}   \\
\hline
\end{tabular}}
\caption{The accuracy and MSE scores of various explanations on variants of HemaX.}\label{tab:exp_var_hmx}
\end{minipage}
\end{table}

\begin{figure*}[!htb]
   \begin{minipage}{0.33\linewidth}
     \centering
     \includegraphics[width=\linewidth]{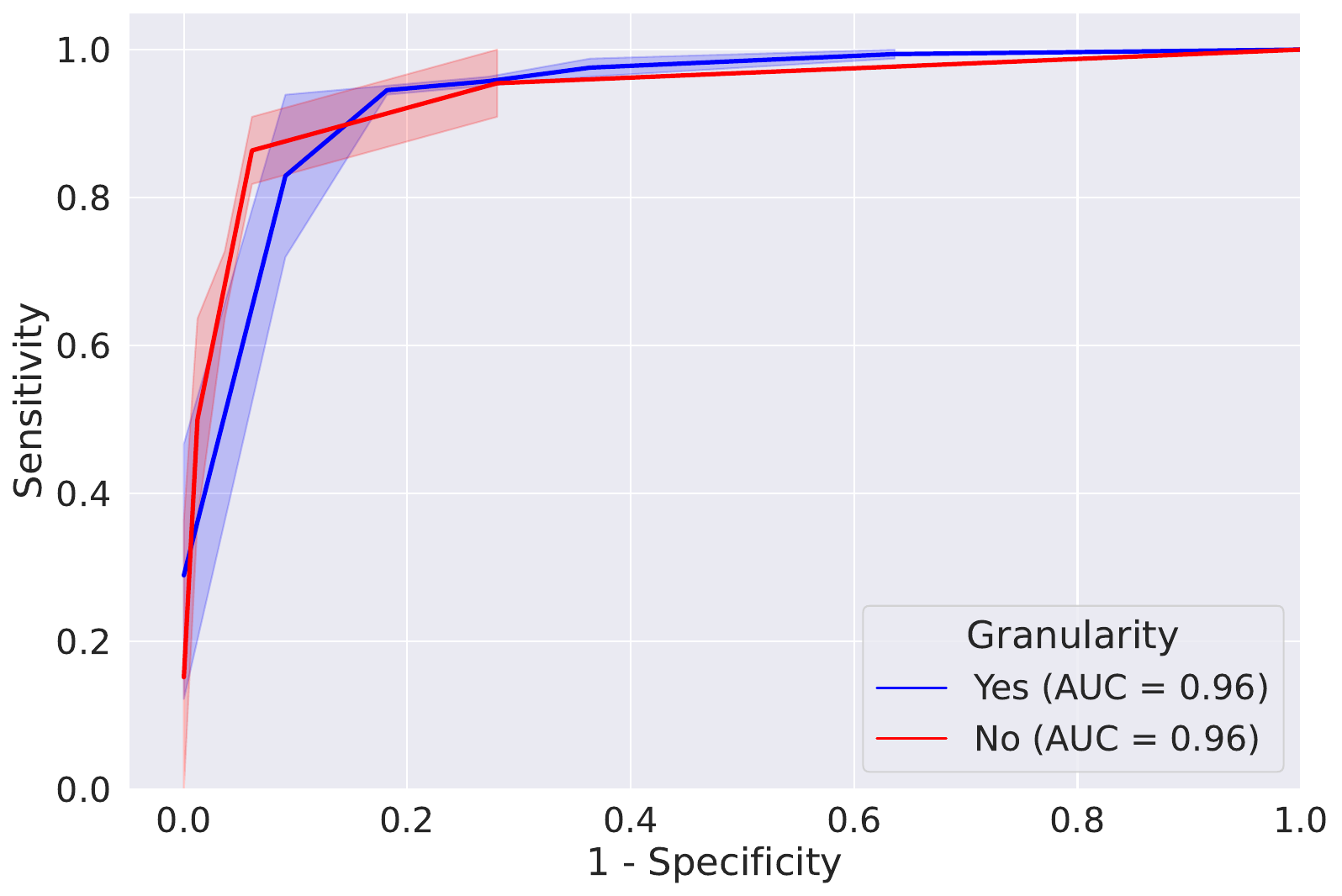}
     \caption*{(a)} \label{a1}
   \end{minipage}\hfill
   \begin{minipage}{0.33\linewidth}
     \centering
     \includegraphics[width=\linewidth]{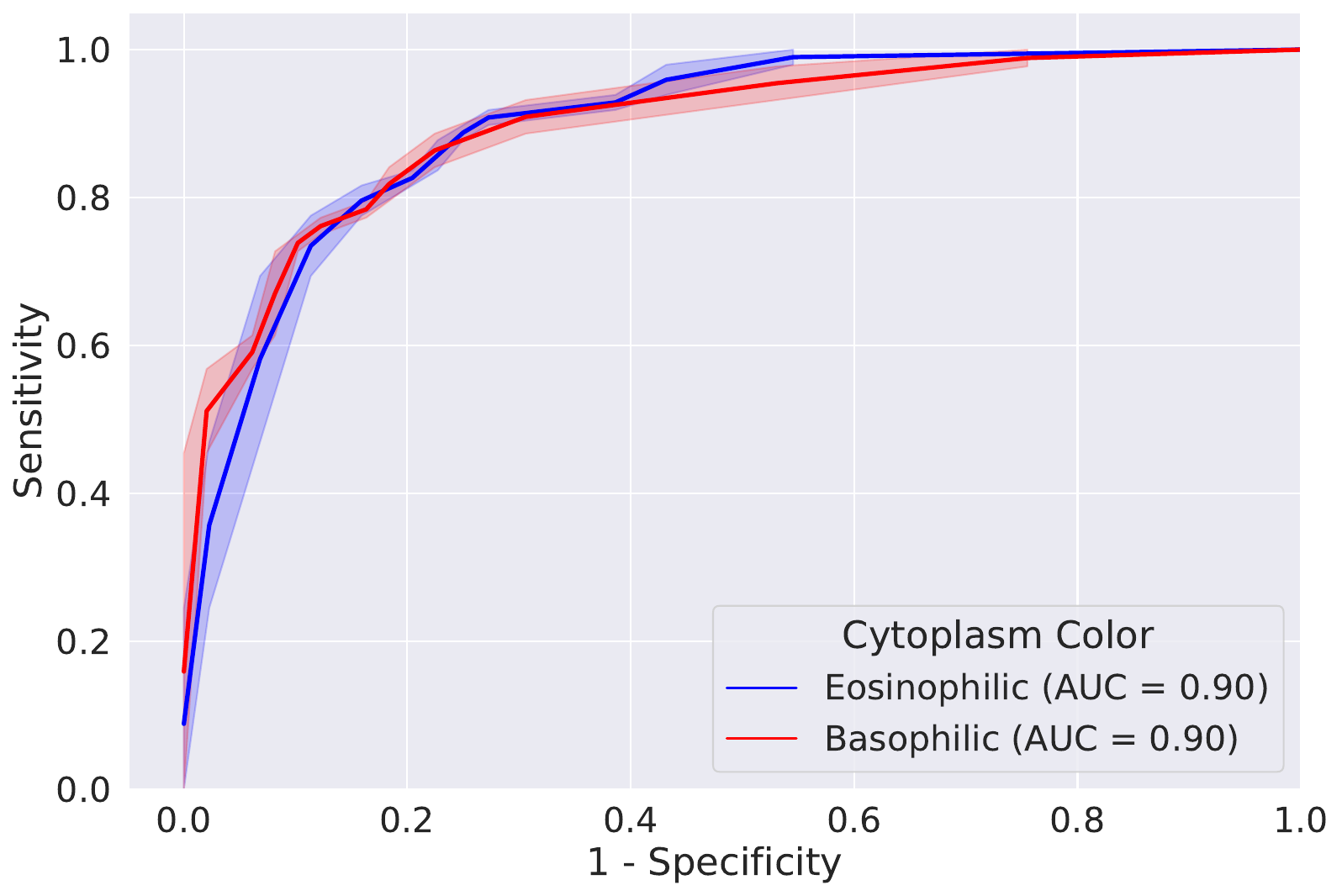}
     \caption*{(b)}
   \end{minipage}
   \begin{minipage}{0.33\linewidth}
     \centering
     \includegraphics[width=\linewidth]{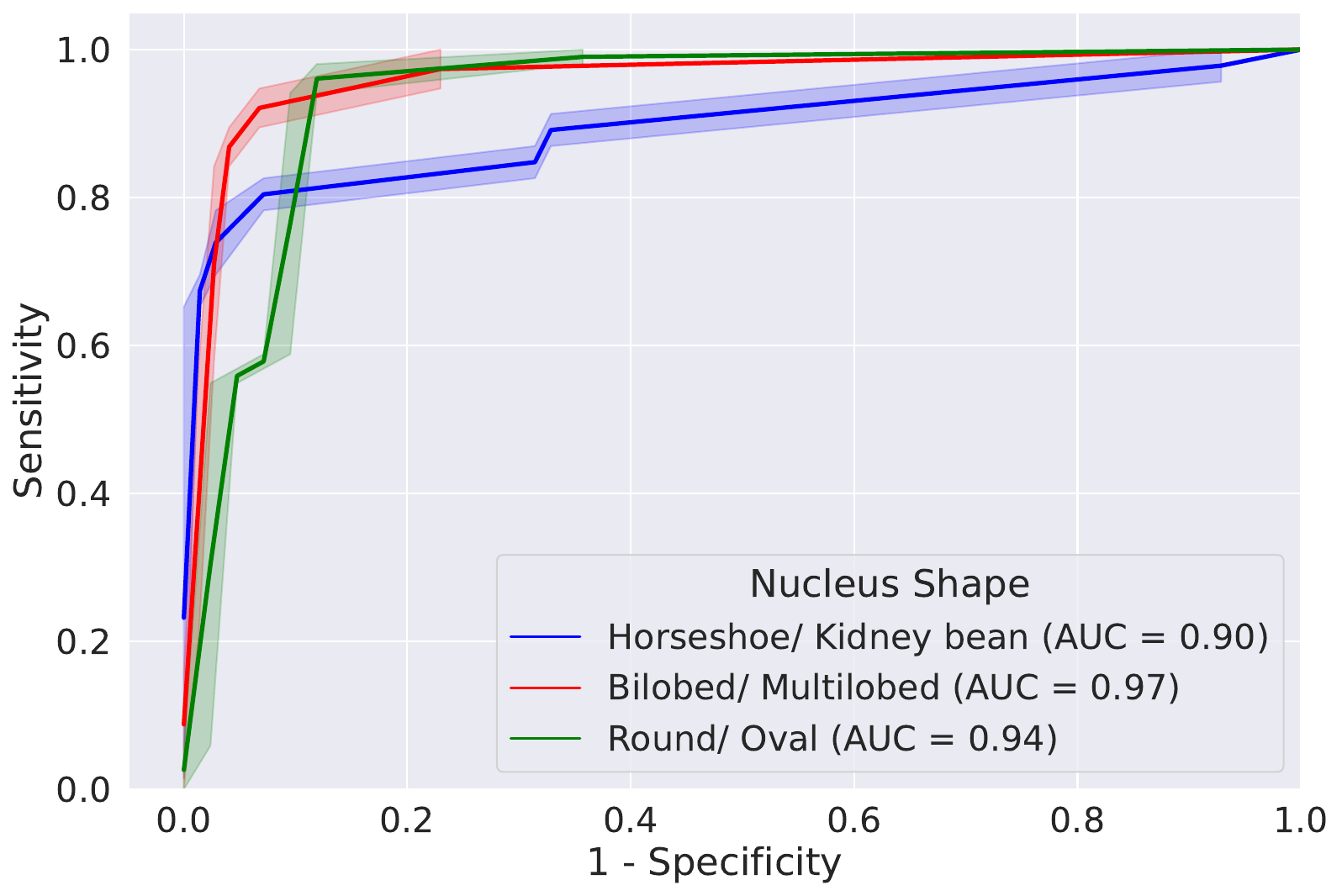}
     \caption*{(c)}
   \end{minipage}\hfill
   \begin{center}
   \begin{minipage}{0.33\linewidth}
       \centering
       \includegraphics[width=\linewidth]{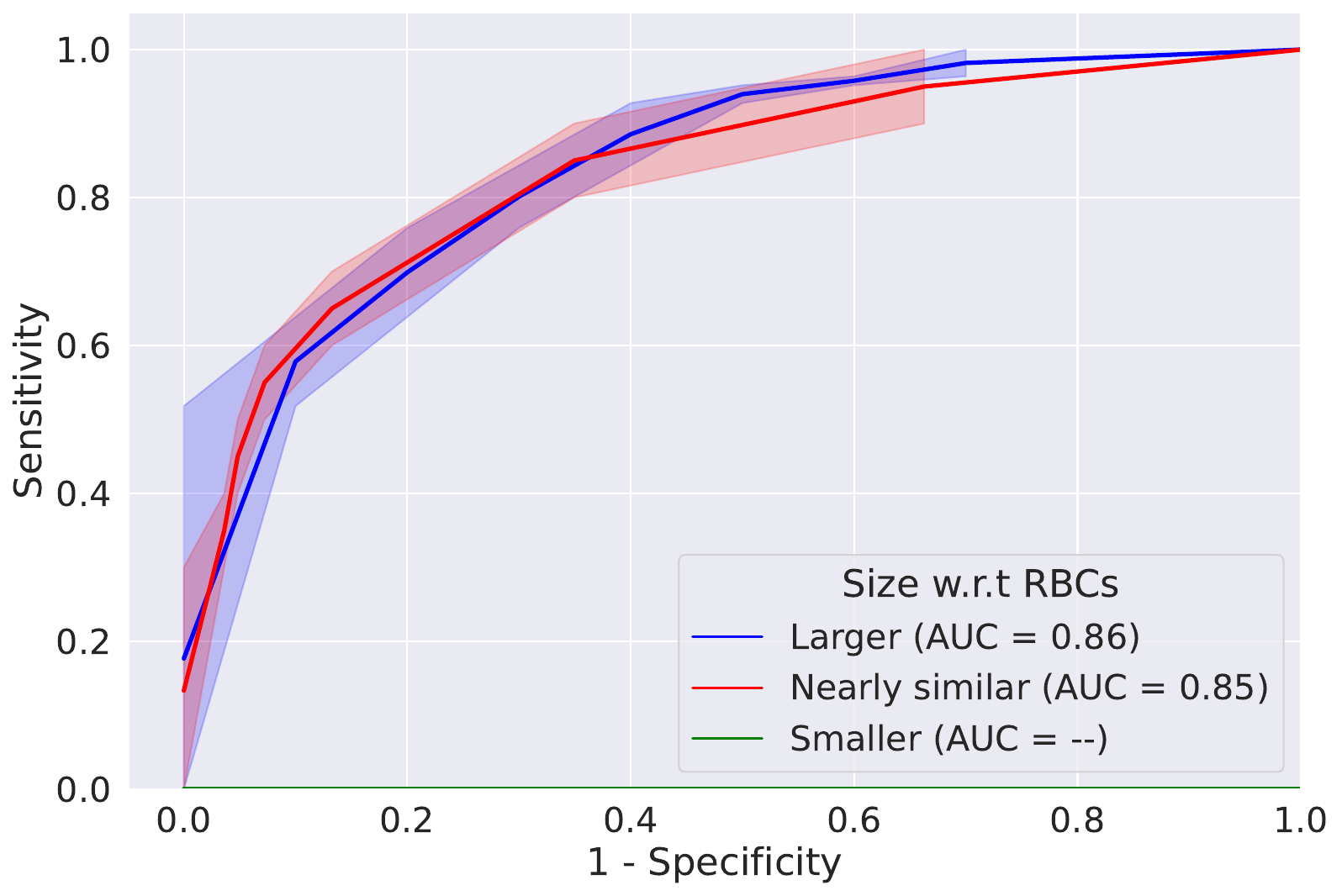}
       \caption*{(d)}
   \end{minipage}
     \begin{minipage}{0.33\linewidth}
       \centering
       \includegraphics[width=\linewidth]{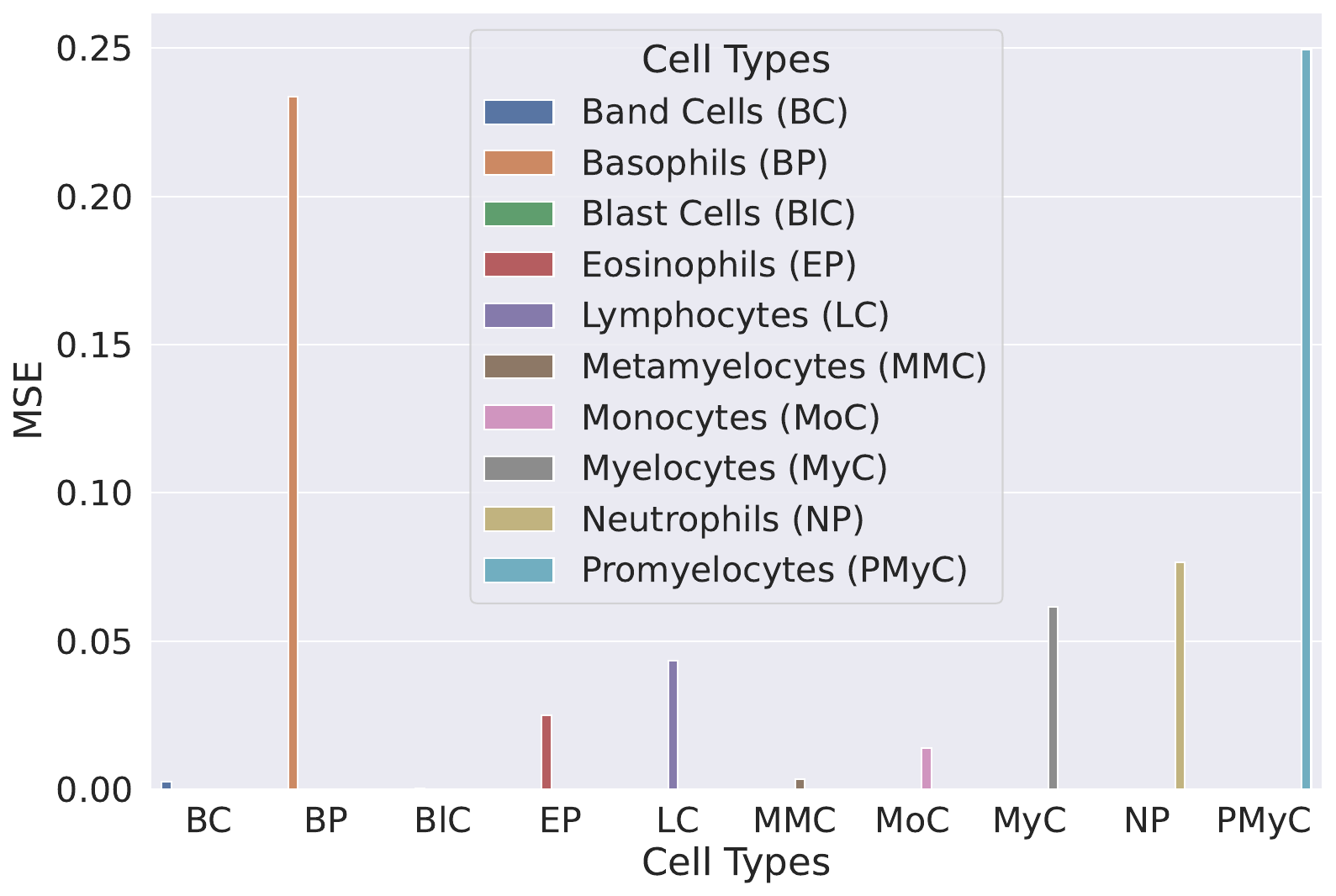}
       \caption*{(e)}
     \end{minipage}
    \end{center}
    \caption{(a), (b), (c) and (d) shows the category-wise ROC curves for granularity, cytoplasm color, nucleus shape, and size w.r.t RBCs respectively. (e) shows the classwise MSE loss for 10 cell types.}\label{Fig:ROCs}
\end{figure*}
We experimented with the proposed network by adding a different number of hidden layers to its explanation prediction branch and the results are analyzed using the metrics outlined in Section \ref{sec:eval_mets}. The classification and segmentation performance of HemaX with various configurations in the explanation prediction branch is shown in Table \ref{tab:cls_var_hmx}. The performance of HemaX's explanation prediction is illustrated in Table \ref{tab:exp_var_hmx}. The best result corresponding to each evaluation metric is marked in bold. The comparison between the variants of HemaX shows that by varying the number of hidden layers, we achieve similar performance in terms of classification metrics (Dice score, Jaccard index, classification accuracy, precision, and $F_1$ score). However on decreasing the number of hidden layers from four to zero, HemaX achieves a boost of $3\%$ accuracy in nucleus shape prediction and an increment of $\approx 1\%$ accuracy in predicting cytoplasm color. The HemaX variant with no hidden layers in the explanation prediction branch is selected as the best-performing model based on the $F_1$ score. Thus, we have empirically demonstrated that HemaX possesses the capability to decipher the characteristics of colored WBC images through five explanations: granularity, cytoplasm color, nucleus shape, size relative to red blood cells, and the nucleus-to-cytoplasm ratio. An illustrative outcome of HemaX, applied to two sample images, is presented in Fig.~\ref{fig:HemaX_ouput}. The segmentation masks predicted by HemaX showcase its capability to discern the nucleus from the cytoplasm, despite the granular nature of the cells. Additionally, we have utilized the top-performing HemaX model to generate Receiver Operator Characteristic (ROC) curves for the four explanations (Figures~\ref{Fig:ROCs}.(a),~\ref{Fig:ROCs}.(b),~\ref{Fig:ROCs}.(c), and~\ref{Fig:ROCs}.(d)). The ROC curve plots demonstrate that HemaX achieves an area under the curve (AUC) greater than 0.85 for each sub-explanation, indicating the model's precise ability to distinguish between different explanations. Additionally, we plotted the classwise normalized mean square error (MSE) for the N:C ratio (Figure~\ref{Fig:ROCs}.(e)). The classwise normalized MSE scores for the N:C ratio reveal that blast cells have the lowest MSE (0.0002), while promyelocytes have the highest MSE (0.2494).\par

\subsection{Comparison with other Leukocyte Classification Models}\label{sec:comp_sota}

\begin{table}[ht!]
\begin{minipage}[c]{\linewidth}
\resizebox{\textwidth}{!}{
\begin{tabular}{|c|c|c|c|}
\hline
\textbf{Architecture} &  \textbf{Classification acc.} & \textbf{Precision} & \textbf{F$_1$ score}\\
\hline
\hline
\textbf{Faster-RCNN~\cite{ren2015faster}} & 0.7226 $\pm$ 0.0749 & 0.7494 $\pm$ 0.0850 & 0.7130 $\pm$ 0.0897 \\
\hline
\textbf{Tavakoli's algorithm~\cite{tavakoli2021new}} & 0.5441 $\pm$  0.0285 & 0.5452 $\pm$  0.0356 & 0.5372 $\pm$ 0.0288 \\
\hline
\textbf{CUSS-Net~\cite{zhou2023cuss}} &  0.7161 $\pm$ 0.0364  & 0.7340 $\pm$ 0.0511  & 0.7077 $\pm$ 0.0418  \\
\hline
\textbf{HemaX (0-Hidden Layers)} & \textbf{0.8108 $\pm$ 0.0161} & \textbf{0.8351 $\pm$ 0.0088} & \textbf{0.8089 $\pm$ 0.0162} \\
\hline
\end{tabular}}
\caption{The classification metrics corresponding to various techniques on LeukoX dataset}\label{tab:cls_sota}
\end{minipage}
\end{table}
We continue to assess the capabilities of HemaX by comparing it with the state-of-the-art classification techniques, including Faster RCNN, CUSS-Net, and Tavakoli's algorithm. Metrics corresponding to the four classification techniques are presented in Table \ref{tab:cls_sota}. HemaX stands out with an classification accuracy of 0.8108, precision of 0.8351, and $F_1$-score of 0.8089, surpassing the performance of the other three best known techniques. Furthermore, this experiment demonstrates that HemaX can simultaneously classify cells and provide explanations without experiencing a significant drop in performance.

\begin{table}[ht!]
\begin{minipage}[c]{\linewidth}
\resizebox{\textwidth}{!}{
\begin{tabular}{|c|c|c|c|}
\hline
\textbf{Architecture} &  \textbf{Classification acc.} & \textbf{Precision} & \textbf{F$_1$ score}\\
\hline
\hline
\textbf{Faster-RCNN~\cite{ren2015faster}} & 0.9039 $\pm$ 0.0192 & 0.9840 $\pm$ 0.0161 & 0.9357 $\pm$ 0.0194 \\
\hline
\textbf{Tavakoli's algorithm~\cite{tavakoli2021new}} & 0.9423 $\pm$  0.0192 & 0.9506 $\pm$  0.0175 & 0.9404 $\pm$ 0.0208 \\
\hline
\textbf{CUSS-Net~\cite{zhou2023cuss}} &  0.9896 $\pm$ 0.0181  & 0.9913 $\pm$ 0.0150  & 0.9895 $\pm$ 0.0182  \\
\hline
\textbf{HemaX (No explanation branch)} &  0.9808 $\pm$ 0.0192 & 0.9839 $\pm$ 0.0161 & 0.9806 $\pm$ 0.0195 \\
\hline
\end{tabular}}
\caption{The classification metrics corresponding to various techniques on LISC dataset.} \label{tab:cls_var_lisc}
\end{minipage}
\end{table}

\begin{table}[ht!]
\begin{minipage}[c]{\linewidth}
\resizebox{\textwidth}{!}{
\begin{tabular}{|c|c|c|c|}
\hline
\textbf{Architecture} &  \textbf{Classification acc.} & \textbf{Precision} & \textbf{F$_1$ score}\\
\hline
\hline
\textbf{Faster-RCNN~\cite{ren2015faster}} & 0.9335 $\pm$ 0.0168 & 0.9371 $\pm$ 0.0157 & 0.9331 $\pm$ 0.0170 \\
\hline
\textbf{Tavakoli's algorithm~\cite{tavakoli2021new}} & 0.8809 $\pm$  0.0131 & 0.8852 $\pm$  0.0127 & 0.8813 $\pm$ 0.0129 \\
\hline
\textbf{CUSS-Net~\cite{zhou2023cuss}} &  0.9614 $\pm$ 0.0068  & 0.9618 $\pm$ 0.0071  & 0.9612 $\pm$ 0.0071  \\
\hline
\textbf{HemaX (No explanation branch)} & 0.9596 $\pm$ 0.0048 & 0.9603 $\pm$ 0.0047 & 0.9595 $\pm$ 0.0047 \\
\hline
\end{tabular}}
\caption{The classification metrics corresponding to various techniques on Raabin dataset.}\label{tab:exp_var_raabin}
\end{minipage}
\end{table}

To demonstrate its generalizability, we compared HemaX (without its explanation prediction branch) with three other state-of-the-art techniques using two publicly available datasets: LISC~\cite{rezatofighi2011automatic} and Raabin~\cite{kouzehkanan2022large}. Detailed information about these datasets is provided in Section~\ref{sec:dataset_training}. The performance of different techniques on the LISC dataset is summarized in Table~\ref{tab:cls_var_lisc}, while Table~\ref{tab:exp_var_raabin} presents the performance of various techniques on the Raabin dataset.\par
Compared to other state-of-the-art approaches (Faster R-CNN, CUSS-Net, and Tavakoli's algorithm), HemaX exhibits several advantages. Contrary to Faster RCNN's two-step process, which involves first proposing regions followed by detecting objects, HemaX accomplishes one-step detection by directly predicting a group of candidate objects and their corresponding bounding box for the given input image. The experimental results highlight that this single-phase detection approach leads to an $8.32\%$, $4.49\%$, and $2.64\%$ improvement in $F_1$-score on the LeukoX, LISC, and Raabin datasets, respectively, compared to Faster RCNN. Moreover, HemaX leverages attention maps to emphasize relevant parts of the input image, resulting in superior performance compared to CUSS-Net in the LeukoX dataset and similar performance on the two publicly available datasets.\par
Our study presents significant advancements compared to prior research, contributing both in terms of translation and methodology. From a translational perspective, we have demonstrated HemaX's strong generalization capabilities using two publicly available WBC datasets. This robust performance in WBC classification, coupled with its ability to provide explanations, is expected to bolster pathologists' confidence in the reliability of HemaX. As a result, there is a higher likelihood of its widespread adoption. This is in contrast to previous studies that solely evaluated AI systems for WBC classification, often leaving their inner workings opaque. In terms of methodology, HemaX is on par with previously developed AI systems, which were primarily designed for WBC classification. The key advantage of HemaX lies in its unified approach, enabling it to seamlessly perform multiple tasks without necessitating fundamental architectural changes. This distinguishes HemaX from the aforementioned prior systems and highlights its versatility.

\subsection{Faithfulness of Predictions}
\begin{figure}[ht!]
   \begin{minipage}{0.48\linewidth}
     \centering
     \includegraphics[width=\linewidth]{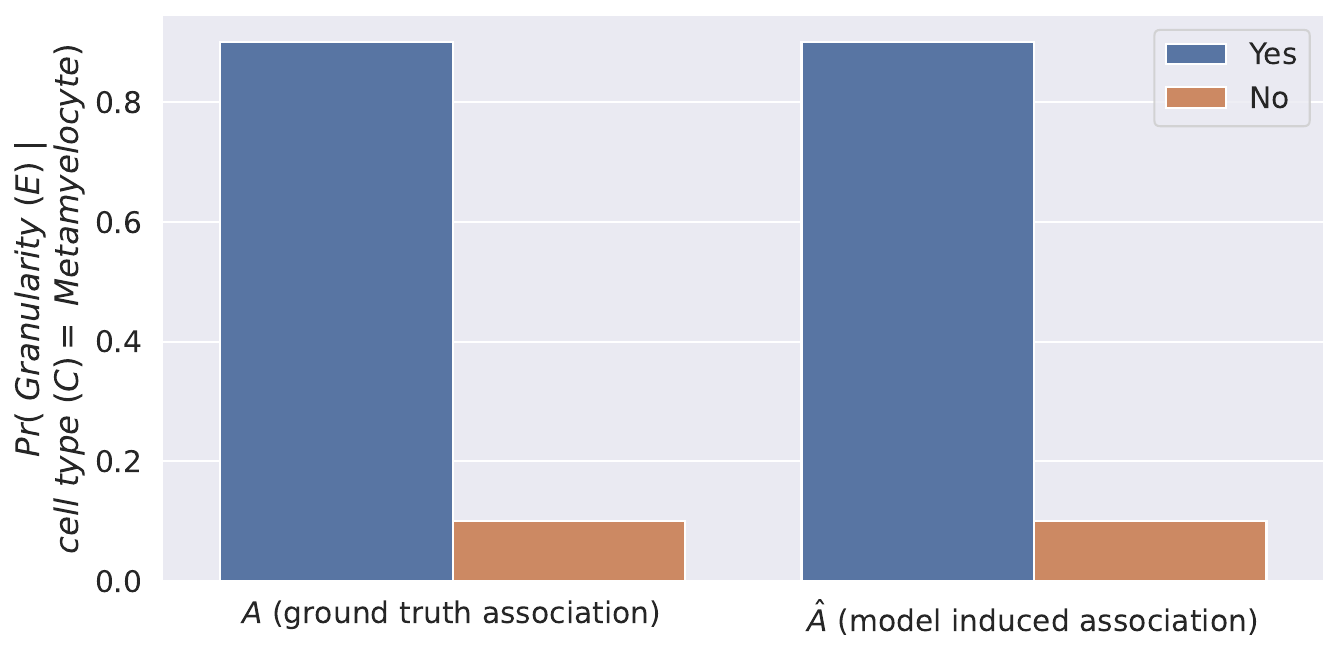}
   \end{minipage}\hfill
   \begin{minipage}{0.48\linewidth}
     \centering
     \includegraphics[width=\linewidth]{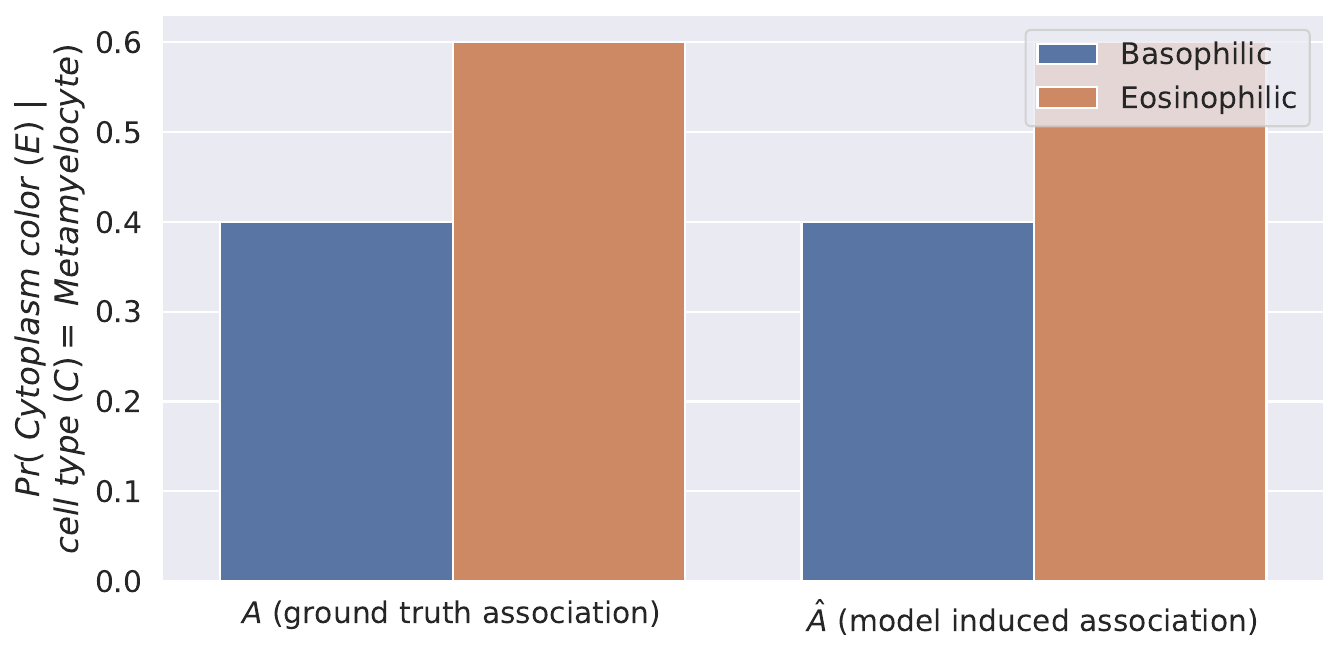}
   \end{minipage}
   \begin{minipage}{0.48\linewidth}
     \centering
     \includegraphics[width=\linewidth]{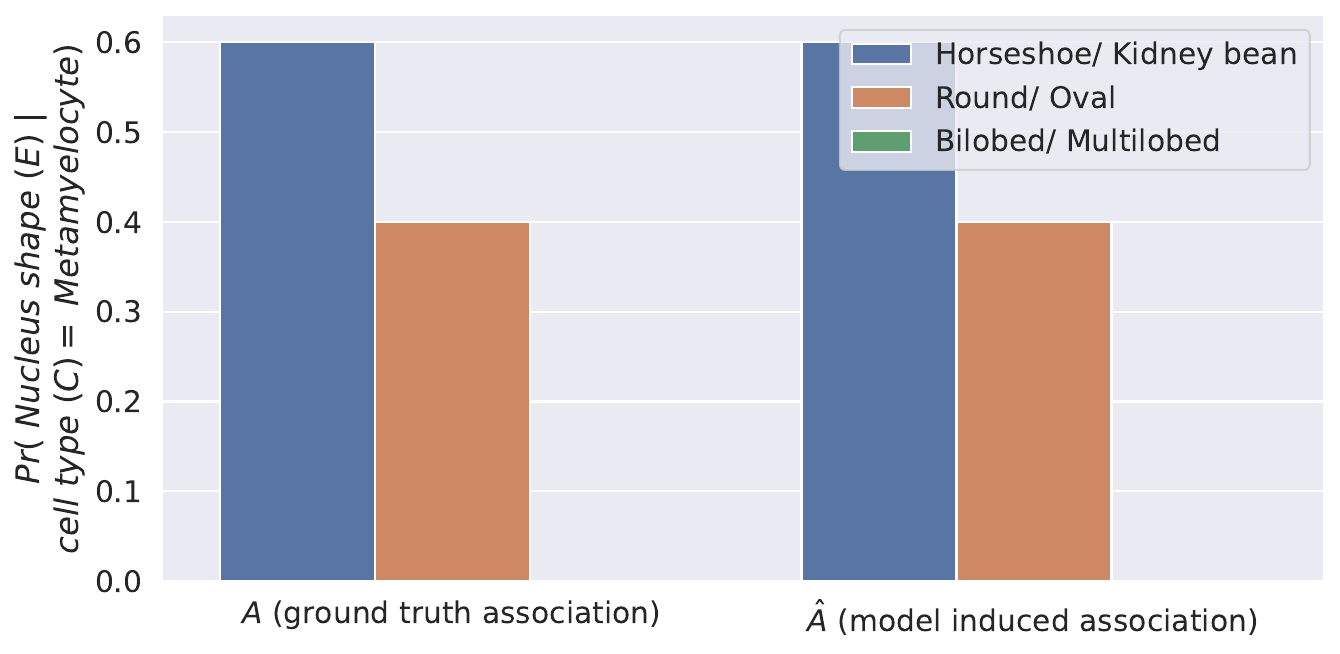}
   \end{minipage}\hfill
   \begin{minipage}{0.48\linewidth}
     \centering
     \includegraphics[width=\linewidth]{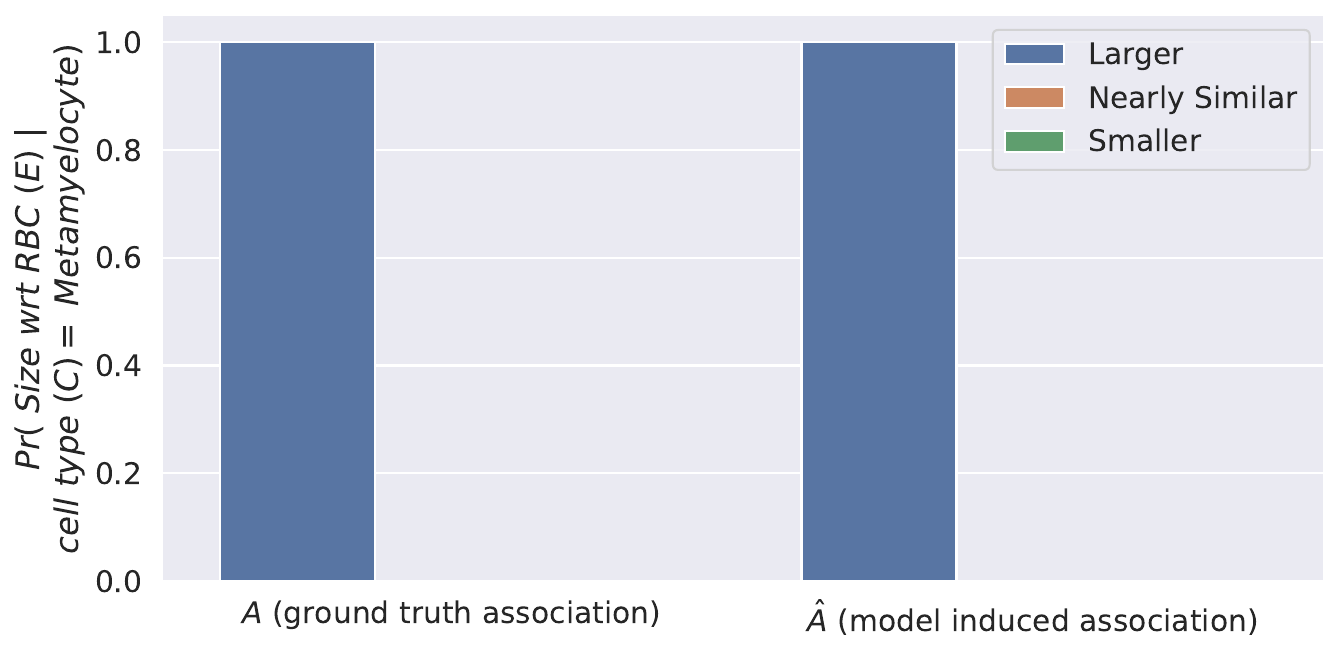}
   \end{minipage}
   \caption{Comparison between model-induced association ($\hat{A}$) and ground truth association ($A$) for all four explanations (`Granularity', `Cytoplasm Color', `Nucleus Shape' and `Size w.r.t. RBC') on `Metamyelocytes' cell type.}
   \label{fig:faithfulness_metamyelocyte}
\end{figure}
\begin{figure}[ht!]
   \begin{minipage}{0.48\linewidth}
     \centering
     \includegraphics[width=\linewidth]{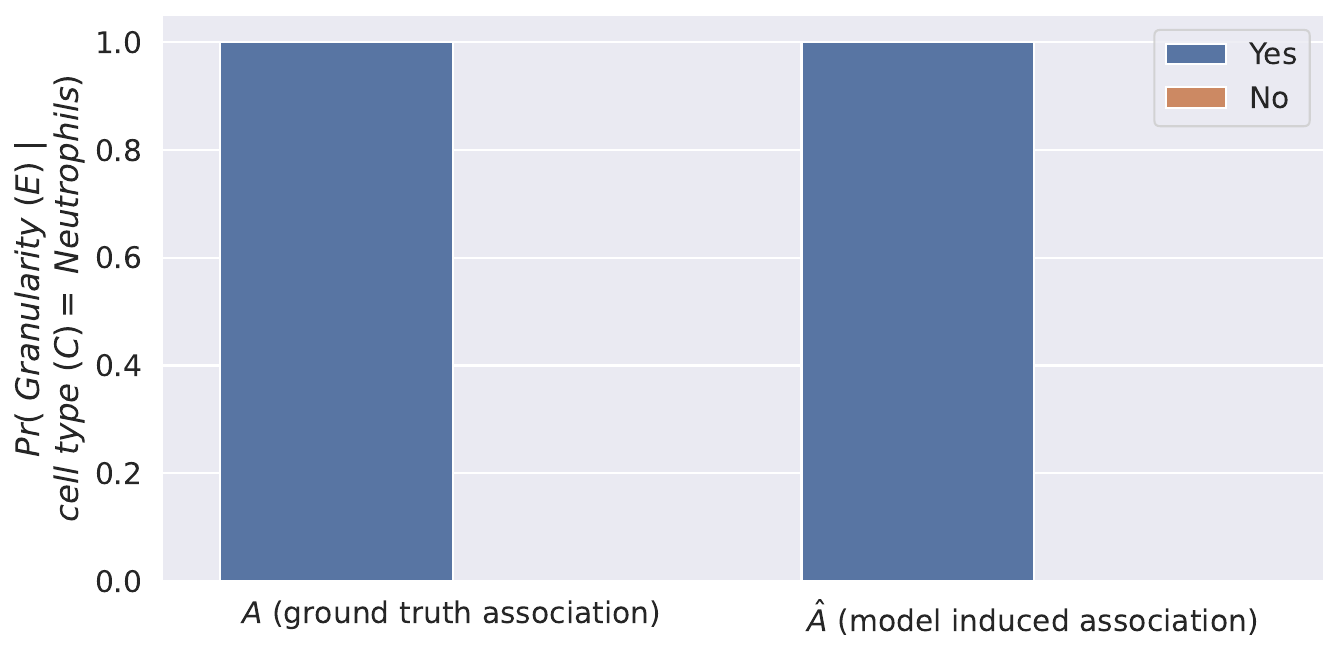}
   \end{minipage}\hfill
   \begin{minipage}{0.48\linewidth}
     \centering
     \includegraphics[width=\linewidth]{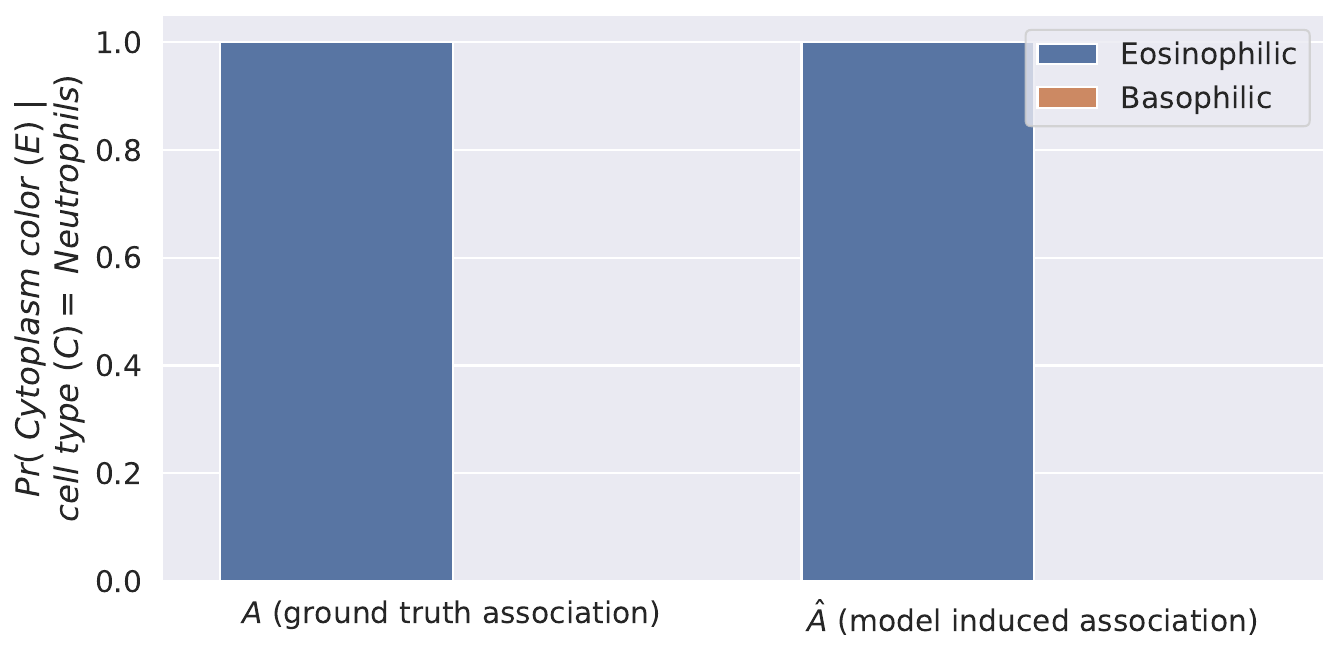}
   \end{minipage}
   \begin{minipage}{0.48\linewidth}
     \centering
     \includegraphics[width=\linewidth]{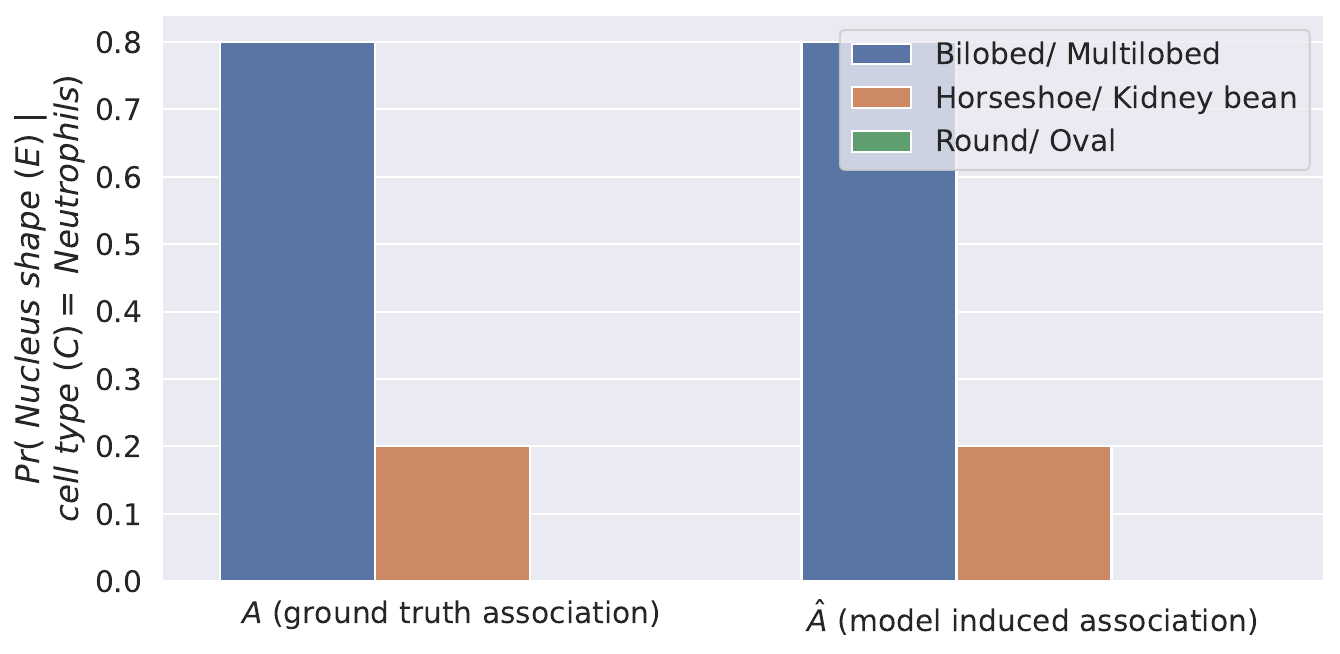}
   \end{minipage}\hfill
   \begin{minipage}{0.48\linewidth}
     \centering
     \includegraphics[width=\linewidth]{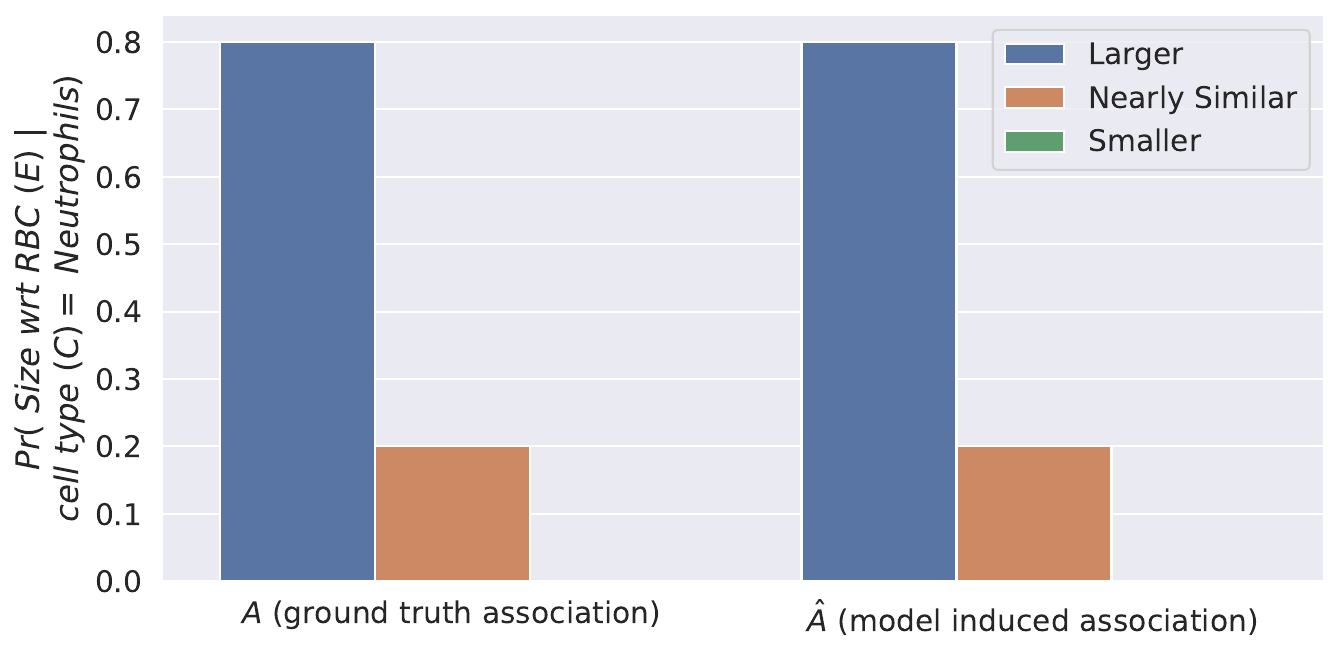}
   \end{minipage}
   \caption{Comparison between model-induced association ($\hat{A}$) and ground truth association ($A$) for all four explanations (`Granularity', `Cytoplasm Color', `Nucleus Shape' and `Size w.r.t. RBC') on `Neutrophils' cell type.}
   \label{fig:faithfulness_neutrophils}
\end{figure}
\begin{figure}[ht!]
   \begin{minipage}{0.48\linewidth}
     \centering
     \includegraphics[width=\linewidth]{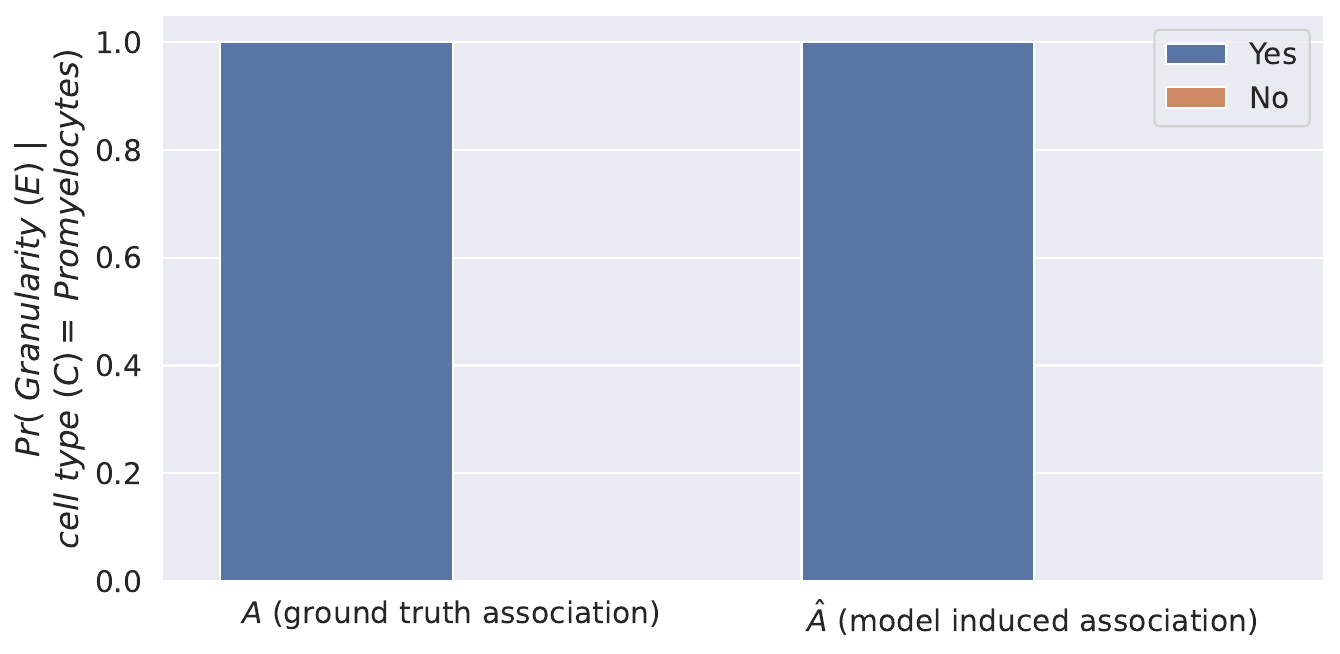}
   \end{minipage}\hfill
   \begin{minipage}{0.48\linewidth}
     \centering
     \includegraphics[width=\linewidth]{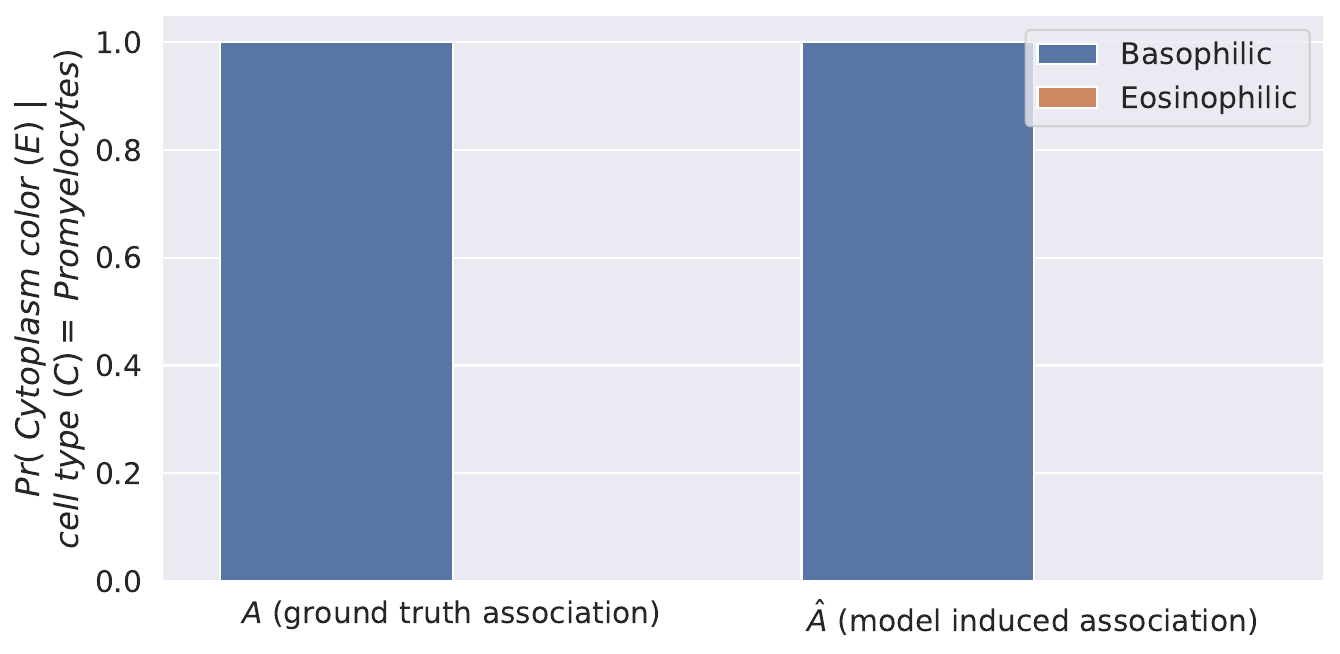}
   \end{minipage}
   \begin{minipage}{0.48\linewidth}
     \centering
     \includegraphics[width=\linewidth]{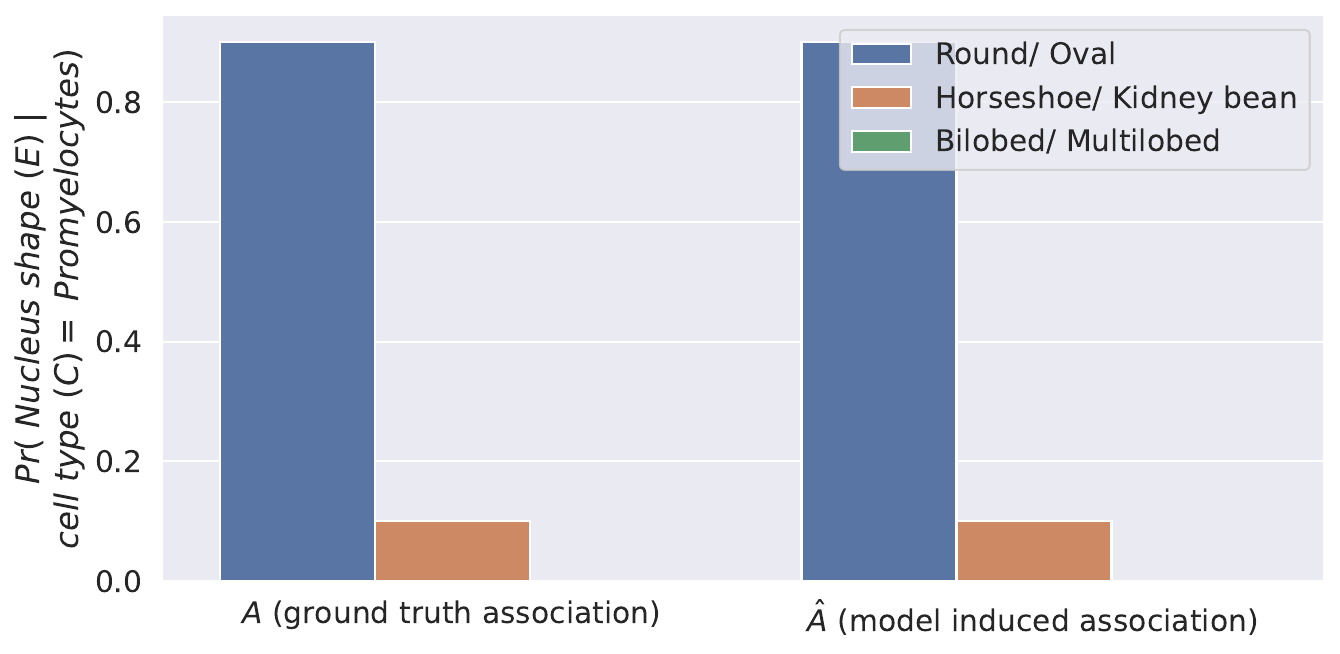}
   \end{minipage}\hfill
   \begin{minipage}{0.48\linewidth}
     \centering
     \includegraphics[width=\linewidth]{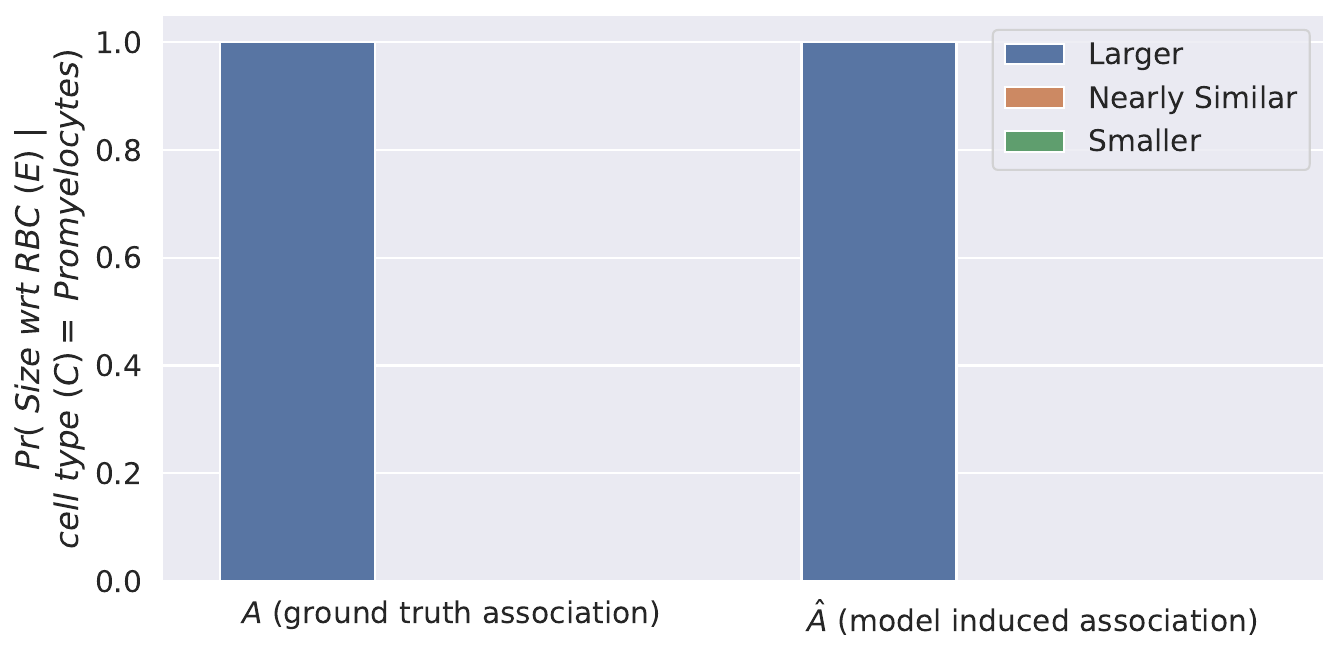}
   \end{minipage}
   \caption{Comparison between model-induced association ($\hat{A}$) and ground truth association ($A$) for all four explanations (`Granularity', `Cytoplasm Color', `Nucleus Shape' and `Size w.r.t. RBC') on `Promyelocyte' cell type.}
   \label{fig:faithfulness_promyelocyte}
\end{figure}
We have established the faithfulness of explanation predictions by the best HemaX model using the technique described in Section~\ref{faithfulness}. The Figures~\ref{fig:faithfulness_metamyelocyte},~\ref{fig:faithfulness_neutrophils} and~\ref{fig:faithfulness_promyelocyte} depicts $\hat{A}$ and $A$ for explanations based on four explanations: `Granularity', `Cytoplasm color', `Nucleus shape' and `Size w.r.t. RBC', focusing on \textit{Metamyelocyte}, \textit{Neutrophil} and \textit{Promyelocyte} cell types. The figure clearly illustrates that both the model-induced and ground truth associations ($p_{\hat{A}}$ and $p_A$) exhibit a high degree of similarity. This resemblance provides evidence for the faithfulness of the four explanations (`Granularity', `Cytoplasm color', `Nucleus shape' and `Size w.r.t. RBC') generated by HemaX. Similar results have been achieved for other cell types on four explanations (listed in the supplemental note). 

\subsection{Independence of Cell Classification and Explanation }
We have also conducted an investigation to ascertain whether the explanations were influenced by the predicted cell type, using two methods: comparing the Area Under the Curves (AUCs) for various explanations on the LeukoX dataset and validating the results with an expert hematologist. First, the AUCs for each explanation were computed separately for both correctly and incorrectly classified WBCs, as outlined in Table~\ref{tab:cls_rocs_auc}. This analysis yielded two significant observations: (1) The AUCs for the explanations of correctly classified WBCs were consistently higher than the explanations for the misclassified ones and (2) Even when the cell type prediction was inaccurate, all the explanations still exhibited AUC values exceeding 0.60. This indicates the model's proficiency in accurately capturing the explanatory aspects, regardless of the classification outcome.
\begin{table}[ht!]
\centering
\resizebox{\linewidth}{!}{
\begin{tabular}{|c|c|c|c|}
\hline
\multirow{2}{*}{\textbf{Explanation Type}} &  \multirow{2}{*}{\textbf{Explanation}} &\multicolumn{2}{c|}{\textbf{AUC for}} \\ 
\hhline{~~--}
 & &\textbf{Classified WBCs} & \textbf{Misclassified WBCs}\\
\hline
\hline
\multirow{2}{*}{\textbf{Granularity}}      & Yes                    & 0.99 & 0.60 \\  
                                           & No                     & 0.99 & 0.60 \\
\hline
\multirow{2}{*}{\textbf{Cytoplasm Color}}  & Eosinophilic           & 0.93 & 0.77 \\
                                           & Basophilic             & 0.93 & 0.77 \\
\hline
\multirow{3}{*}{\textbf{Nucleus Shape}}    & Horseshoe/ Kidney Bean & 0.93 & 0.84 \\
                                           & Bilobed/ Multilobed    & 0.97 & 0.95 \\
                                           & Round/ Oval            & 0.95 & 0.92 \\
\hline
\multirow{3}{*}{\textbf{Size w.r.t. RBCs}} & Larger                 & 0.88 & 0.82 \\
                                           & Nearly Similar         & 0.89 & 0.68 \\
                                           & Smaller                &  $-$   & $-$    \\
\hline
\end{tabular}}
\caption{The AUC values for each explanation predicted by the best HemaX model}\label{tab:cls_rocs_auc}
\end{table}

Furthermore, explanations from the best HemaX model were validated by a senior hematologist. As per the hematologist's assessment, the correct predictions for the granularity attribute were $90.91\%$, for cytoplasm color they were $84.42\%$, for nucleus shapes they were $93.51\%$, and for size with respect to RBC they were $88.31\%$, all pertaining to correctly classified WBCs. In the case of misclassified WBCs, the correct predictions for granularity, cytoplasm color, nucleus shape, and size with respect to RBC were $87.50\%$, $75.00\%$, $87.50\%$, and $81.25\%$, respectively.
What's particularly noteworthy is that even for misclassified cell types, the system generated explanations of high quality, achieving validation accuracy of over $80\%$ for all four attributes. This observation underscores the model's ability to independently learn both the classification task and explanation prediction, highlighting its robustness and efficacy in handling both objectives concurrently. The N:C ratio is similarly learned independently, as it is determined using the segmentation masks for the nucleus and cytoplasm.\\

\section{Conclusion} \label{sec:conclusion}
In this study, we developed a transformer encoder-decoder based network, HemaX, for classifying WBCs and predicting human-centric explanations for the classified cell simultaneously. The experimental results show that HemaX is quite efficient in serving a dual purpose: classifying White Blood Cells (WBCs) and concurrently predicting human-centric explanations for the classified cells.
As far as we are aware, HemaX stands as the pioneering explainable architecture tailored for white blood cell classification. Given that clinical practice relies heavily on a vast knowledge foundation, any computer-aided application must provide machine decisions in a comprehensible format to gain acceptance among medical professionals. This paper adds novelty to the ongoing research and shows a way of integrating an expert knowledge base into a learning system. The dataset which will be available under open data licensing is another contribution of this research towards explainable artificial intelligence (XAI) in medical imaging. Though further research is needed to improve the accuracy of feature-level interpretation, the system's ability to predict hematologist-like explanations would make it more appealing to medical practitioners.\par
One of the limitations of our research pertains to the utilization of a dataset featuring single-cell WBC images for explanation prediction. To extend the exploration of HemaX's capability for predicting explanations across multiple cell instances and to ensure its robust applicability in pathology laboratories, we encourage the public release of comprehensive explainable WBC datasets. These datasets should encompass multiple cells per image and be sourced from various hospitals and pathological laboratories. By combining such datasets with HemaX, researchers can unlock the potential for generating human-centric explanations on a larger scale, thus fostering more extensive insights.\par
Although our focus has been on employing HemaX to predict the five human-centric explanations for each cell, the potential exists for its application to explore other cell explanations across various cell types. The exploration of additional cell explanations could be realized by curating a dataset with annotations specific to the desired property. We are making our code accessible to enable others to utilize HemaX and gain insights into other cellular properties. Looking ahead, our future focus will revolve around examining the impact of HemaX on clinical stakeholders. These investigations will yield valuable practical insights into the effectiveness of AI-based systems in real-world scenarios, shedding light on their reception and perception by clinical experts. This endeavor promises to enrich our understanding of the translation of AI innovations into clinical practice. 

\section*{Acknowledgment}
This research is supported by the Science and Engineering Research Board (SERB), Dept. of Science and Technology (DST), Govt. of India through Grant File No. SPR/2020/000495.

\bibliographystyle{IEEEtran}
\bibliography{IEEEabrv,adityabibfile}

\begin{thebibliography}{10}
\providecommand{\url}[1]{#1}
\csname url@samestyle\endcsname
\providecommand{\newblock}{\relax}
\providecommand{\bibinfo}[2]{#2}
\providecommand{\BIBentrySTDinterwordspacing}{\spaceskip=0pt\relax}
\providecommand{\BIBentryALTinterwordstretchfactor}{4}
\providecommand{\BIBentryALTinterwordspacing}{\spaceskip=\fontdimen2\font plus
\BIBentryALTinterwordstretchfactor\fontdimen3\font minus \fontdimen4\font\relax}
\providecommand{\BIBforeignlanguage}[2]{{%
\expandafter\ifx\csname l@#1\endcsname\relax
\typeout{** WARNING: IEEEtran.bst: No hyphenation pattern has been}%
\typeout{** loaded for the language `#1'. Using the pattern for}%
\typeout{** the default language instead.}%
\else
\language=\csname l@#1\endcsname
\fi
#2}}
\providecommand{\BIBdecl}{\relax}
\BIBdecl

\bibitem{walker1990clinical}
H.~K. Walker, W.~D. Hall, and J.~W. Hurst, ``Clinical methods: the history, physical, and laboratory examinations,'' 1990.

\bibitem{rezatofighi2011automatic}
S.~H. Rezatofighi and H.~Soltanian-Zadeh, ``Automatic recognition of five types of white blood cells in peripheral blood,'' \emph{Computerized Medical Imaging and Graphics}, vol.~35, no.~4, pp. 333--343, 2011.

\bibitem{gomez2008feature}
P.~G{\'o}mez-Gil, M.~Ram{\'\i}rez-Cort{\'e}s, J.~Gonz{\'a}lez-Bernal, {\'A}.~G. Pedrero, C.~I. Prieto-Castro, D.~Valencia, R.~Lobato, and J.~E. Alonso, ``A feature extraction method based on morphological operators for automatic classification of leukocytes,'' in \emph{2008 Seventh Mexican International Conference on Artificial Intelligence}.\hskip 1em plus 0.5em minus 0.4em\relax IEEE, 2008, pp. 227--232.

\bibitem{ushizima2005support}
D.~M. Ushizima, A.~C. Lorena, and A.~De~Carvalho, ``Support vector machines applied to white blood cell recognition,'' in \emph{Fifth International Conference on Hybrid Intelligent Systems (HIS'05)}.\hskip 1em plus 0.5em minus 0.4em\relax IEEE, 2005, pp. 6--pp.

\bibitem{kulkarni2013classification}
S.~S. Kulkarni, C.~S. Hinge, and A.~G. Ambekar, ``Classification of rbc and wbc in peripheral blood smear using knn,'' \emph{Indian J Res}, 2013.

\bibitem{gupta2019optimized}
D.~Gupta, J.~Arora, U.~Agrawal, A.~Khanna, and V.~H.~C. de~Albuquerque, ``Optimized binary bat algorithm for classification of white blood cells,'' \emph{Measurement}, vol. 143, pp. 180--190, 2019.

\bibitem{tavakoli2021new}
S.~Tavakoli, A.~Ghaffari, Z.~M. Kouzehkanan, and R.~Hosseini, ``New segmentation and feature extraction algorithm for classification of white blood cells in peripheral smear images,'' \emph{Scientific Reports}, vol.~11, no.~1, p. 19428, 2021.

\bibitem{khashman2008ibcis}
A.~Khashman, ``Ibcis: Intelligent blood cell identification system,'' \emph{Progress in Natural Science}, vol.~18, no.~10, pp. 1309--1314, 2008.

\bibitem{wang2016spectral}
Q.~Wang, L.~Chang, M.~Zhou, Q.~Li, H.~Liu, and F.~Guo, ``A spectral and morphologic method for white blood cell classification,'' \emph{Optics \& Laser Technology}, vol.~84, pp. 144--148, 2016.

\bibitem{li2023rethinking}
J.~Li, A.~Liu, Y.~Li, W.~Wei, R.~Qian, Q.~Xie, B.~Qiu, and X.~Chen, ``Rethinking deep supervision for brain tumor segmentation,'' \emph{IEEE Transactions on Artificial Intelligence}, 2023.

\bibitem{abubaker2022detection}
M.~B. Abubaker and B.~Babayi{\u{g}}it, ``Detection of cardiovascular diseases in ecg images using machine learning and deep learning methods,'' \emph{IEEE Transactions on Artificial Intelligence}, vol.~4, no.~2, pp. 373--382, 2022.

\bibitem{mitani2020detection}
A.~Mitani, A.~Huang, S.~Venugopalan, G.~S. Corrado, L.~Peng, D.~R. Webster, N.~Hammel, Y.~Liu, and A.~V. Varadarajan, ``Detection of anaemia from retinal fundus images via deep learning,'' \emph{Nature Biomedical Engineering}, vol.~4, no.~1, pp. 18--27, 2020.

\bibitem{shahin2019white}
A.~I. Shahin, Y.~Guo, K.~M. Amin, and A.~A. Sharawi, ``White blood cells identification system based on convolutional deep neural learning networks,'' \emph{Computer methods and programs in biomedicine}, vol. 168, pp. 69--80, 2019.

\bibitem{yao2021classification}
X.~Yao, K.~Sun, X.~Bu, C.~Zhao, and Y.~Jin, ``Classification of white blood cells using weighted optimized deformable convolutional neural networks,'' \emph{Artificial Cells, Nanomedicine, and Biotechnology}, vol.~49, no.~1, pp. 147--155, 2021.

\bibitem{al2021improving}
R.~Al-Qudah and C.~Y. Suen, ``Improving blood cells classification in peripheral blood smears using enhanced incremental training,'' \emph{Computers in Biology and Medicine}, vol. 131, p. 104265, 2021.

\bibitem{zhou2023cuss}
X.~Zhou, Z.~Li, Y.~Xue, S.~Chen, M.~Zheng, C.~Chen, Y.~Yu, X.~Nie, X.~Lin, L.~Wang \emph{et~al.}, ``Cuss-net: a cascaded unsupervised-based strategy and supervised network for biomedical image diagnosis and segmentation,'' \emph{IEEE Journal of Biomedical and Health Informatics}, 2023.

\bibitem{li2023deep}
M.~Li, C.~Lin, P.~Ge, L.~Li, S.~Song, H.~Zhang, L.~Lu, X.~Liu, F.~Zheng, S.~Zhang \emph{et~al.}, ``A deep learning model for detection of leukocytes under various interference factors,'' \emph{Scientific Reports}, vol.~13, no.~1, p. 2160, 2023.

\bibitem{tahiri2023white}
M.~A. Tahiri, A.~Bencherqui, H.~Karmouni, H.~Amakdouf, M.~Sayyouri, H.~Qjidaa \emph{et~al.}, ``White blood cell automatic classification using deep learning and optimized quaternion hybrid moments,'' \emph{Biomedical Signal Processing and Control}, vol.~86, p. 105128, 2023.

\bibitem{leng2023deep}
B.~Leng, C.~Wang, M.~Leng, M.~Ge, and W.~Dong, ``Deep learning detection network for peripheral blood leukocytes based on improved detection transformer,'' \emph{Biomedical Signal Processing and Control}, vol.~82, p. 104518, 2023.

\bibitem{tougaccar2020classification}
M.~To{\u{g}}a{\c{c}}ar, B.~Ergen, and Z.~C{\"o}mert, ``Classification of white blood cells using deep features obtained from convolutional neural network models based on the combination of feature selection methods,'' \emph{Applied Soft Computing}, vol.~97, p. 106810, 2020.

\bibitem{han2023one}
Z.~Han, H.~Huang, D.~Lu, Q.~Fan, C.~Ma, X.~Chen, Q.~Gu, and Q.~Chen, ``One-stage and lightweight cnn detection approach with attention: Application to wbc detection of microscopic images,'' \emph{Computers in Biology and Medicine}, vol. 154, p. 106606, 2023.

\bibitem{rawal2021recent}
A.~Rawal, J.~McCoy, D.~B. Rawat, B.~M. Sadler, and R.~S. Amant, ``Recent advances in trustworthy explainable artificial intelligence: Status, challenges, and perspectives,'' \emph{IEEE Transactions on Artificial Intelligence}, vol.~3, no.~6, pp. 852--866, 2021.

\bibitem{tizhoosh2018artificial}
H.~R. Tizhoosh and L.~Pantanowitz, ``Artificial intelligence and digital pathology: challenges and opportunities,'' \emph{Journal of pathology informatics}, vol.~9, no.~1, p.~38, 2018.

\bibitem{wu2021interpretable}
H.~Wu, W.~Ruan, J.~Wang, D.~Zheng, B.~Liu, Y.~Geng, X.~Chai, J.~Chen, K.~Li, S.~Li \emph{et~al.}, ``Interpretable machine learning for covid-19: An empirical study on severity prediction task,'' \emph{IEEE Transactions on Artificial Intelligence}, 2021.

\bibitem{Niazi2019-td}
M.~K.~K. Niazi, A.~V. Parwani, and M.~N. Gurcan, ``\BIBforeignlanguage{en}{Digital pathology and artificial intelligence},'' \emph{\BIBforeignlanguage{en}{Lancet Oncol.}}, vol.~20, no.~5, pp. e253--e261, May 2019.

\bibitem{esteva2019guide}
A.~Esteva, A.~Robicquet, B.~Ramsundar, V.~Kuleshov, M.~DePristo, K.~Chou, C.~Cui, G.~Corrado, S.~Thrun, and J.~Dean, ``A guide to deep learning in healthcare,'' \emph{Nature medicine}, vol.~25, no.~1, pp. 24--29, 2019.

\bibitem{ren2015faster}
S.~Ren, K.~He, R.~Girshick, and J.~Sun, ``Faster r-cnn: Towards real-time object detection with region proposal networks,'' \emph{Advances in neural information processing systems}, vol.~28, 2015.

\bibitem{kouzehkanan2022large}
Z.~M. Kouzehkanan, S.~Saghari, S.~Tavakoli, P.~Rostami, M.~Abaszadeh, F.~Mirzadeh, E.~S. Satlsar, M.~Gheidishahran, F.~Gorgi, S.~Mohammadi \emph{et~al.}, ``A large dataset of white blood cells containing cell locations and types, along with segmented nuclei and cytoplasm,'' \emph{Scientific Reports}, vol.~12, no.~1, pp. 1--14, 2022.

\bibitem{ozcan2014mobile}
A.~Ozcan, ``Mobile phones democratize and cultivate next-generation imaging, diagnostics and measurement tools,'' \emph{Lab on a Chip}, vol.~14, no.~17, pp. 3187--3194, 2014.

\bibitem{kwon2016medical}
L.~Kwon, K.~Long, Y.~Wan, H.~Yu, and B.~Cunningham, ``Medical diagnostics with mobile devices: Comparison of intrinsic and extrinsic sensing,'' \emph{Biotechnology advances}, vol.~34, no.~3, pp. 291--304, 2016.

\bibitem{kim2017smartphone}
H.~Kim, Y.~Jung, I.-J. Doh, R.~A. Lozano-Mahecha, B.~Applegate, and E.~Bae, ``Smartphone-based low light detection for bioluminescence application,'' \emph{Scientific reports}, vol.~7, no.~1, p. 40203, 2017.

\bibitem{jung2015smartphone}
Y.~Jung, J.~Kim, O.~Awofeso, H.~Kim, F.~Regnier, and E.~Bae, ``Smartphone-based colorimetric analysis for detection of saliva alcohol concentration,'' \emph{Applied optics}, vol.~54, no.~31, pp. 9183--9189, 2015.

\bibitem{de2020smartphone}
I.~M.~P. de~Vargas-Sansalvador, M.~M. Erenas, A.~Mart{\'\i}nez-Olmos, F.~Mirza-Montoro, D.~Diamond, and L.~F. Capitan-Vallvey, ``Smartphone based meat freshness detection,'' \emph{Talanta}, vol. 216, p. 120985, 2020.

\bibitem{dutta2015dye}
S.~Dutta, D.~Sarma, A.~Patel, and P.~Nath, ``Dye-assisted ph sensing using a smartphone,'' \emph{IEEE Photonics Technology Letters}, vol.~27, no.~22, pp. 2363--2366, 2015.

\bibitem{chen2023joint}
F.~Chen, H.~Han, P.~Wan, H.~Liao, C.~Liu, and D.~Zhang, ``Joint segmentation and differential diagnosis of thyroid nodule in contrast-enhanced ultrasound images,'' \emph{IEEE Transactions on Biomedical Engineering}, 2023.

\bibitem{kiyasseh2023vision}
D.~Kiyasseh, R.~Ma, T.~F. Haque, B.~J. Miles, C.~Wagner, D.~A. Donoho, A.~Anandkumar, and A.~J. Hung, ``A vision transformer for decoding surgeon activity from surgical videos,'' \emph{Nature Biomedical Engineering}, pp. 1--17, 2023.

\bibitem{phan2022sleeptransformer}
H.~Phan, K.~Mikkelsen, O.~Y. Ch{\'e}n, P.~Koch, A.~Mertins, and M.~De~Vos, ``Sleeptransformer: Automatic sleep staging with interpretability and uncertainty quantification,'' \emph{IEEE Transactions on Biomedical Engineering}, vol.~69, no.~8, pp. 2456--2467, 2022.

\bibitem{shokouhmand2023diagnosis}
A.~Shokouhmand, H.~Wen, S.~Khan, J.~A. Puma, A.~Patel, P.~Green, F.~Ayazi, and N.~Ebadi, ``Diagnosis of coexisting valvular heart diseases using image-to-sequence translation of contact microphone recordings,'' \emph{IEEE Transactions on Biomedical Engineering}, 2023.

\bibitem{wang2022medical}
Z.~Wang, M.~Tang, L.~Wang, X.~Li, and L.~Zhou, ``A medical semantic-assisted transformer for radiographic report generation,'' in \emph{Medical Image Computing and Computer Assisted Intervention--MICCAI 2022: 25th International Conference, Singapore, September 18--22, 2022, Proceedings, Part III}.\hskip 1em plus 0.5em minus 0.4em\relax Springer, 2022, pp. 655--664.

\bibitem{you2021aligntransformer}
D.~You, F.~Liu, S.~Ge, X.~Xie, J.~Zhang, and X.~Wu, ``Aligntransformer: Hierarchical alignment of visual regions and disease tags for medical report generation,'' in \emph{Medical Image Computing and Computer Assisted Intervention--MICCAI 2021: 24th International Conference, Strasbourg, France, September 27--October 1, 2021, Proceedings, Part III 24}.\hskip 1em plus 0.5em minus 0.4em\relax Springer, 2021, pp. 72--82.

\bibitem{carion2020end}
N.~Carion, F.~Massa, G.~Synnaeve, N.~Usunier, A.~Kirillov, and S.~Zagoruyko, ``End-to-end object detection with transformers,'' in \emph{Computer Vision--ECCV 2020: 16th European Conference, Glasgow, UK, August 23--28, 2020, Proceedings, Part I 16}.\hskip 1em plus 0.5em minus 0.4em\relax Springer, 2020, pp. 213--229.

\bibitem{kuhn1955hungarian}
H.~W. Kuhn, ``The hungarian method for the assignment problem,'' \emph{Naval research logistics quarterly}, vol.~2, no. 1-2, pp. 83--97, 1955.

\bibitem{rezatofighi2019generalized}
H.~Rezatofighi, N.~Tsoi, J.~Gwak, A.~Sadeghian, I.~Reid, and S.~Savarese, ``Generalized intersection over union: A metric and a loss for bounding box regression,'' in \emph{Proceedings of the IEEE/CVF conference on computer vision and pattern recognition}, 2019, pp. 658--666.

\bibitem{jadon2020survey}
S.~Jadon, ``A survey of loss functions for semantic segmentation,'' in \emph{2020 IEEE conference on computational intelligence in bioinformatics and computational biology (CIBCB)}.\hskip 1em plus 0.5em minus 0.4em\relax IEEE, 2020, pp. 1--7.

\bibitem{Agrawal1993Association}
R.~Agrawal, T.~Imielinski, and A.~N. Swami, ``Mining association rules between sets of items in large databases,'' in \emph{ACM SIGMOD}, 1993.

\bibitem{loshchilov2017decoupled}
I.~Loshchilov and F.~Hutter, ``Decoupled weight decay regularization,'' \emph{arXiv preprint arXiv:1711.05101}, 2017.

\bibitem{jaccard1901distribution}
P.~Jaccard, ``Distribution de la flore alpine dans le bassin des dranses et dans quelques r{\'e}gions voisines,'' \emph{Bull Soc Vaudoise Sci Nat}, vol.~37, pp. 241--272, 1901.

\bibitem{Srensen1948AMO}
T.~S{\o}rensen, T.~S{\o}rensen, T.~Biering-S{\o}rensen, T.~S{\o}rensen, and J.~T. Sorensen, ``A method of establishing group of equal amplitude in plant sociobiology based on similarity of species content and its application to analyses of the vegetation on danish commons,'' in \emph{Kongelige Danske Videnskabernes Selskab}, 1948.

\end{thebibliography}

\title{\begin{center}\underline{Supplementary Material}\end{center} Pathologist-Like Explanations Unveiled: an Explainable Deep Learning System for White Blood Cell Classification}
\maketitle
\section*{Part-1: Graphs for Faithfulness of Explanations}
\begin{figure}[ht!]
   \begin{minipage}{0.48\linewidth}
     \centering
     \includegraphics[width=\linewidth]{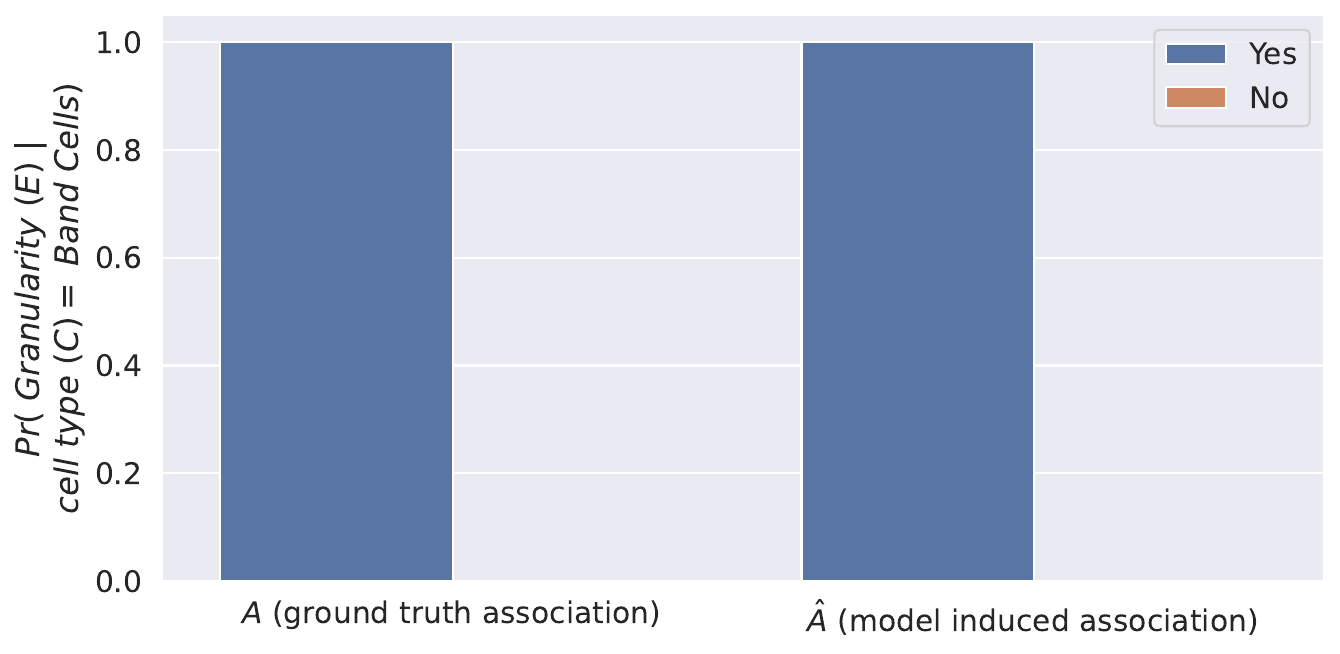}
   \end{minipage}\hfill
   \begin{minipage}{0.48\linewidth}
     \centering
     \includegraphics[width=\linewidth]{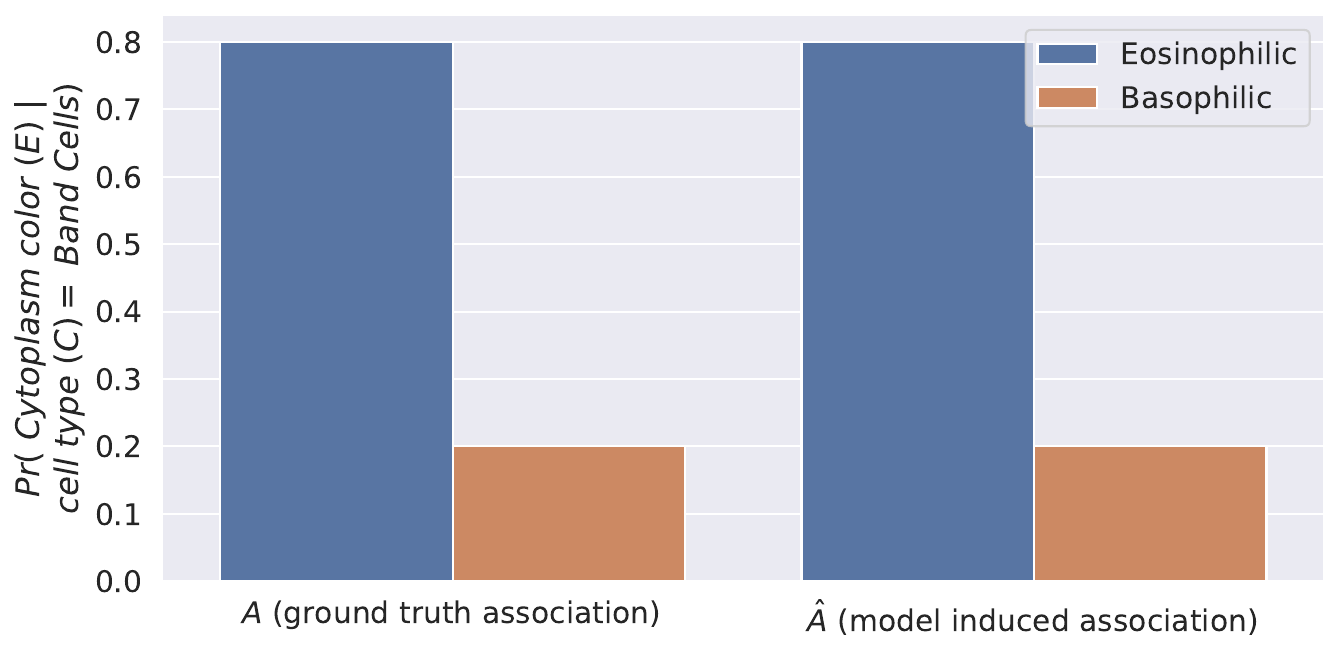}
   \end{minipage}
   \begin{minipage}{0.48\linewidth}
     \centering
     \includegraphics[width=\linewidth]{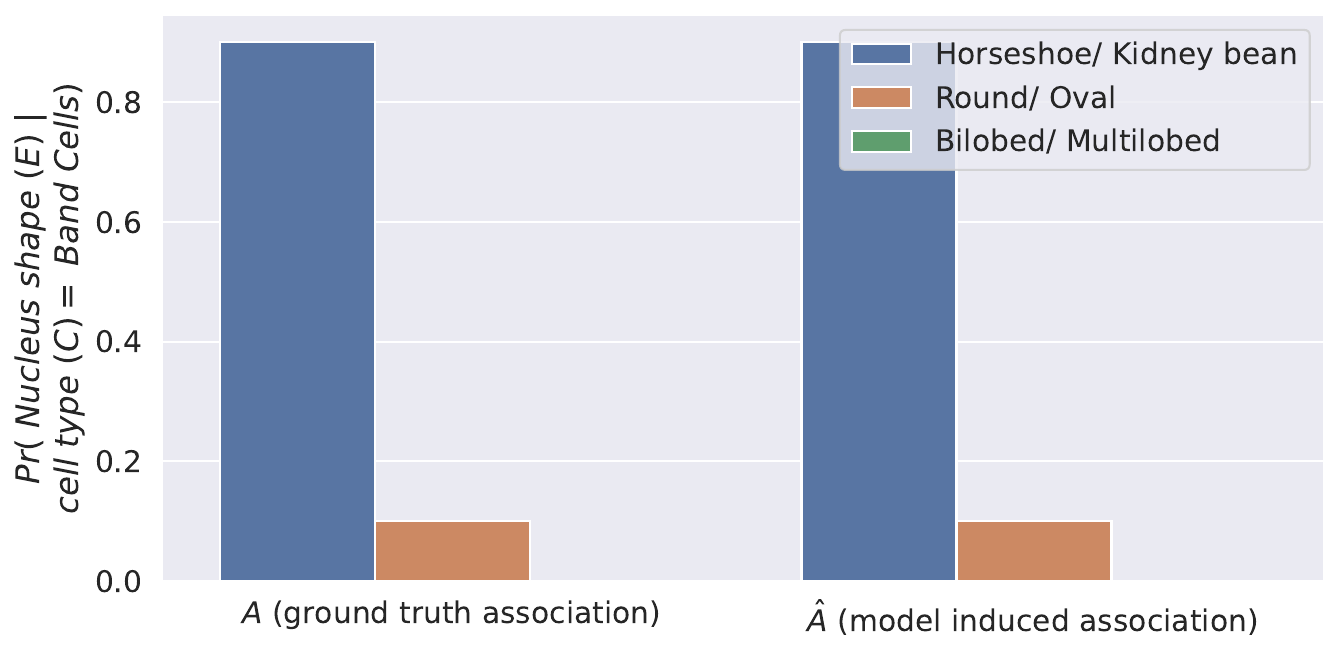}
   \end{minipage}\hfill
   \begin{minipage}{0.48\linewidth}
     \centering
     \includegraphics[width=\linewidth]{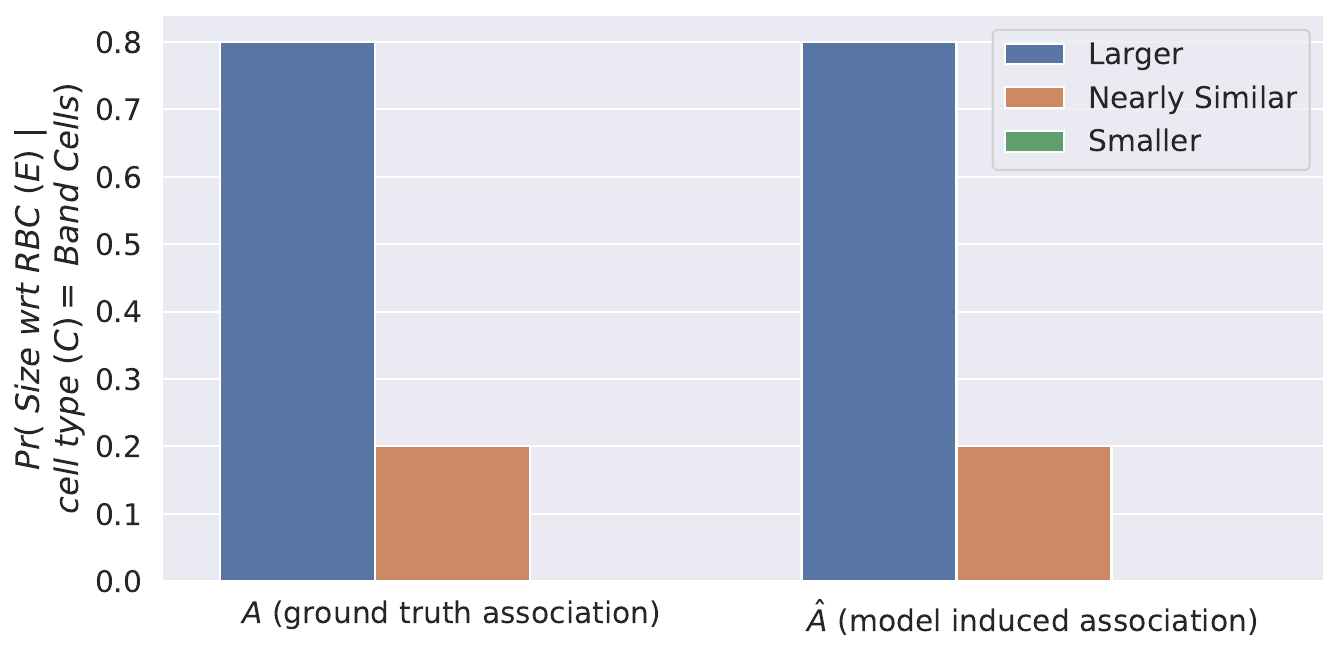}
   \end{minipage}
   \caption{Comparison between model-induced association ($\hat{A}$) and ground truth association ($A$) for all four explanations (`Granularity', `Cytoplasm Color', `Nucleus Shape' and `Size w.r.t. RBC') on `Band Cell' cell type.}
   \label{fig:faithfulness_bandcell}
\end{figure}

\begin{figure}[ht!]
   \begin{minipage}{0.48\linewidth}
     \centering
     \includegraphics[width=\linewidth]{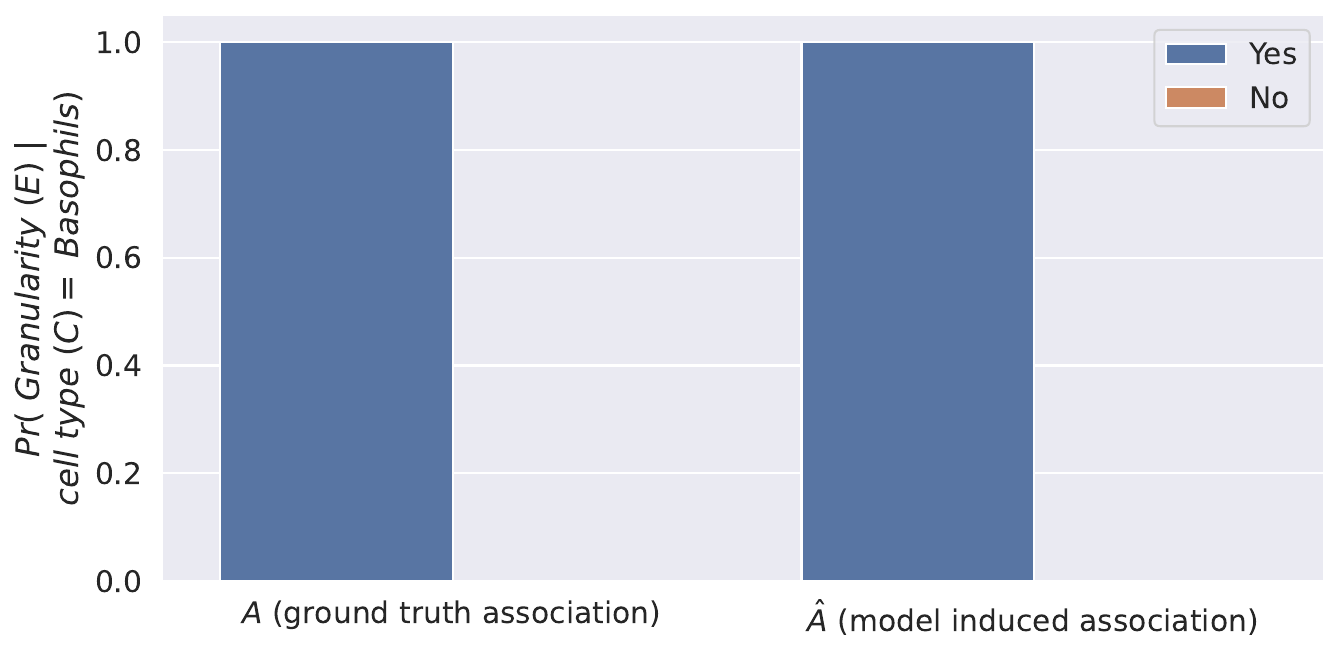}
   \end{minipage}\hfill
   \begin{minipage}{0.48\linewidth}
     \centering
     \includegraphics[width=\linewidth]{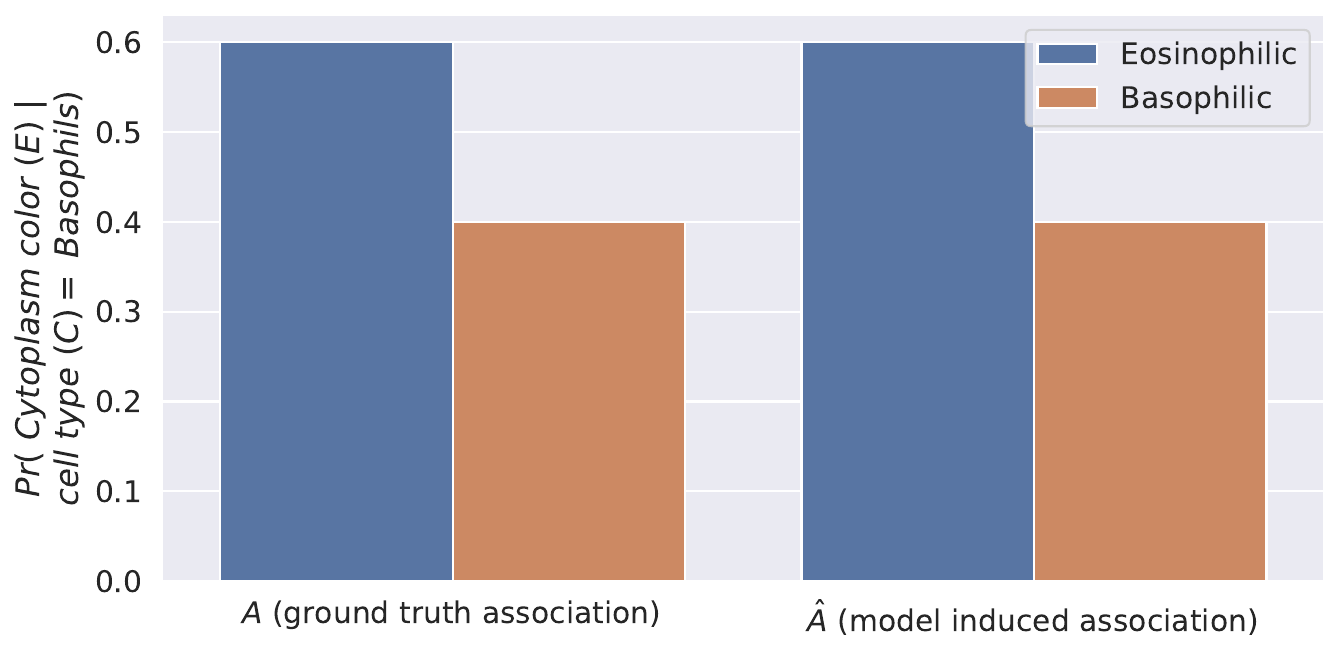}
   \end{minipage}
   \begin{minipage}{0.48\linewidth}
     \centering
     \includegraphics[width=\linewidth]{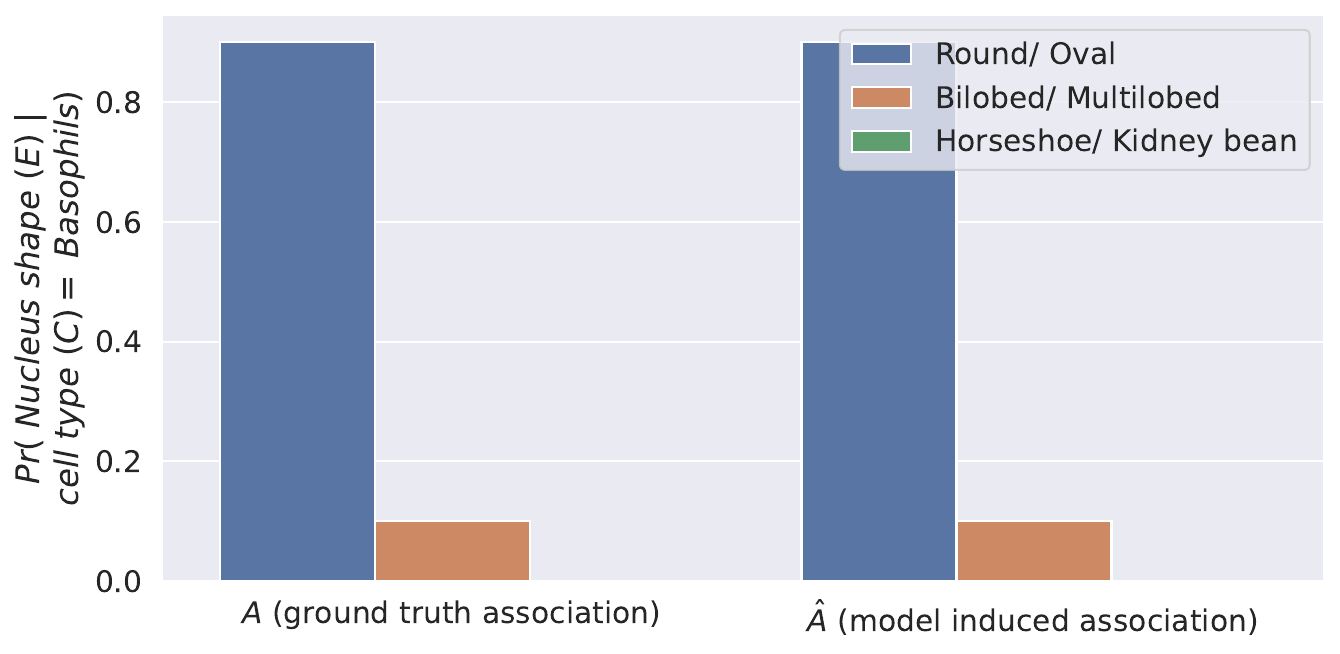}
   \end{minipage}\hfill
   \begin{minipage}{0.48\linewidth}
     \centering
     \includegraphics[width=\linewidth]{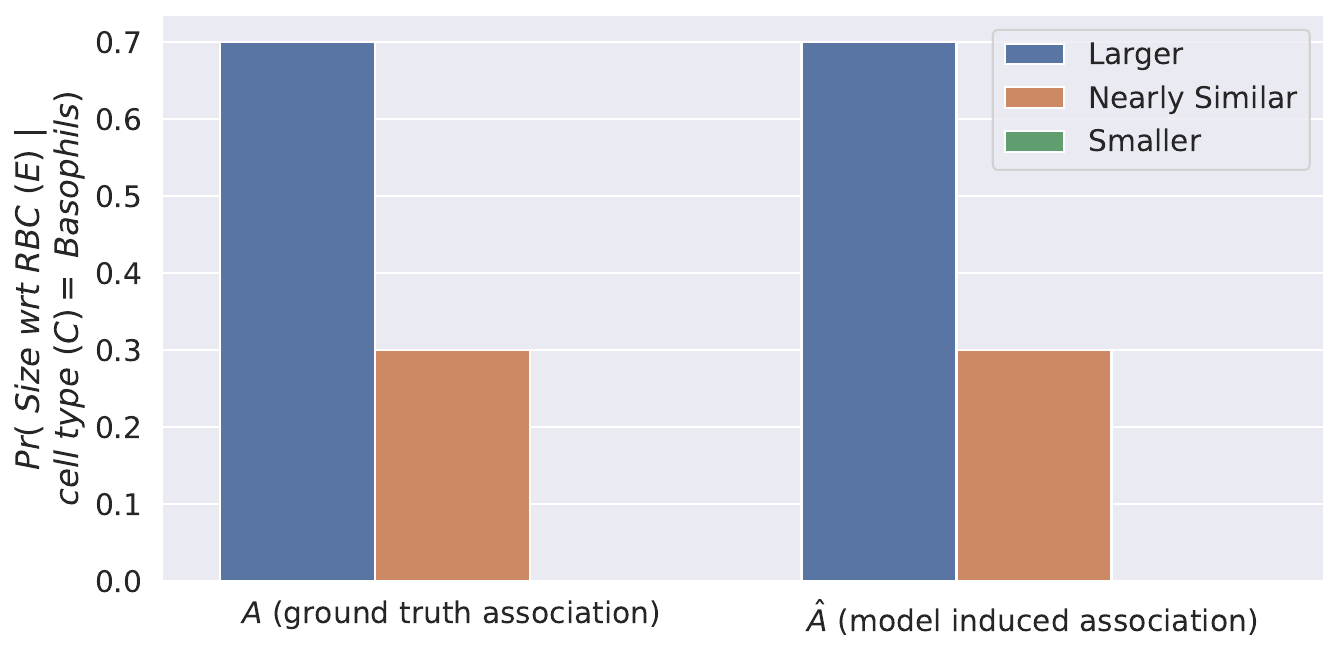}
   \end{minipage}
   \caption{Comparison between model-induced association ($\hat{A}$) and ground truth association ($A$) for all four explanations (`Granularity', `Cytoplasm Color', `Nucleus Shape' and `Size w.r.t. RBC') on `Basophil' cell type.}
   \label{fig:faithfulness_basophil}
\end{figure}

\begin{figure}[ht!]
   \begin{minipage}{0.48\linewidth}
     \centering
     \includegraphics[width=\linewidth]{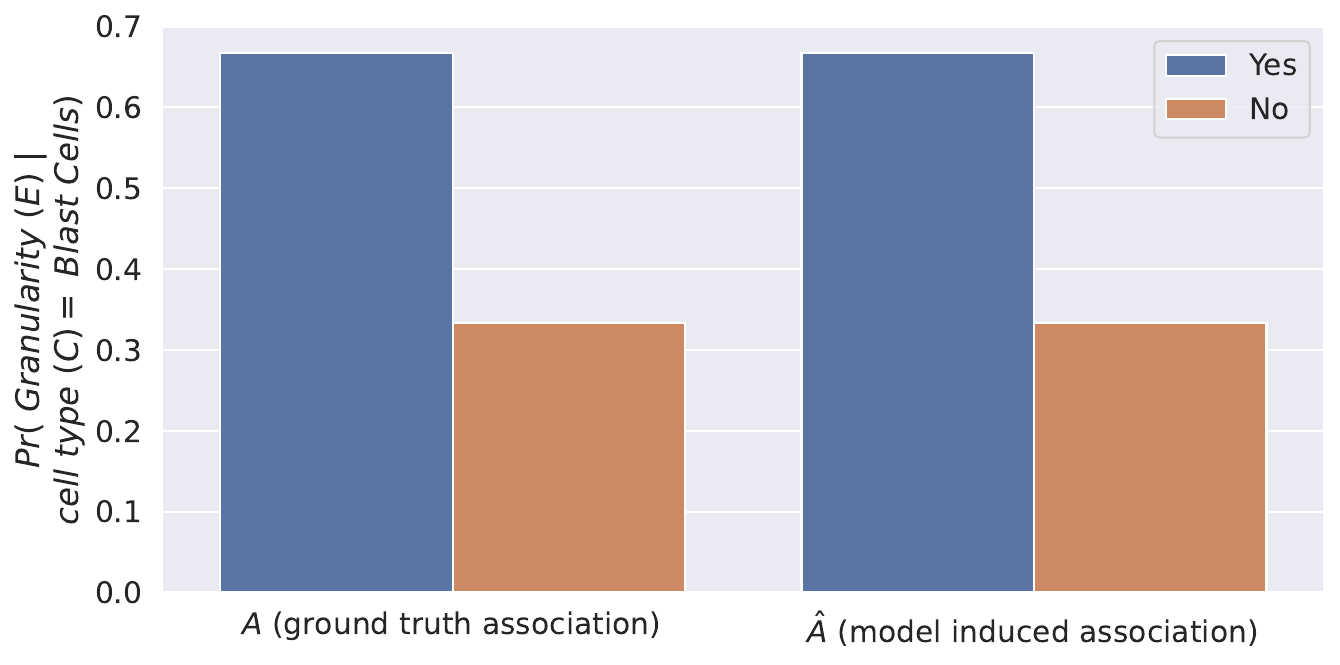}
   \end{minipage}\hfill
   \begin{minipage}{0.48\linewidth}
     \centering
     \includegraphics[width=\linewidth]{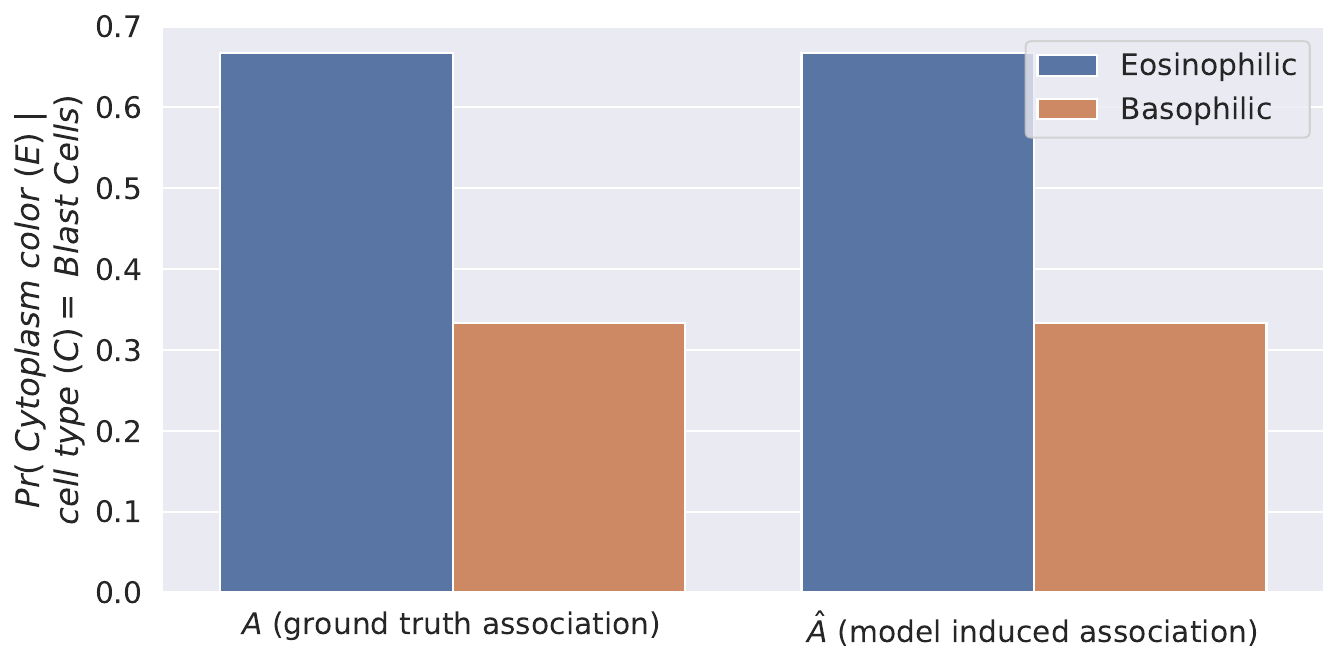}
   \end{minipage}
   \begin{minipage}{0.48\linewidth}
     \centering
     \includegraphics[width=\linewidth]{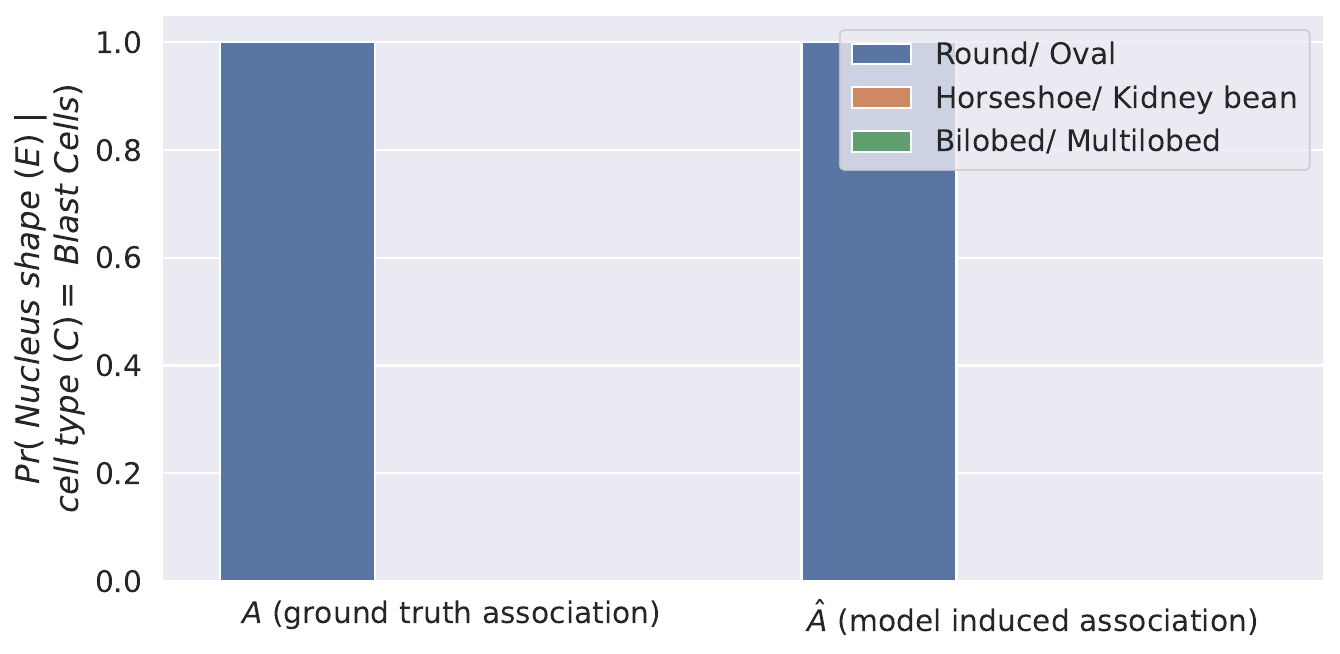}
   \end{minipage}\hfill
   \begin{minipage}{0.48\linewidth}
     \centering
     \includegraphics[width=\linewidth]{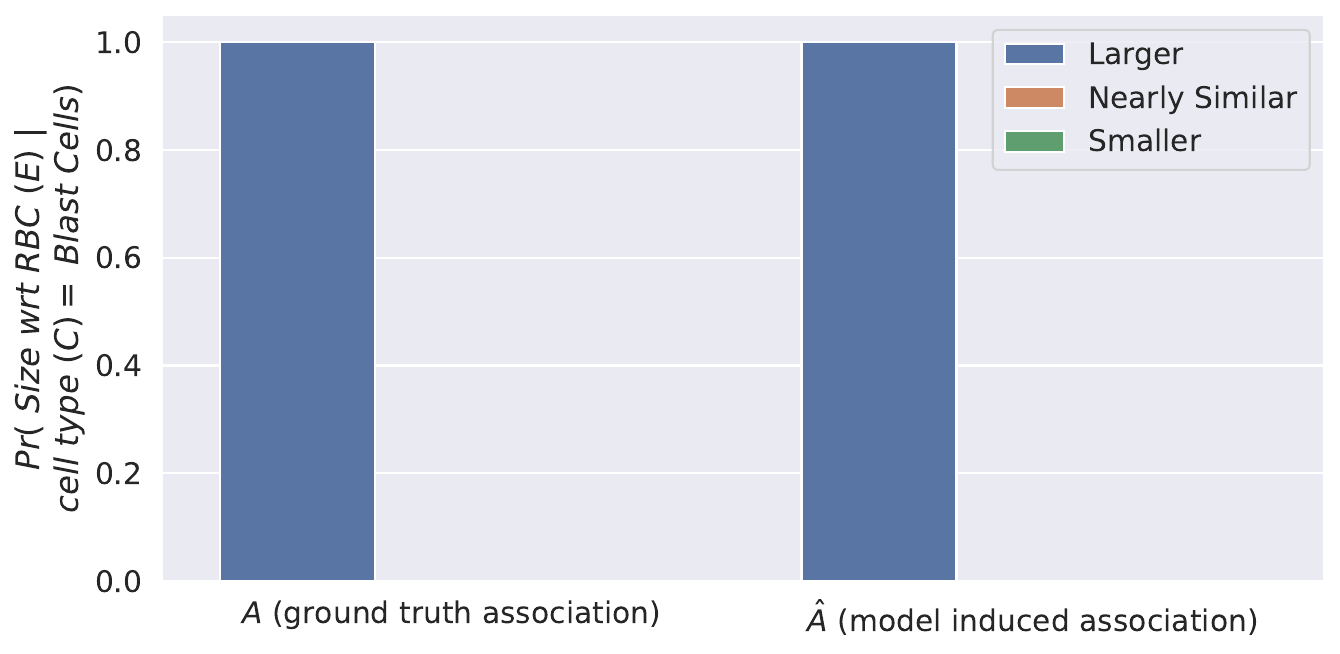}
   \end{minipage}
   \caption{Comparison between model-induced association ($\hat{A}$) and ground truth association ($A$) for all four explanations (`Granularity', `Cytoplasm Color', `Nucleus Shape' and `Size w.r.t. RBC') on `Blast Cell' cell type.}
   \label{fig:faithfulness_blastcell}
\end{figure}

\begin{figure}[ht!]
   \begin{minipage}{0.48\linewidth}
     \centering
     \includegraphics[width=\linewidth]{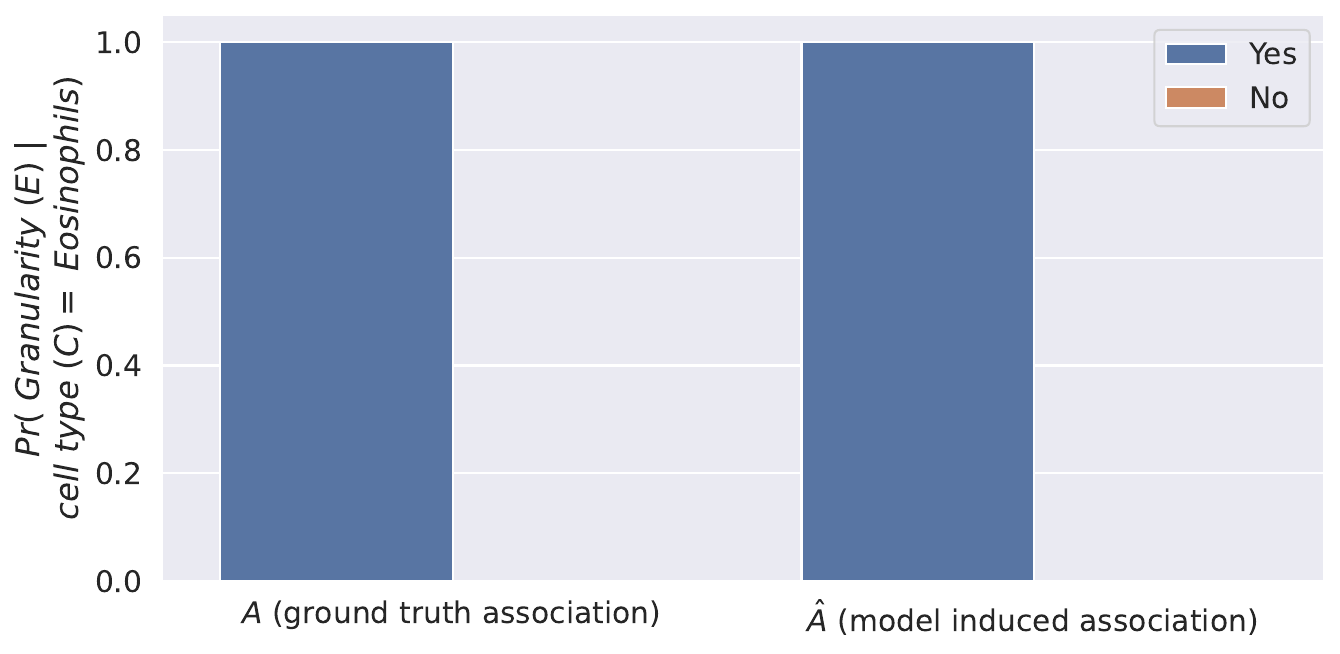}
   \end{minipage}\hfill
   \begin{minipage}{0.48\linewidth}
     \centering
     \includegraphics[width=\linewidth]{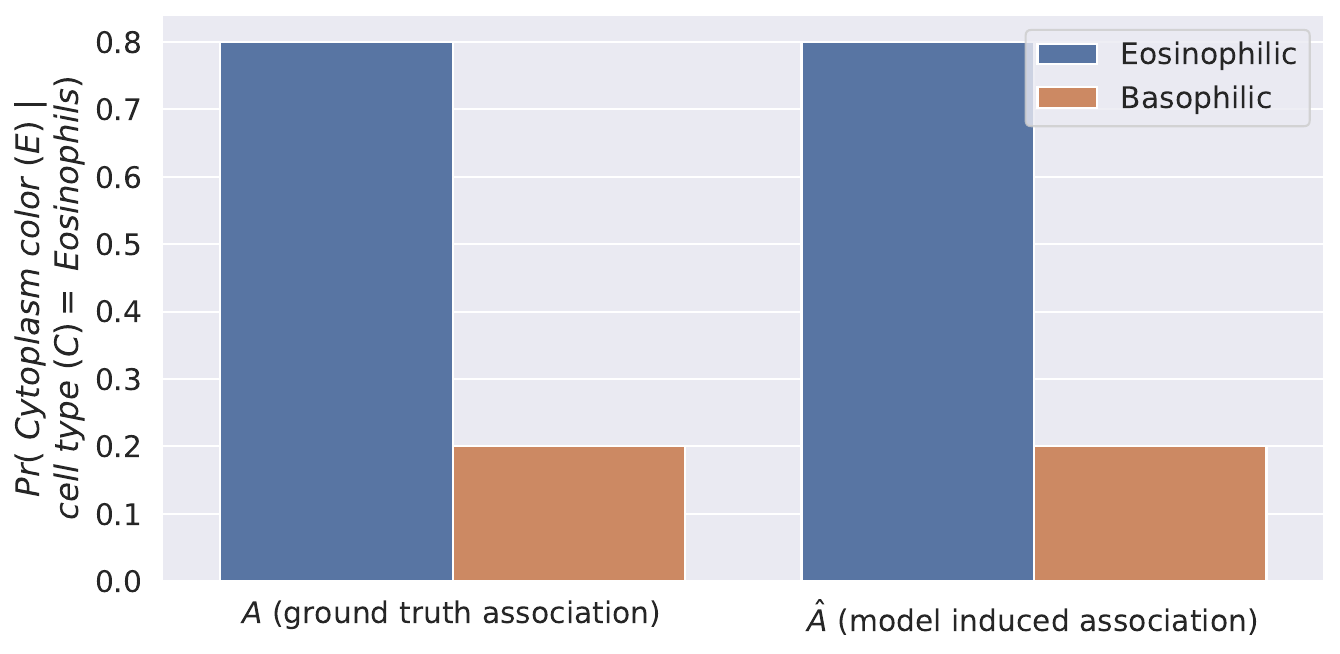}
   \end{minipage}
   \begin{minipage}{0.48\linewidth}
     \centering
     \includegraphics[width=\linewidth]{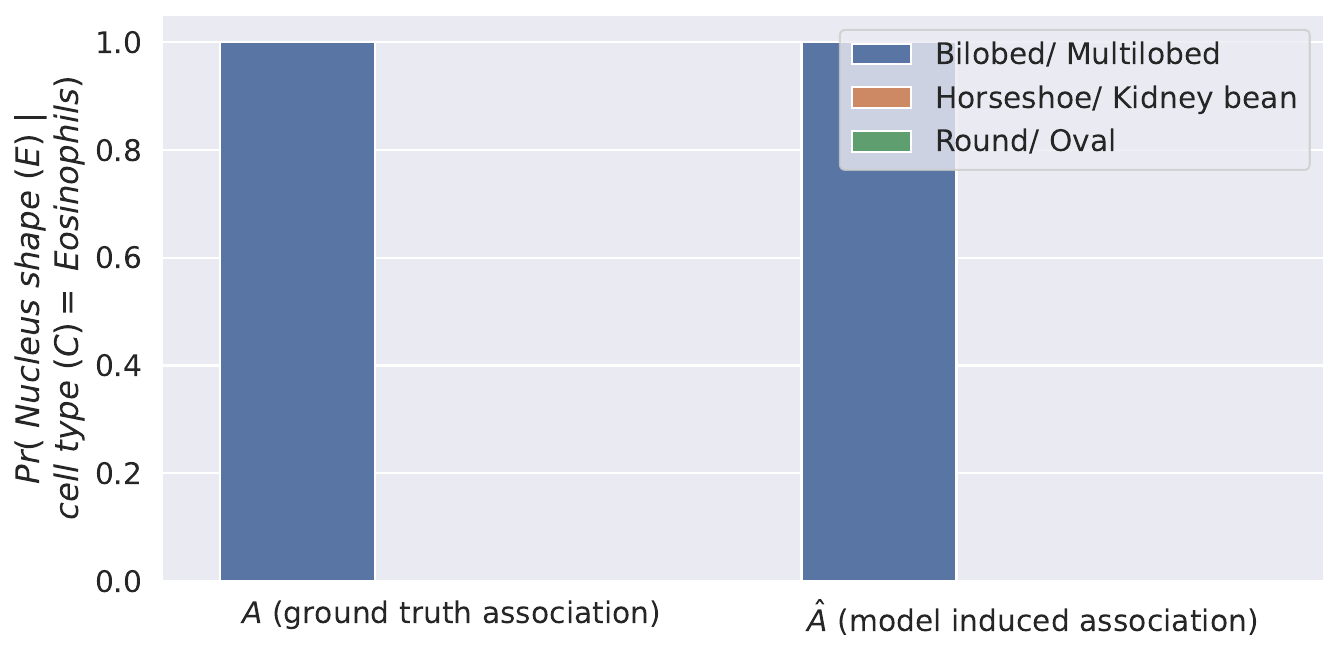}
   \end{minipage}\hfill
   \begin{minipage}{0.48\linewidth}
     \centering
     \includegraphics[width=\linewidth]{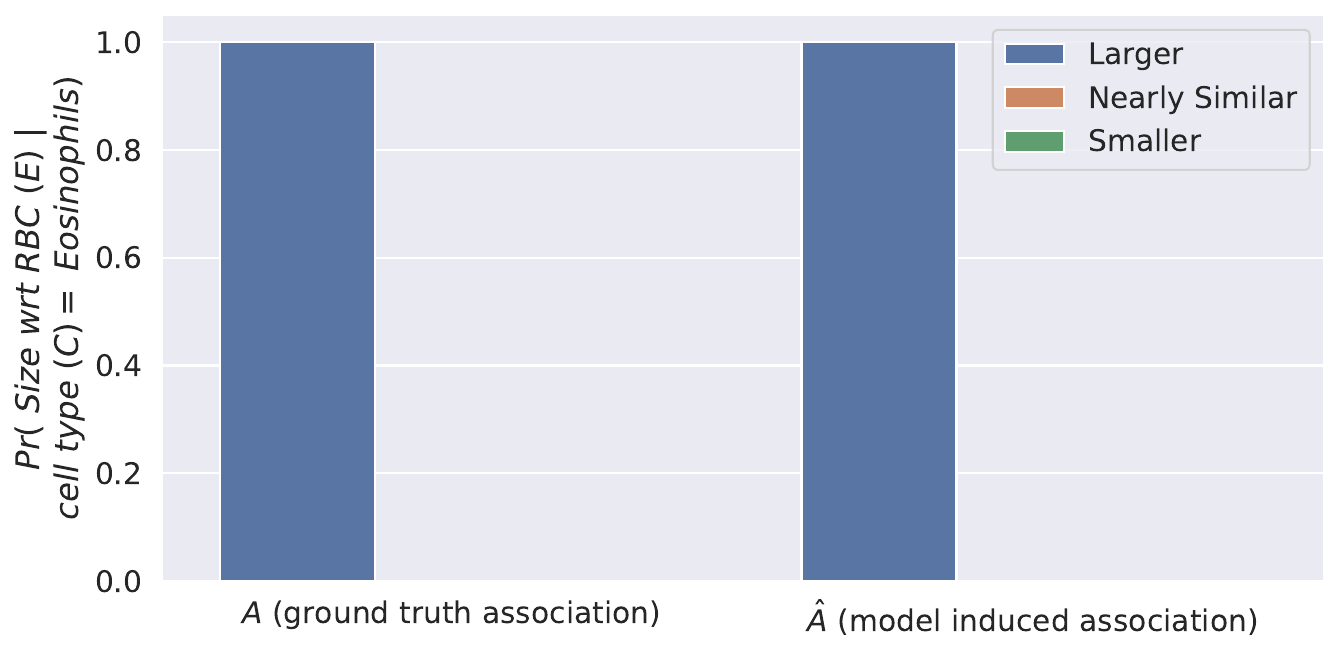}
   \end{minipage}
   \caption{Comparison between model-induced association ($\hat{A}$) and ground truth association ($A$) for all four explanations (`Granularity', `Cytoplasm Color', `Nucleus Shape' and `Size w.r.t. RBC') on `Eosinophil' cell type.}
   \label{fig:faithfulness_eosinophil}
\end{figure}

\begin{figure}[ht!]
   \begin{minipage}{0.48\linewidth}
     \centering
     \includegraphics[width=\linewidth]{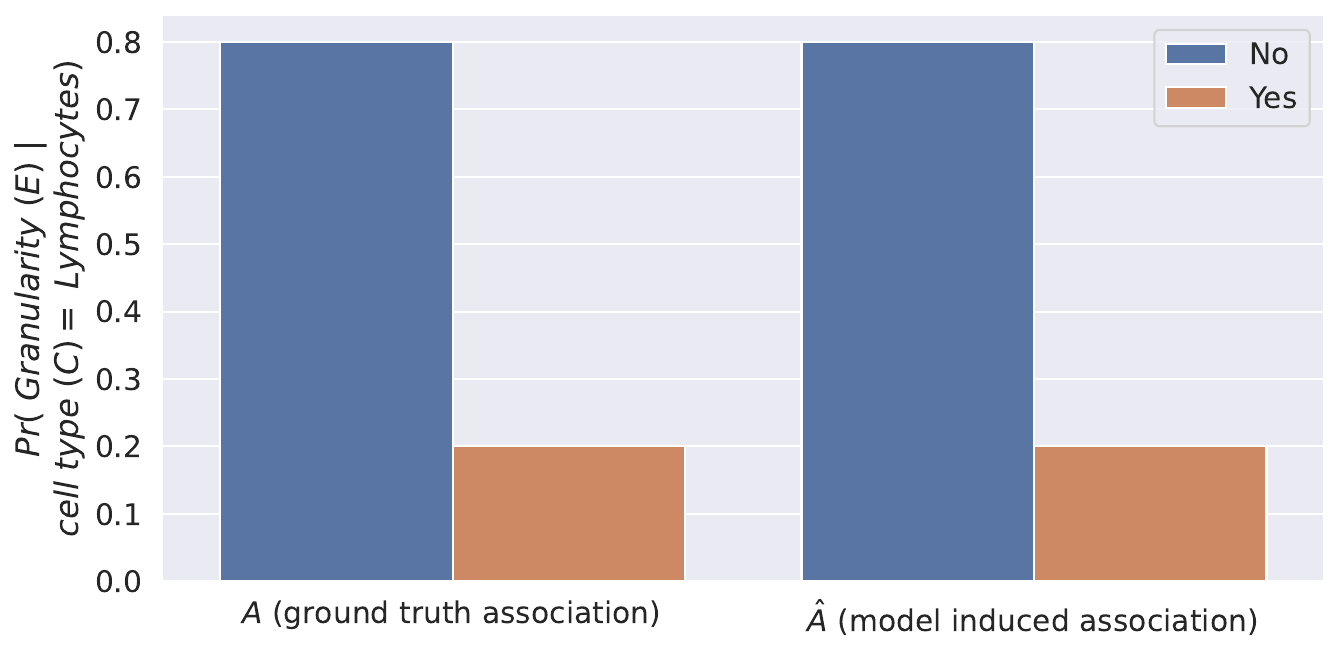}
   \end{minipage}\hfill
   \begin{minipage}{0.48\linewidth}
     \centering
     \includegraphics[width=\linewidth]{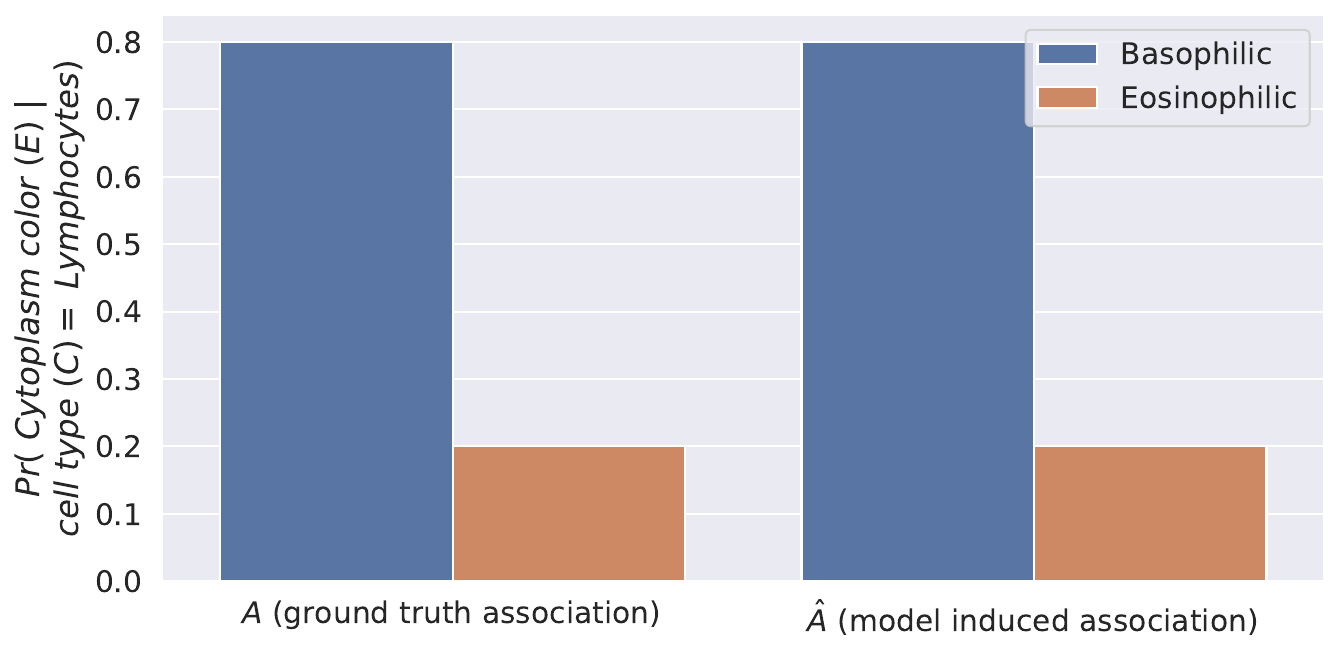}
   \end{minipage}
   \begin{minipage}{0.48\linewidth}
     \centering
     \includegraphics[width=\linewidth]{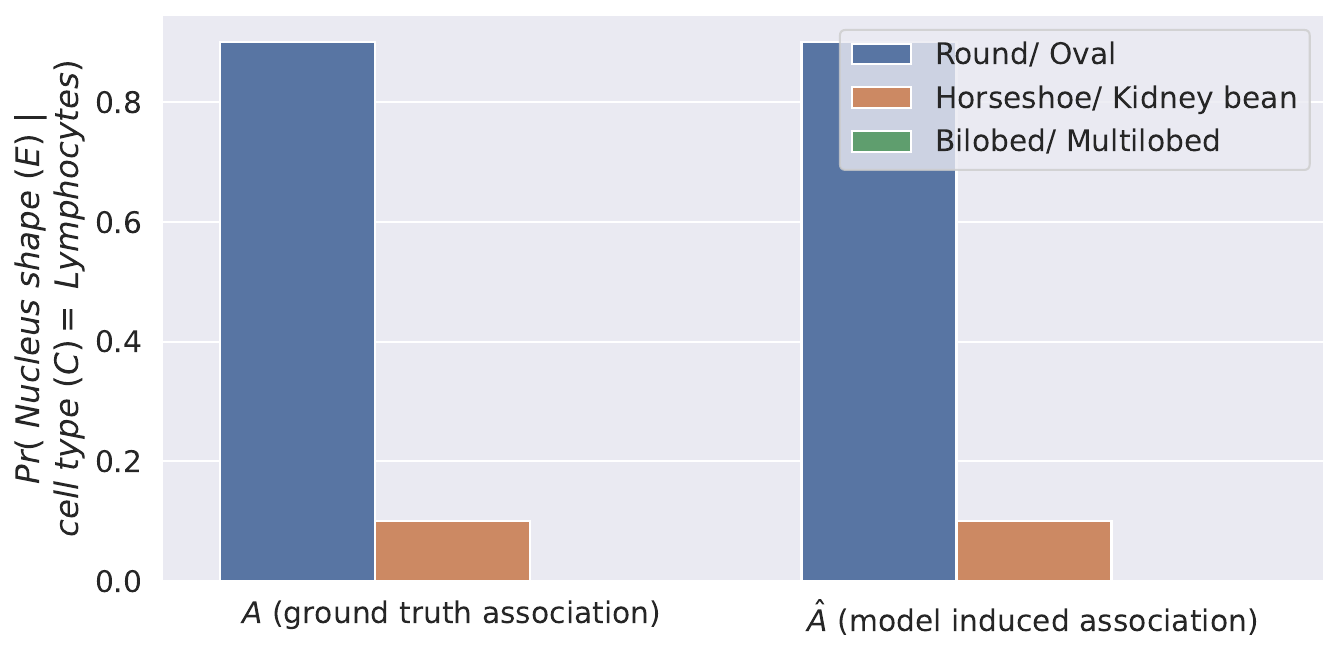}
   \end{minipage}\hfill
   \begin{minipage}{0.48\linewidth}
     \centering
     \includegraphics[width=\linewidth]{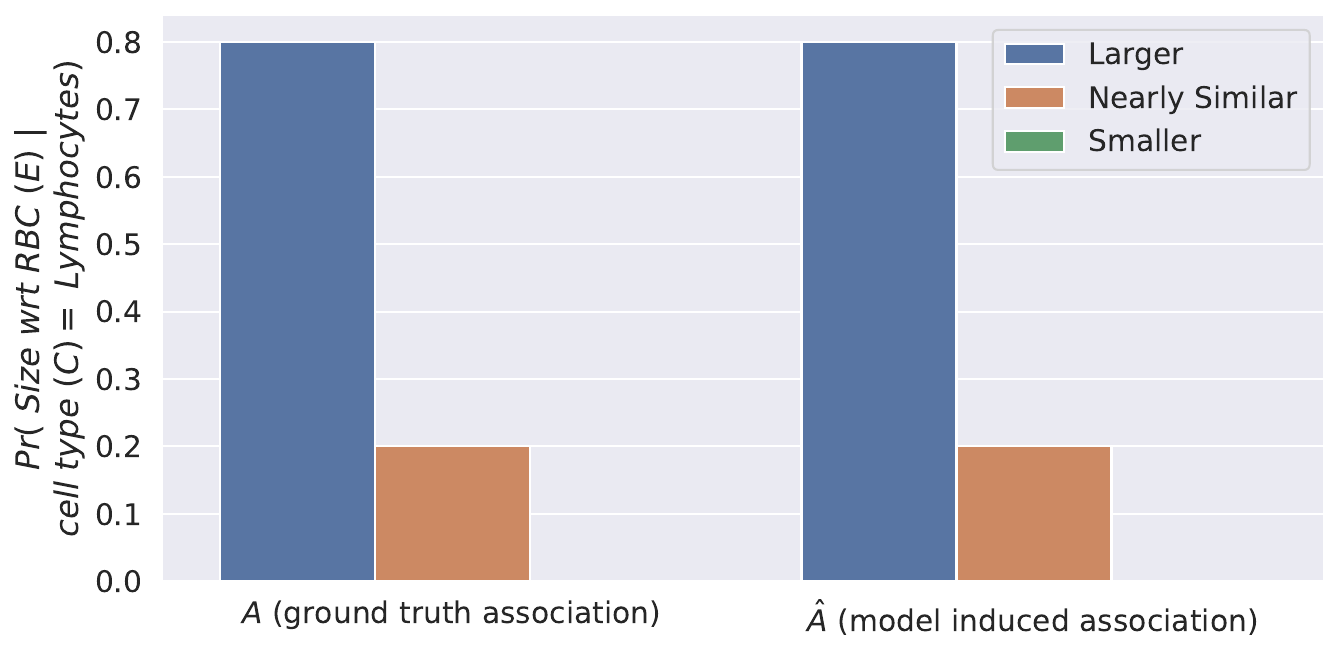}
   \end{minipage}
   \caption{Comparison between model-induced association ($\hat{A}$) and ground truth association ($A$) for all four explanations (`Granularity', `Cytoplasm Color', `Nucleus Shape' and `Size w.r.t. RBC') on `Lymphocyte' cell type.}
   \label{fig:faithfulness_lymphocyte}
\end{figure}

\begin{figure}[ht!]
   \begin{minipage}{0.48\linewidth}
     \centering
     \includegraphics[width=\linewidth]{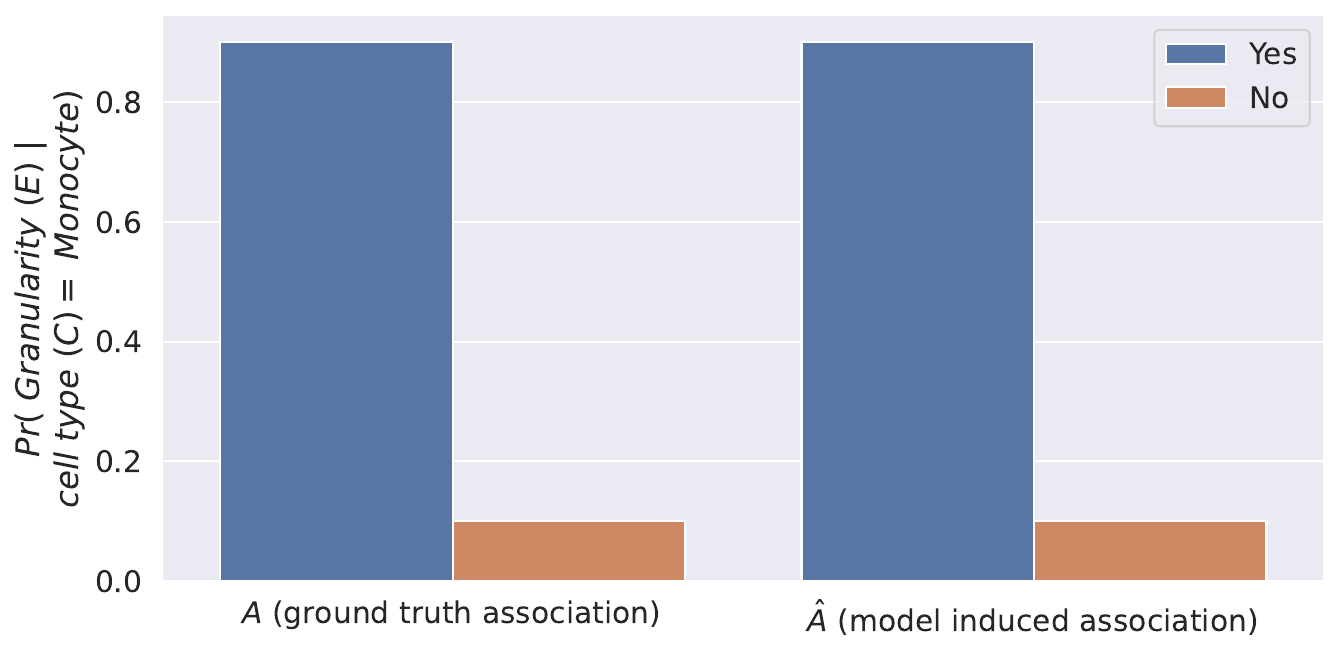}
   \end{minipage}\hfill
   \begin{minipage}{0.48\linewidth}
     \centering
     \includegraphics[width=\linewidth]{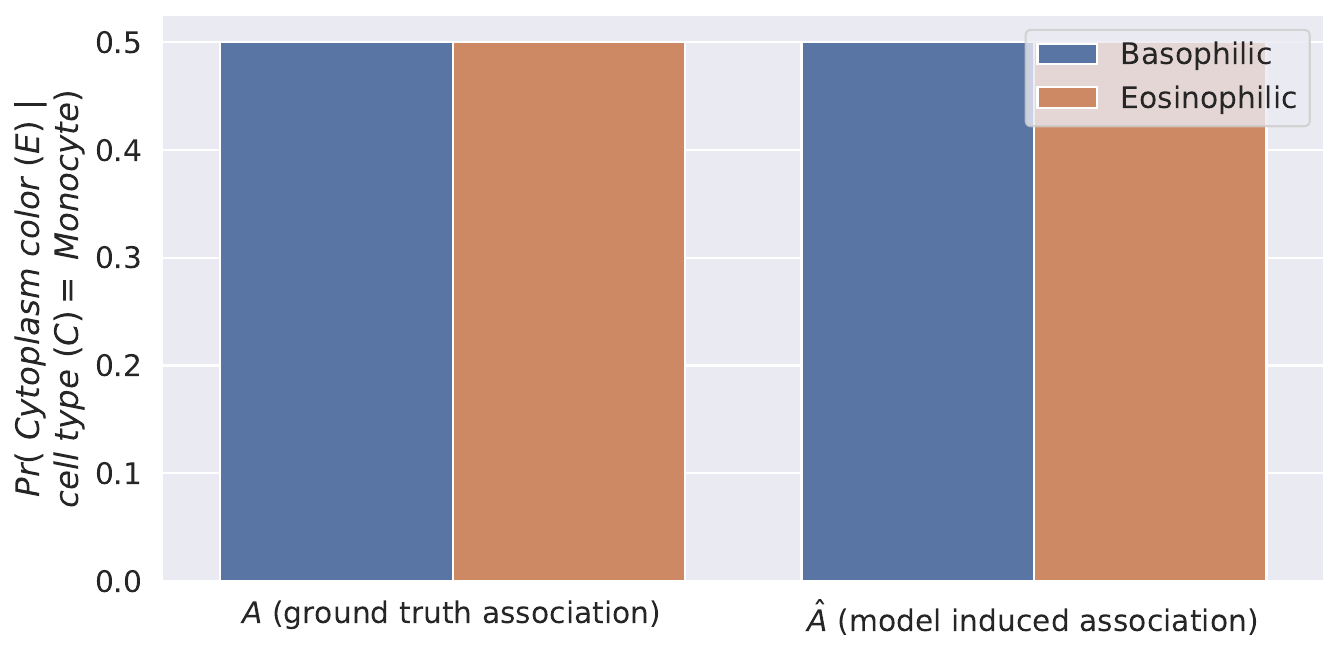}
   \end{minipage}
   \begin{minipage}{0.48\linewidth}
     \centering
     \includegraphics[width=\linewidth]{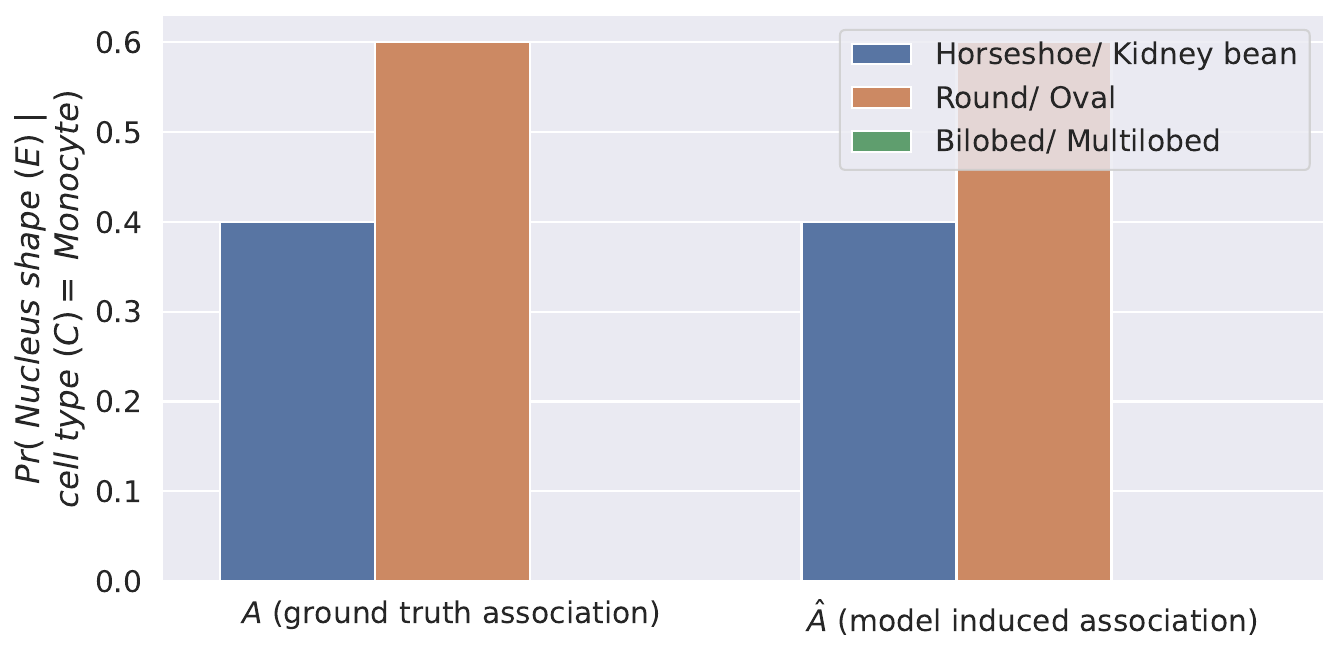}
   \end{minipage}\hfill
   \begin{minipage}{0.48\linewidth}
     \centering
     \includegraphics[width=\linewidth]{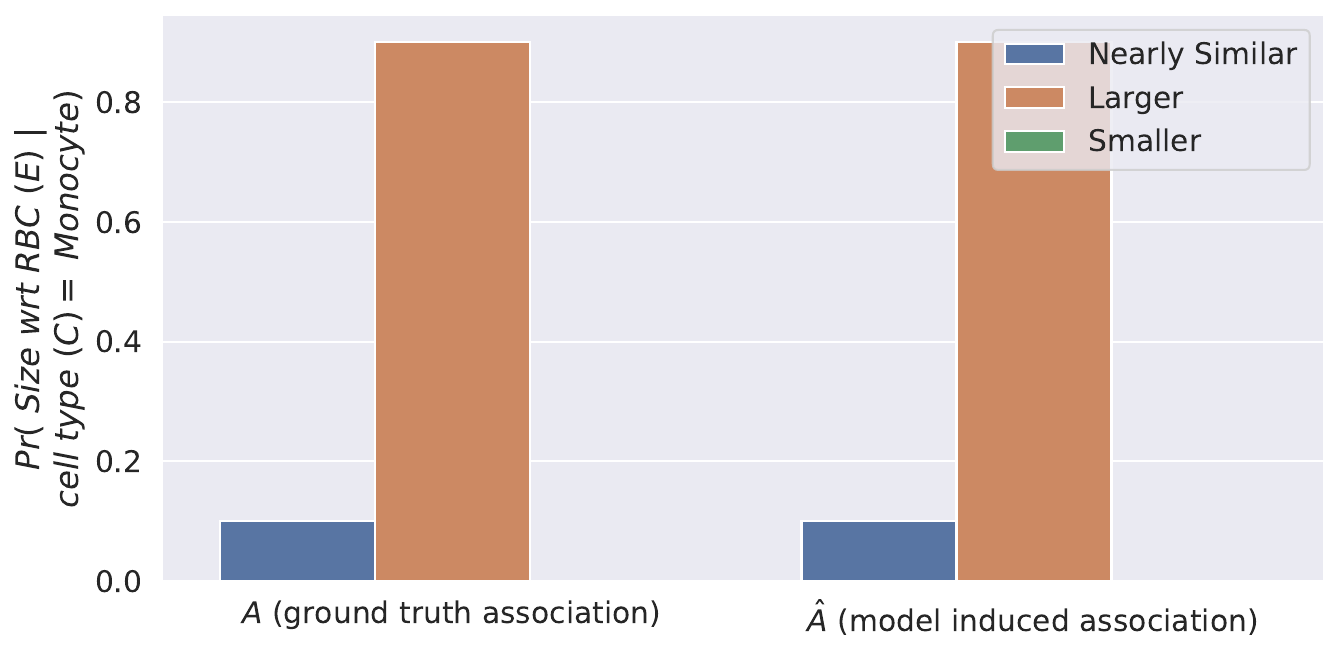}
   \end{minipage}
   \caption{Comparison between model-induced association ($\hat{A}$) and ground truth association ($A$) for all four explanations (`Granularity', `Cytoplasm Color', `Nucleus Shape' and `Size w.r.t. RBC') on `Monocyte' cell type.}
   \label{fig:faithfulness_monocyte}
\end{figure}

\begin{figure}[ht!]
   \begin{minipage}{0.48\linewidth}
     \centering
     \includegraphics[width=\linewidth]{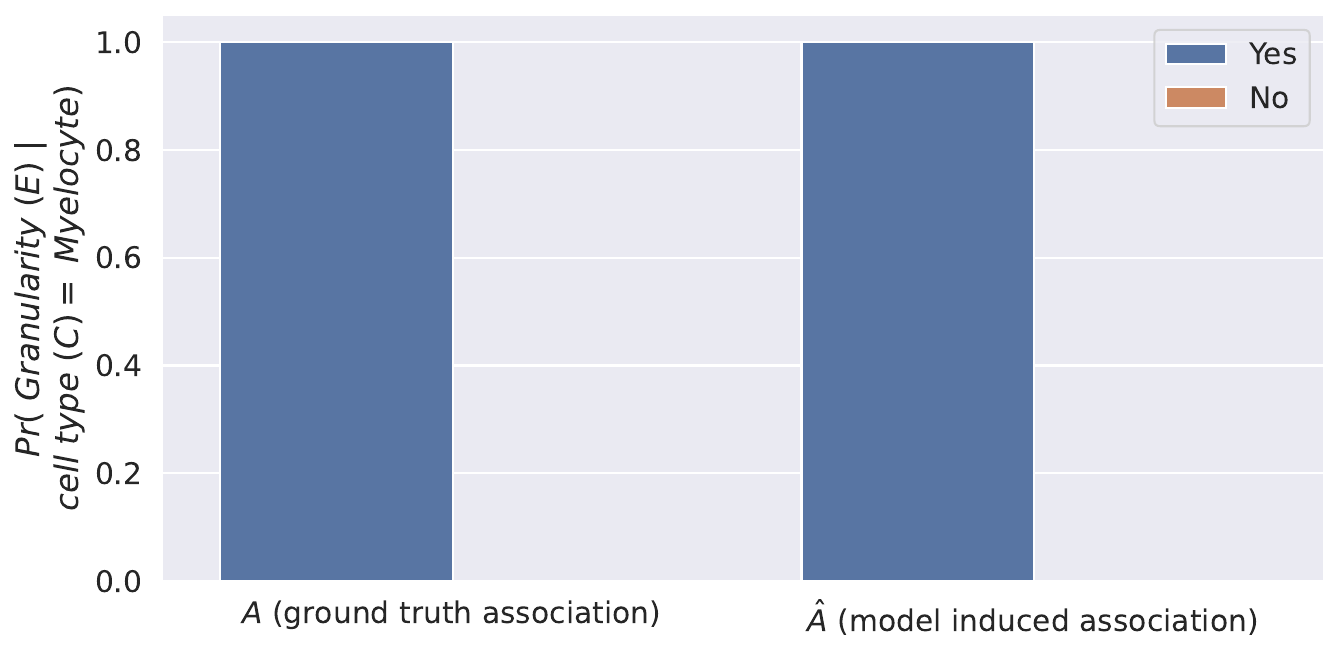}
   \end{minipage}\hfill
   \begin{minipage}{0.48\linewidth}
     \centering
     \includegraphics[width=\linewidth]{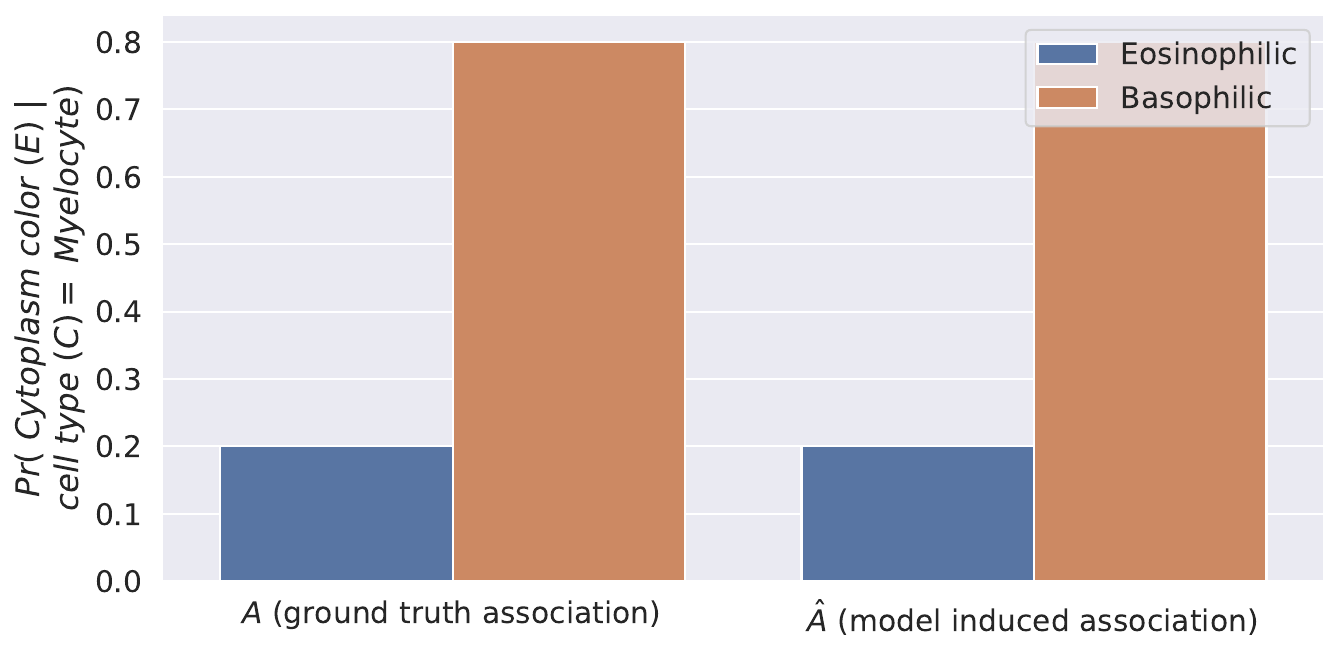}
   \end{minipage}
   \begin{minipage}{0.48\linewidth}
     \centering
     \includegraphics[width=\linewidth]{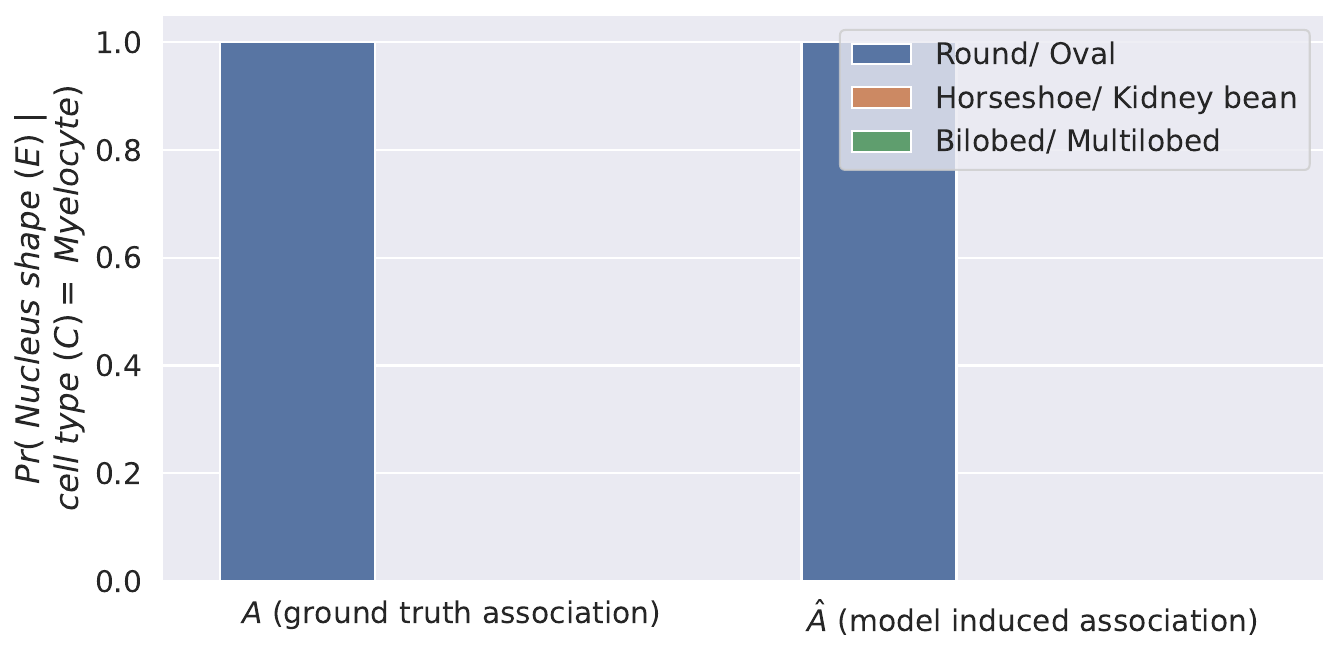}
   \end{minipage}\hfill
   \begin{minipage}{0.48\linewidth}
     \centering
     \includegraphics[width=\linewidth]{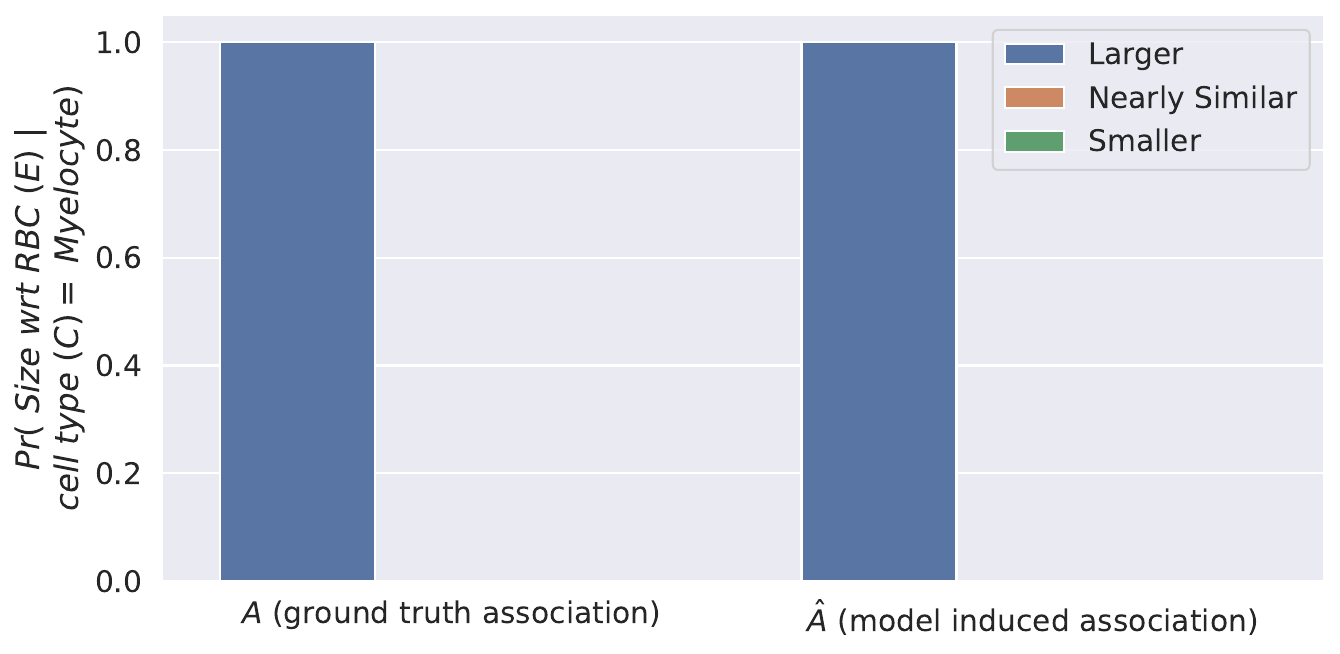}
   \end{minipage}
   \caption{Comparison between model-induced association ($\hat{A}$) and ground truth association ($A$) for all four explanations (`Granularity', `Cytoplasm Color', `Nucleus Shape' and `Size w.r.t. RBC') on `Myelocyte' cell type.}
   \label{fig:faithfulness_myelocyte}
\end{figure}

In the discussion section of the paper, we show the comparison between model-induced association ($\hat{A}$) and ground truth association ($A$) for four explanations only on two among the ten cell types.
Here we provide such comparision for rest of the eight (8) cell type, namely---`Band Cells' (Fig.~\ref{fig:faithfulness_bandcell}), `Basophils' (Fig.~\ref{fig:faithfulness_basophil}), `Blast Cells' (Fig.~\ref{fig:faithfulness_blastcell}), `Eosinophils' (Fig.~\ref{fig:faithfulness_eosinophil}), `Lymphocytes' (Fig.~\ref{fig:faithfulness_lymphocyte}), `Monocytes' (Fig.~\ref{fig:faithfulness_monocyte}), and `Myelocytes' (Fig.~\ref{fig:faithfulness_myelocyte}).
All these figures support the conclusion that we reached in the discussion section regarding the faithfulness of explanation prediction.

\vspace*{20cm}
\section*{Part-2: HemaX Results}
\begin{figure*}[ht!]
   \begin{minipage}{\textwidth}
     \centering
     \includegraphics[width=\linewidth]{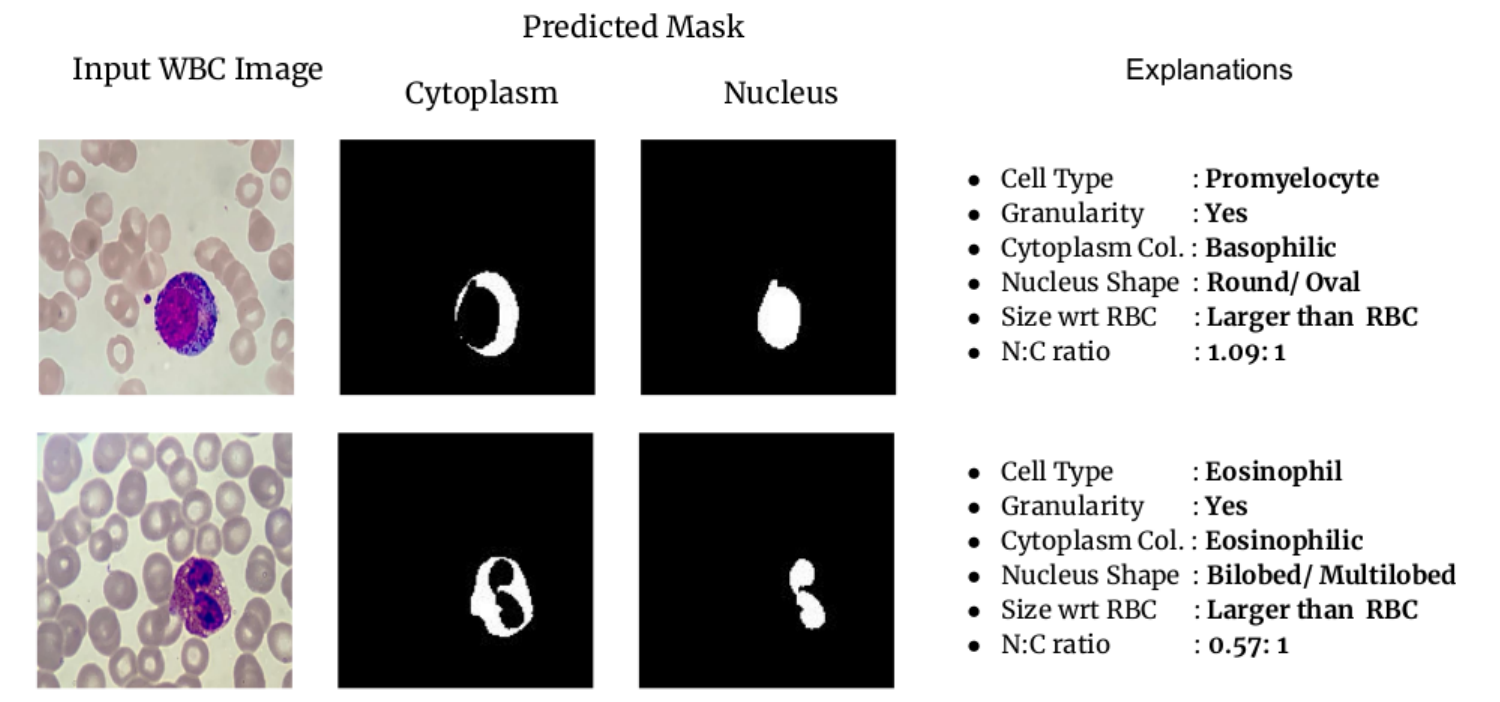}
   \end{minipage}
   \begin{minipage}{\textwidth}
     \centering
     \includegraphics[width=\linewidth]{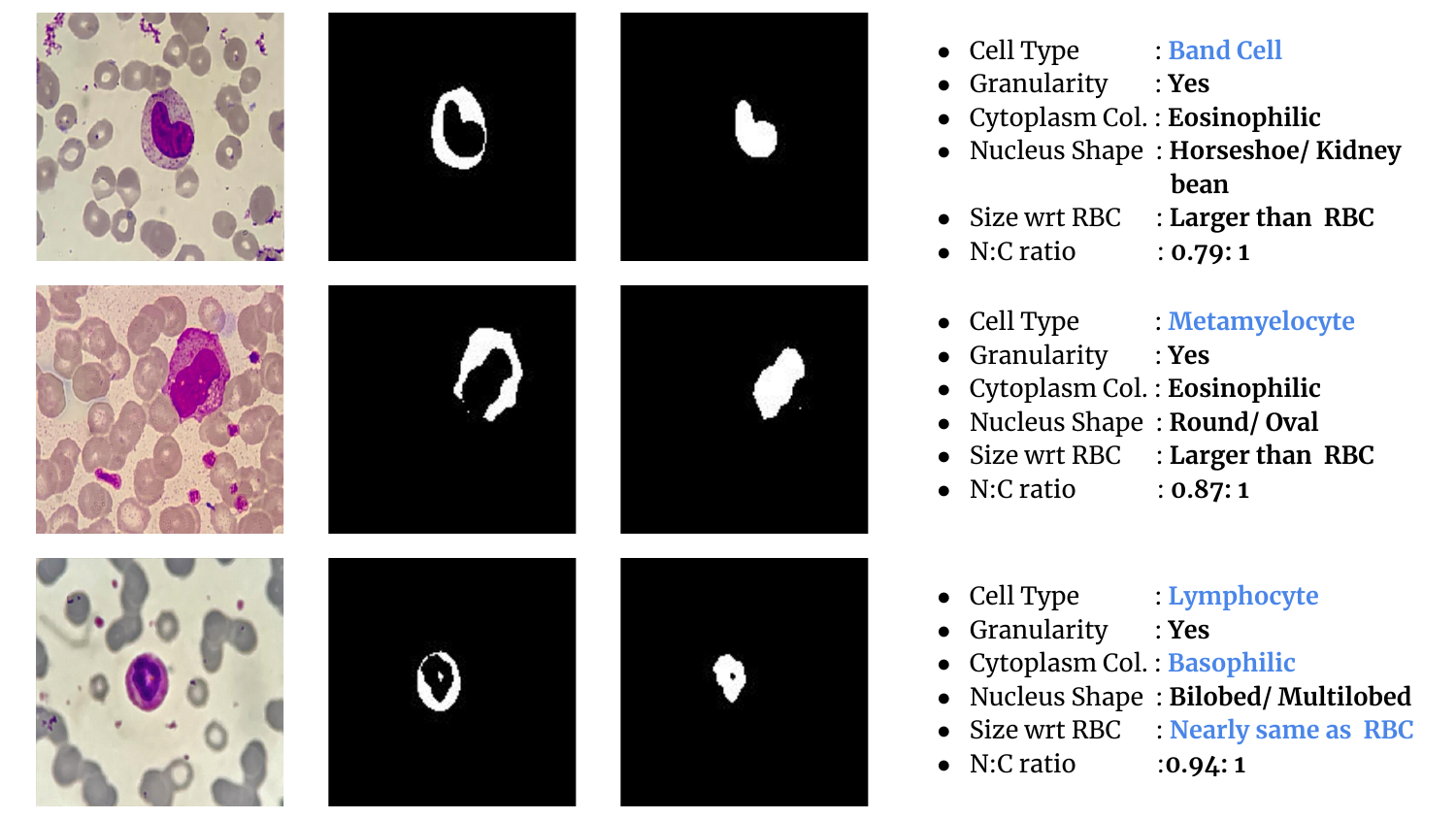}
   \end{minipage}
   \caption{\textbf{Results on LeukoX dataset:} We present HemaX's output on sample images from LeukoX dataset. The misclassified results are marked in blue.}\label{fig:add_res}
\end{figure*}
Figure~\ref{fig:add_res} shows some additional results on LeukoX produced by HemaX. The initial two rows exhibit accurately predicted WBCs with their accompanying explanations, while the subsequent three rows demonstrate instances where the model failed. In the third and fourth rows, examples are presented where the top-performing HemaX model misclassified labels but still provided accurate explanations for the cells.

\end{document}